\RequirePackage{silence}
\documentclass[fleqn,usenatbib,useAMS]{mnras} 
\usepackage{graphicx}
\usepackage{amsmath}

\usepackage{amsfonts}
\usepackage{float}
\usepackage{bm}
\usepackage{ae,aecompl}
\usepackage{array}
\usepackage{soul}
\usepackage{mathtools}
\usepackage{multirow}
\usepackage[utf8]{inputenc}
\usepackage{booktabs}
\usepackage{graphicx}

\newcommand{\appropto}{\mathrel{\vcenter{
		\offinterlineskip\halign{\hfil$##$\cr 
			\propto\cr\noalign{\kern2pt}\sim\cr\noalign{\kern-2pt}}}}}

\DeclareRobustCommand*{\matr}[1]{\mathbfss{#1}}

\title[The tidal stability of Fornax Cluster dwarfs]{The distribution and morphologies of Fornax Cluster dwarf galaxies suggest they lack dark matter} 

\author[E. Asencio et al.]{Elena Asencio$^{1}$\thanks{Email: \href{mailto:s6elena@uni-bonn.de}{s6elena@uni-bonn.de} (Elena Asencio), \newline $~~~~~~~~~~$ \href{mailto:ib45@st-andrews.ac.uk}{ib45@st-andrews.ac.uk} (Indranil Banik)}, Indranil Banik$^{2}$, Steffen Mieske$^{3}$, Aku Venhola$^{4}$, Pavel Kroupa$^{1,5}$
\newauthor
and Hongsheng Zhao$^{2}$ \vspace{10pt} \\
$^{1}$Helmholtz-Institut f\"ur Strahlen und Kernphysik (HISKP), University of Bonn, Nussallee 14$-$16, D-53115 Bonn, Germany \\
$^{2}$Scottish Universities Physics Alliance, University of Saint Andrews, North Haugh, Saint Andrews, Fife, KY16 9SS, UK \\
$^{3}$European Southern Observatory, Alonso de Cordova 3107, Vitacura, Santiago, Chile \\
$^{4}$Space Physics and Astronomy Research Unit, University of Oulu, Pentti Kaiteran katu 1, 90014 Oulu, Finland \\
$^{5}$Astronomical Institute, Faculty of Mathematics and Physics, Charles University, V Hole\v{s}ovi\v{c}k\'ach 2, CZ-180 00 Praha 8, Czech Republic}

\pubyear{2022}
\begin{document}
\label{firstpage}
\pagerange{\pageref{firstpage}--\pageref{lastpage}}

\maketitle

\begin{abstract} 
Due to their low surface brightness, dwarf galaxies are particularly susceptible to tidal forces. The expected degree of disturbance depends on the assumed gravity law and whether they have a dominant dark halo. This makes dwarf galaxies useful for testing different gravity models. In this project, we use the Fornax Deep Survey (FDS) dwarf galaxy catalogue to compare the properties of dwarf galaxies in the Fornax Cluster with those predicted by the Lambda cold dark matter ($\Lambda$CDM) standard model of cosmology and Milgromian dynamics (MOND). We construct a test particle simulation of the Fornax system. We then use the MCMC method to fit this to the FDS distribution of tidal susceptibility $\eta$ (half-mass radius divided by theoretical tidal radius), the fraction of dwarfs that visually appear disturbed as a function of $\eta$, and the distribution of projected separation from the cluster centre. This allows us to constrain the $\eta$ value at which dwarfs should get destroyed by tides. Accounting for an $r'$-band surface brightness limit of 27.8 magnitudes per square arcsec, the required stability threshold is $\eta_{\textrm{destr}} = 0.25^{+0.07}_{-0.03}$ in $\Lambda$CDM and $ 1.88^{+0.85}_{-0.53}$ in MOND. The $\Lambda$CDM value is in tension with previous \textit{N}-body dwarf galaxy simulations, which indicate that $\eta_{\textrm{destr}} \approx 1$. Our MOND \textit{N}-body simulations indicate that $\eta_{\textrm{destr}} = 1.70 \pm 0.30$, which agrees well with our MCMC analysis of the FDS. We therefore conclude that the observed deformations of dwarf galaxies in the Fornax Cluster and the lack of low surface brightness dwarfs towards its centre are incompatible with $\Lambda$CDM expectations but well consistent with MOND.

\end{abstract}


\section{Introduction}
\label{Introduction}

Dwarf galaxies are the smallest and most common type of galaxy. They are characterized by their low mass ($M < 10^9 \, M_{\odot}$) and low metallicity. Most dwarfs are found in galaxy clusters or near a larger galaxy, making them potentially susceptible to the gravitational effect of these larger structures. The currently standard Lambda-cold dark matter ($\Lambda$CDM) cosmological model \citep{Efstathiou_1990, Ostriker_1995} provides two different scenarios by which dwarf galaxies can form \citep[the Dual Dwarf Galaxy Theorem;][]{Kroupa_2012}:
\begin{enumerate}
    \item From the collapse of dark matter particles into haloes, which then accrete baryonic matter into their potential wells \citep{White_1978}. Such dwarfs are known as `primordial dwarf galaxies' and are expected to be dark matter-dominated; and
    \item From the collapse of overdense regions in tidal tails generated by an interaction between larger, gas-rich galaxies. These so-called `tidal dwarf galaxies' (TDGs) must be free of dark matter as the velocity dispersion of the dark matter particles surrounding the host galaxy is too high to allow for their efficient capture by the shallow potential wells of substructures in the tidal tail \citep{Barnes_1992, Wetzstein_2007}. In recent years, cosmological $\Lambda$CDM simulations have advanced to the point where they can resolve TDGs \citep{Ploeckinger_2018, Haslbauer_2019_TDG}.
\end{enumerate}

Dwarf galaxies can also be classified according to their morphology into early and late types depending on whether they have star-forming regions, which are present only for late-type dwarfs. This category includes blue compact dwarfs and dwarf irregular galaxies like the Magellanic Clouds, while early-type dwarfs include dwarf elliptical (dE) and dwarf spheroidal (dSph) galaxies, with dSphs generally having a lower stellar mass ($M_{\star}$). The lowest $M_{\star}$ dwarfs tend to have velocity dispersions ($\sigma$) which are too high if one assumes virial equilibrium, with $\sigma$ sometimes even exceeding the escape velocity \citep{Aaronson_1983, Grebel_2001}.

This discrepancy relies on the validity of General Relativity and our ability to detect nearly all the matter. $\Lambda$CDM is a cosmological model based on General Relativity in which the addition of the dark matter component was motivated by the mismatch between the observed baryonic mass and the mass calculated dynamically from the observed $\sigma$ assuming the virial theorem \citep{Zwicky_1933}. Such acceleration discrepancies are also apparent in the gravity between the Milky Way (MW) and Andromeda \citep[M31;][]{Kahn_Woltjer_1959} and in the outer rotation curves of galaxies \citep[e.g.,][]{Babcock_1939, Rubin_1970, Rogstad_1972, Roberts_1975, Bosma_1978, Bosma_1981}, as reviewed in \citet{Faber_1979}. Therefore, the natural $\Lambda$CDM explanation for dSphs having such high $\sigma$ is to assume that most of their mass is in the form of dark matter, in which case they must be primordial dwarfs.

$\Lambda$CDM predicts that primordial dwarfs should be distributed nearly isotropically around galaxies \citep{Moore_1999, Gao_2004}. However, the dwarf satellite galaxies of the MW, M31, and Centaurus A preferentially align in flattened planes \citep{Lynden_Bell_1976, Ibata_2013, Tully_2015_Cen_A, Muller_2018}. This is in significant tension with the $\Lambda$CDM model \citep{Kroupa_2005}. While it was later shown that the distribution of dark matter subhaloes is not supposed to be exactly isotropic due to the preferential accretion of subhaloes along cosmic filaments and the intrinsic triaxiality of dark matter haloes \citep{Libeskind_2005, Zentner_2005}, the mild expected flattening is not sufficient to explain the strong correlation in position and velocity space observed in nearby satellite systems \citep{Ibata_2014, Pawlowski_2014, Pawlowski_2020, Pawlowski_Sohn_2021, Muller_2021}. The satellite plane problem is reviewed in \citet{Pawlowski_2021}, which also considers tentative evidence for more satellite planes beyond the three mentioned above. The Local Group (LG) satellite planes are each in $3.55\sigma$ tension with $\Lambda$CDM \citep[table 3 of][and references therein]{Banik_2021_backsplash}, while the satellite plane around Centaurus A is only 0.2\% ($3.09\sigma$) likely to arise in this paradigm \citep{Muller_2021}. These are the only three host galaxies near enough for us to reliably know the phase-space distribution of their satellites. We can approximately combine their low likelihoods in $\Lambda$CDM using Gaussian statistics. Since we effectively have $\chi^2 = 3.55^2 + 3.55^2 + 3.09^2 = 34.75$, the combined tension can be estimated as the likelihood of the $\chi^2$ statistic exceeding this value for three degrees of freedom. This suggests that the LG and Centaurus A satellite planes combined cause a tension of $1.40\times 10^{-7}$ ($5.27\sigma$). A new interpretation is thus needed to explain the origin of the observed satellite galaxy planes.

Another less widely known problem is the distorted morphologies of MW satellites, which strongly imply that they have been affected by tidal forces \citep{Kleyna_1998, Walcher_2003, Sand_2012}. Because the inner region of a satellite galaxy can hardly be affected by tides if it is protected by a dominant dark matter halo \citep{Kazantzidis_2004}, $\la 10\%$ of the MW satellites are expected to be distorted in this paradigm \citep{Kroupa_2012}. However, \citet{McGaugh_Wolf_2010} found that the majority of the MW satellites present signs of being disturbed, both in their elevated $\sigma$ and in their observed ellipticity. More recently, \citet{Hammer_2020} pointed out that the high $\sigma$ of dSphs surrounding the MW and their proximity to perigalacticon makes it extremely unlikely for them to be dark matter dominated.

An alternative explanation for the planar distribution of the satellite galaxies is that they are of tidal origin. This is because TDGs are expected to be phase-space correlated \citep{Pawlowski_2011, Kroupa_2012, Pawlowski_2018, Haslbauer_2019_TDG}. But if the observed satellites are of tidal origin, they would be dark matter free, in which case their high $\sigma$ for their low $M_{\star}$ should be explained in a different way. \citet{Kroupa_1997} proposed that due to close encounters of the TDGs with their parent galaxy, the TDGs are highly perturbed. As a result, they should be significantly anisotropic both in terms of their internal structure and their velocity dispersion tensor. More generally, they should not be in dynamical equilibrium, making it incorrect to directly apply the virial theorem to infer the mass from $\sigma$ as this could cause a significant overestimate. However, purely baryonic dwarfs would be very fragile and easily destroyed, making it unlikely that so many of them exist in the LG right now \citep{Haslbauer_2019_DF2, Haslbauer_2019_TDG}. Even if this scenario can explain the high $\sigma$ of all observed dSphs, $\Lambda$CDM would still struggle to explain why almost all observed dwarf satellites of the MW, M31, and Centaurus A are of tidal origin $-$ the quenching mechanisms invoked to solve the missing substructure problem are not expected to be so destructive as to get rid of all observable primordial dwarfs \citep{Kim_2018, Read_2019_missing_satellites, Webb_2020}.

Given these difficulties, it is important to note that the properties of both primordial and tidal dSphs can be explained without resorting to the assumption of a surrounding dark matter halo. This entails discarding the $\Lambda$CDM cosmological model and using instead an alternative framework, the currently leading contender being Milgromian dynamics \citep[MOND;][]{Milgrom_1983}. MOND proposes that the deviations from Newtonian behaviour in the rotation curves of galaxies should be attributed to a departure from Newtonian gravity in the regime of weak gravitational fields \citep[$g \la a_{_0} = 1.2 \times 10^{-10}$~m/s$^2 = 3.9$~pc/Myr$^2$;][]{Begeman_1991, Gentile_2011, McGaugh_Lelli_2016}. The gravity boost that dwarf galaxies experience in this regime would explain their high $\sigma$ \citep{McGaugh_Wolf_2010, McGaugh_2013a, McGaugh_2013b, McGaugh_2021}. It would also make the dwarfs less vulnerable to tides and stellar feedback than Newtonian TDGs, which are expected to be extremely fragile. Moreover, MOND offers an elegant scenario for the origin of the LG satellite planes by means of a past flyby encounter between M31 and the MW $9 \pm 2$~Gyr ago, which is required in MOND \citep{Zhao_2013} and seems to reproduce important aspects of their satellite planes \citep{Banik_Ryan_2018, Bilek_2018, Bilek_2021, Banik_2022_satellite_plane}. Therefore, we will focus mainly on $\Lambda$CDM and MOND in this contribution.

The planes of satellites problem is one of the most well-known challenges to $\Lambda$CDM on galaxy scales \citep{Kroupa_2005, Pawlowski_2018, Pawlowski_2021_Nature_Astronomy, Pawlowski_2021}. It provides a compelling motivation to further investigate dwarf galaxies and question their very nature. Fortunately, the properties of dwarf galaxies make them very suitable for testing different gravity theories. Due to their low mass and especially their low surface brightness, dwarf galaxies can be very susceptible to the effects of gravitational tides. Depending on whether we assume the $\Lambda$CDM or MOND model to be valid significantly affects the expected influence of tides on dwarfs. These expectations can then be compared with observations to try and distinguish the models.

Since MOND is a non-linear theory of gravity, the internal dynamics of an object can be affected by the presence of an external field \citep{Bekenstein_1984}. This is because the enhancement to the self-gravity depends on the total strength of $g$, including any external sources. In a dwarf galaxy that experiences a strong gravitational field (usually from a nearby massive galaxy), the MOND boost to the self-gravity will be limited by the dominant external field from the larger central galaxy. This effect becomes stronger as the dwarf gets closer to the central galaxy, to the point that the dwarf can become almost fully Newtonian. Because of this, dwarfs are expected to be more vulnerable to tides in MOND than in $\Lambda$CDM, where they would be shielded by their dark matter halo throughout their whole trajectory \citep{Brada_2000_tides}.\footnote{For an isolated dwarf, the dark matter halo in $\Lambda$CDM and the correction to Newtonian gravity in MOND both provide a similar enhancement to the self-gravity.}

In this project, we use the Fornax Deep Survey (FDS) dwarf galaxy catalogue \citep{Venhola_2018, Venhola_2019} to compare the observed morphological properties of Fornax Cluster dwarf galaxies with the properties predicted by $\Lambda$CDM and MOND. Our aim is to find out if the observed level of disturbance in the Fornax dwarfs is similar to that expected in $\Lambda$CDM or MOND, or if neither model works well. $\Lambda$CDM could provide too much protection against tides such that it under-predicts the observed level of disturbance in the Fornax dwarfs population. Meanwhile, the lack of protective dark matter haloes around all dwarf galaxies and their reduced self-gravity due to the background cluster gravity could mean that in the MOND scenario, dwarfs are too fragile to survive in the harsh Fornax Cluster environment. Determining which of these scenarios is more likely would help to clarify the physics governing the formation and dynamics of galaxies, whose dominant source of gravity remains unknown.

The layout of this paper is as follows: In Section~\ref{Fornax}, we describe the FDS dwarf galaxy catalogue and the selection criteria that we apply to it (Section~\ref{data_sel}). In Section~\ref{effects_gravi}, we explain the relevant types of gravitational interactions that dwarfs might experience in this cluster: disruption from cluster tides (Section~\ref{cluster_tides}) and galaxy-galaxy harassment (Section~\ref{harassment}). These sections consider only Newtonian gravity $-$ the generalization to MOND is presented in Section~\ref{MOND}. In Section~\ref{tidal_sus}, we provide the equations describing the susceptibility of dwarfs to tidal forces in the $\Lambda$CDM and MOND models, obtain the tidal susceptibility of the dwarfs in the FDS catalogue for each model (Section~\ref{tidal_sus_Fornax}), and show how this theoretical quantity is related to the distribution of the dwarfs (Section~\ref{surf_dens_dwarfs}) and whether their observed morphology appears disturbed or undisturbed (Section~\ref{comparison_disturbance}). In Section~\ref{test_mass}, we construct a test particle simulation of the orbits of Fornax dwarfs and, using the MCMC method, fit it to the real Fornax system using the FDS catalogue. In Section~\ref{Results}, we present the results obtained from our MCMC analysis and how they compare to the results of \textit{N}-body simulations, which we complement with our own \textit{N}-body simulations of a typical Fornax dwarf in MOND (Section~\ref{Nbody_sim}). We then discuss our results in Section~\ref{discussion} before concluding in Section~\ref{conclusions}.

\section{The Fornax Deep Survey (FDS)}
\label{Fornax}

The Fornax Cluster is one of the nearest galaxy clusters \citep[$d_{\textrm{Fornax}} = 20.0 \pm 0.3$~Mpc;][]{Blakeslee_2009}. It is named after its sky position in the southern hemisphere constellation of Fornax. The cluster is structured into two main components: the main Fornax Cluster centred on NGC 1399, and an infalling subcluster (Fornax A) centred $3^\circ$ to the south-west in which NGC 1316 is the central galaxy \citep*{Drinkwater_2001}. The Fornax Cluster contains a significant number of dwarf galaxies with different luminosities, colours, shapes, sizes, and distances to the cluster centre, making it very valuable for studying the properties of dwarf galaxies.

The FDS is the most recent survey of the Fornax Cluster. It includes the main Fornax Cluster and part of the Fornax A subcluster, with a total sky coverage of $26 \, \textrm{deg}^2$ \citep{Venhola_2018}. The FDS represents a significant improvement in resolution and image depth with respect to the previous spatially complete Fornax Cluster Catalogue \citep[FCC;][]{Ferguson_1989}. This has allowed the FDS to identify a large number of previously unknown faint galaxies, which can be useful to test the effects of the cluster environment on smaller, more vulnerable galaxies. The FDS reaches the 50\% completeness limit at an apparent (absolute) magnitude in the red band of $M_{r'} = -10.5$ ($m_{r'} = 21$), while the corresponding surface brightness limit is $\mu_{e,r'} = 26 \, \textrm{mag} \, \textrm{arcsec}^{-2}$. However, the FDS can still clearly detect some dwarf galaxies down to $M_{r'} = -9$ and $\mu_{e,r'} = 27.8 \, \textrm{mag} \, \textrm{arcsec}^{-2}$ \citep{Venhola_2018}.

\begin{figure}
	\centering
	\includegraphics[width = 8.5cm]{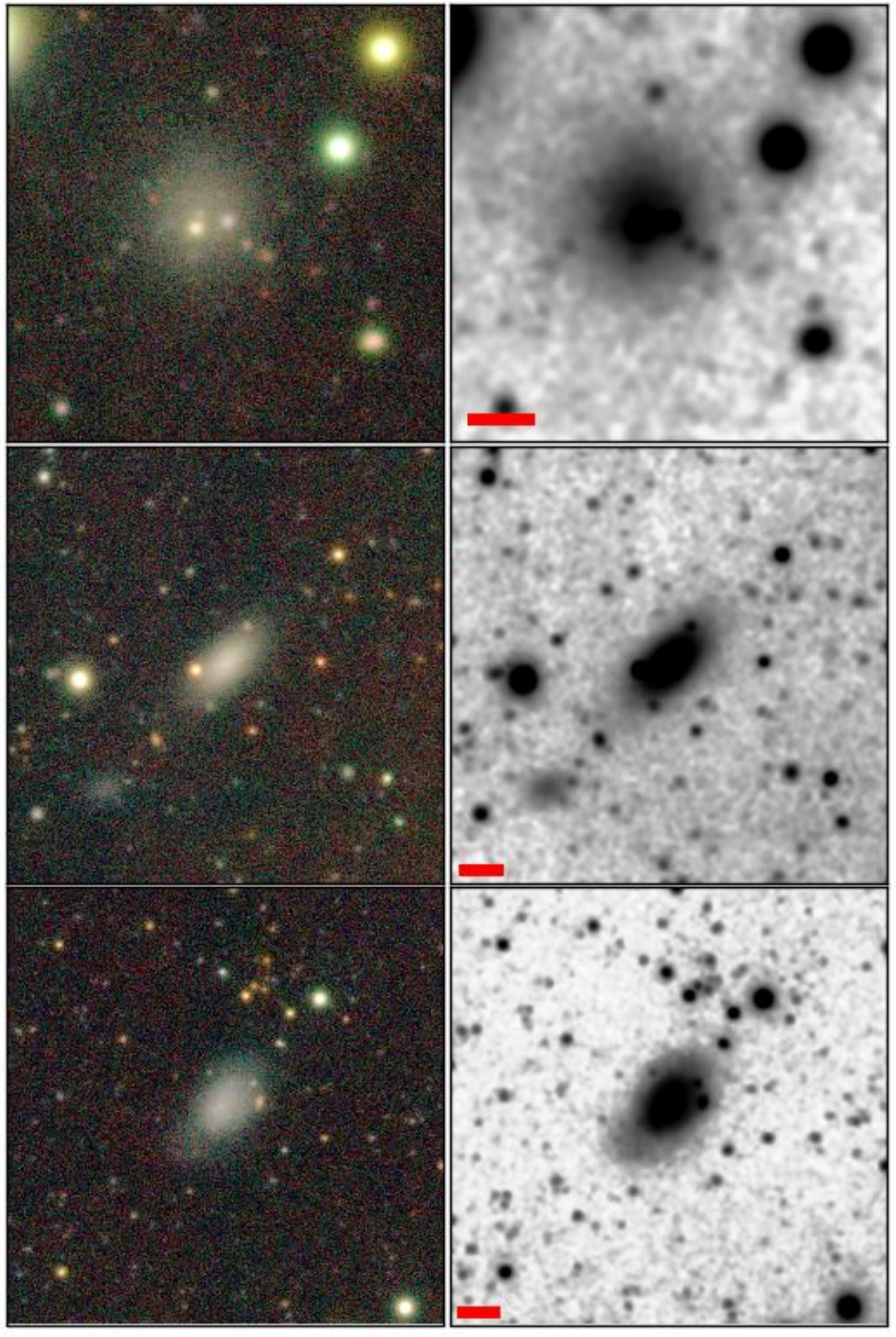}
	\caption{Images of three FDS dwarfs presenting different levels of disturbance in different colour bands and filters. Each row shows the same dwarf as a red-green-blue colour image (left column) and in the $r'$ band with a filter enhancing the dwarf's low surface brightness features (right column). The dwarf in the first, second, and third row is classified as ``undisturbed'', ``mildly disturbed'', and ``very disturbed'', respectively. The horizontal red lines show an angular scale of $10\arcsec$, which corresponds to 970~pc at the 20~Mpc distance to the Fornax Cluster.}
	\label{fig:tid_morph_class}
\end{figure}

The FDS catalogue of dwarf galaxies \citep{Venhola_2017, Venhola_2018, Venhola_2019} includes 564 dwarf galaxies with $2 \times 10^5 < M_{\star}/M_{\odot} < 2 \times 10^9$, some in the main Fornax Cluster and others in the infalling subcluster. As in other galaxy clusters, dEs and dSphs are the most common types of dwarf galaxy that can be found in the Fornax Cluster. These are estimated to have an age of $t_{\textrm{Fornax}} = 10 \pm 1$~Gyr \citep{Rakos_2001}, where $t_{\textrm{Fornax}}$ is the age of the elliptical galaxies in Fornax, which we assume to have a similar age to that of the dwarf galaxies. Because of the similarities in some of their morphological properties, the FDS classifies dE and dSph galaxies as one single type, dE. The FDS catalogue also provides information about other properties of the dwarfs. The ones which are relevant for this project are: $M_{\star}$, the effective radius, the right ascension and declination, the apparent surface brightness in the $r'$ band, the S{\'e}rsic index of the surface brightness profile \citep{Sersic_1963}, the morphological type, the nucleated flag indicating if the dwarf is nucleated or non-nucleated, and the tidal morphology \citep[undisturbed, possibly/mildly disturbed, very disturbed, or unclear;][]{Venhola_2022}. The effective radius, the S{\'e}rsic index, and the apparent brightness in the $r'$-band are obtained by fitting the data to a 2D S{\'e}rsic profile \citep{Venhola_2018} using the \textsc{galfit} software \citep{Peng_2002}. $M_{\star}$ is obtained from the empirical relation between the $g'-i'$ colour and mass-to-light ($M/L$) ratio (\citealt{Taylor_2011}; for further details, see \citealt{Venhola_2019}). The morphological classifications such as the nucleated flags, the Hubble type \citep{Venhola_2018, Venhola_2019}, and the tidal morphologies are done visually. The tidal morphology is classified in \citet{Venhola_2022} based on the following criteria:
\begin{enumerate}
	\item{Undisturbed}: Dwarf galaxies that do not present irregularities, distortions to their shape, or tidal tails;
	\item{Possibly/mildly disturbed}: Hints of irregularities are present in the outskirts of the dwarf galaxy;
	\item{Very disturbed}: Dwarf galaxies with tidal tails and/or very clear distortion in the shape; and
	\item{Unclear}: Nearby bright objects or data artefacts make the classification difficult.
\end{enumerate}
Fig.~\ref{fig:tid_morph_class} shows some illustrative examples of dwarfs in these categories.

\subsection{Data selection}
\label{data_sel}

From the 564 FDS dwarfs, we remove those which are classified as late-type as there is a high chance that these are not physically in the cluster but instead represent line of sight contamination \citep{Venhola_2019}. We also remove dwarfs which have an `unclear' tidal morphology because they are not useful for the analysis. This leaves us with 456 dwarfs. We then obtain the angular distance between each dwarf and the centre of the Fornax Cluster based on the right ascension (RA) and declination (Dec) of the dwarf and that of the Fornax Cluster, whose sky coordinates are $\textrm{RA}_{\textrm{centre}} = 54.6^\circ$, $\textrm{Dec}_{\textrm{centre}} = -35.5^\circ$ \citep[table D1 of][]{Watson_2009}.
\begin{eqnarray}
	\!\!\!\!\!\! \Delta \textrm{RA} &\equiv& \textrm{RA} - \textrm{RA}_{\textrm{centre}} \, , \\
	\!\!\!\!\!\! \Delta \textrm{Dec} &\equiv& \textrm{Dec} - \textrm{Dec}_{\textrm{centre}} \, , \\
	\!\!\!\!\!\! \Delta' \textrm{RA} &=& \Delta \textrm{RA} \cdot \cos \left( \frac{\textrm{Dec} + \textrm{Dec}_{\textrm{centre}}}{2} \right), \\
	\!\!\!\!\!\! \textrm{Angular distance} &=& \sqrt{\left( \Delta' \textrm{RA} \right)^2 + \left( \Delta \textrm{Dec} \right)^2} \, .
\end{eqnarray}
Expressing this angular distance in radians and multiplying it by the 20~Mpc distance to Fornax \citep{Blakeslee_2009} then gives the dwarf's sky-projected distance $R_{\textrm{sky}}$ from the centre of the Fornax Cluster.
\begin{eqnarray}
    R_{\textrm{sky}} ~=~ d_{\textrm{Fornax}} \times \left( \textrm{Angular distance} \right) \, .
\end{eqnarray}

We remove dwarfs with $R_{\textrm{sky}} > 800$~kpc as dwarfs further out mostly belong to the subcluster Fornax A, so including these would contaminate our sample of dwarfs belonging to the main Fornax Cluster \citep[see fig. 4 of][]{Venhola_2019}. This leaves us with 353 dwarf galaxies.

\section{Effects of gravitational interactions on dwarfs}
\label{effects_gravi}

Before discussing the gravitational perturbations experienced by Fornax Cluster dwarf galaxies, we first discuss why non-gravitational forces are not expected to perturb Fornax Cluster dwarfs today. Old dwarf galaxies in a cluster environment are expected to be gas-poor. Most dwarfs in the FDS catalogue are classified as early-type galaxies, implying that they are dominated by old stellar populations and are not currently forming new stars. The scarcity of star-forming dwarfs in the Fornax Cluster is consistent with the fact that they are likely to be gas-poor. One important reason for this is ram pressure stripping \citep{Gunn_1972}. This takes place when a galaxy containing a large amount of cold gas moves through a galaxy cluster full of hot gas. The temperature difference and motion between the two gas components generate a pressure gradient that strips the cold gas from the galaxy. \citet{Venhola_2019_error} estimated in the left panel of their fig.~21 that ram pressure stripping of Fornax Cluster dwarfs at the low masses relevant to our analysis should have been quite efficient $-$ the vast majority of the dwarfs in our sample have $M_{\star} < 10^8 \, M_{\odot}$ (Section~\ref{data_sel}). The fact that the Fornax dwarfs are gas-poor has been observationally confirmed by \citet{Zabel_2019}, who studied the molecular gas in the Fornax Cluster and showed that its dwarfs are gas deficient. \citet{Loni_2021} showed the same for neutral hydrogen in FDS dwarfs with $M_{\star}$ down to a few times $10^7 \, M_\odot$, below which theoretical arguments indicate that the gas reservoir should have been ram pressure stripped by now \citep[see section 7.3.1 of][]{Venhola_2019}. Moreover, the colours of the FDS dwarfs also suggest a lack of recent star formation (see their fig.~18). Ongoing gas loss is thus very unlikely to explain the observed disturbances to the structures of some Fornax Cluster dwarfs. We therefore conclude that their internal structure is to a good approximation only affected by gravity from surrounding structures.

The main types of gravitational interaction that can disturb and transform the structure of a dwarf galaxy in the Fornax Cluster are tidal disruption from the cluster’s tidal field and galaxy-galaxy harassment due to encounters with the cluster's massive elliptical galaxies \citep[see section~7 of][]{Venhola_2019}. In the following, we discuss these processes in the context of Newtonian gravity before deriving their generalization to MOND (Section~\ref{MOND}).

\subsection{Disruption from cluster tides}
\label{cluster_tides}

In this type of interaction, the structure of a dwarf with mass $M_{\textrm{dwarf}}$ is affected by gravitational tides coming from the overall cluster potential, i.e., from the difference in the cluster gravity across the finite size of the dwarf. We quantify the influence of cluster tides on a dwarf using the concept of its tidal radius $r_{\textrm{tid}}$. This is defined such that if $r_{\textrm{tid}}$ were the dwarf's actual size, then the tidal force of the cluster and the self-gravity of the dwarf would have the same strength. We can intuitively see that
\begin{eqnarray}
	\frac{G M_{\textrm{dwarf}}}{r^2_{\textrm{tid}}} ~&\approx&~ r_{\textrm{tid}} \overbrace{\left( \frac{\Delta g_c}{\Delta R} \right)}^{\textrm{Tidal stress}} \, , \\
	\Rightarrow r_{\textrm{tid}} ~&\approx&~ \left( \frac{G M_{\textrm{dwarf}}}{\Delta g_c / \Delta R} \right)^{1/3} \, ,
    \label{approx_rtid}
\end{eqnarray}
where $G$ is the Newtonian constant of gravitation and $\Delta g_c/\Delta R$ is the tidal stress from the cluster potential, with $g_c$ and $R$ being the cluster gravity and the 3D distance to the cluster centre, respectively. Since we want to find out the maximum degree of disturbance that a dwarf can experience due to the cluster potential, we obtain $g_c$ and its gradient when the dwarf is at pericentre ($R = R_{\textrm{per}}$). In order to obtain $R_{\textrm{per}}$ for each dwarf from its projected distance in the FDS, we use $R_{\textrm{per}} = 0.29 \, R$ (see Appendix~\ref{Rper}), with $R$ obtained by deprojecting $R_{\textrm{sky}}$ using the method described in Appendix~\ref{deproj}.

As in \citet{Venhola_2019}, we assume that the galaxy number density and cluster potential have remained constant over time. This approximation is reasonable because the orbital periods of galaxies in the Fornax Cluster are typically much shorter than a Hubble time: The estimated 1D velocity dispersion of 370~km/s \citep{Drinkwater_2001} combined with a maximum size of 800~kpc (Section~\ref{data_sel}) implies a crossing time of only 1.2~Gyr. We assign the cluster a Newtonian dynamical mass profile given by
\begin{eqnarray}
	M_c \left( < \theta_{\textrm{3D}} \right) ~=~ M_{\textrm{norm}} \left( \frac{\theta_{\textrm{3D}}}{\theta_{\textrm{norm}}}\right)^{\alpha} \, ,
	\label{M_cluster}
\end{eqnarray}
where $\theta_{\textrm{3D}} \equiv R/d_{\textrm{Fornax}}$ is the 3D angular distance to the Fornax Cluster centre. The parameters are: $M_{\textrm{norm}} = 3 \times 10^{10}~M_{\odot}$, $\theta_{\textrm{norm}} = 10\arcsec$, and $\alpha = 1.1$. This radial mass dependency is obtained from fitting the above power-law to the mass profile derived in fig.~17b of \citet{Paolillo_2002}, which uses the X-ray surface brightness distribution of the central Fornax galaxy and its gas temperature profile to find the gas density distribution. The mass profile is then derived assuming hydrostatic equilibrium by applying the spherical Jeans equation. Note that the mass derived here is a Newtonian dynamical mass. A more model-independent way to describe the observations is in terms of the cluster gravity $g_c \equiv GM_c/R^2$. This method of obtaining $g_c$ relies on the well-understood physical process of thermal X-ray emission from hot gas. Its temperature and density profile require a particular radial run of $g_c$ regardless of the gravity law. Therefore, it is not relevant whether $g_c$ has been enhanced by a dark matter halo or by MOND (or indeed by some elements of both, as argued in Section~\ref{MOND}). Consequently, $g_c$ will be the same in the $\Lambda$CDM and MOND scenarios, as will the resulting tidal stress on each dwarf.

This is not the case for $M_{\textrm{dwarf}}$. The FDS catalogue gives only $M_{\star}$ for each dwarf. This can be equated with $M_{\textrm{dwarf}}$ in MOND, but not in $\Lambda$CDM where each dwarf is expected to have a substantial dark halo of mass $M_{\textrm{halo}}$. We find this using the same abundance matching procedure as \citet{Venhola_2019}. We first find $M_{\textrm{halo}}$ from the relation between $M_{\star}$ and $M_{\textrm{halo}}$ given in equation 2 of \citet{Moster_2010}:
\begin{eqnarray}
    \label{m/M}
	&& \frac{M_{\star}}{M_{\textrm{halo}}} \\
	&& = 2 \left( \frac{M_{\star}}{M_{\textrm{halo}}} \right)_0 \left[ \left( \frac{M_{\textrm{halo}}}{M_1} \right)^{-\beta} + \left( \frac{M_{\textrm{halo}}}{M_1} \right)^{-\gamma} \right]^{-1} \, . \nonumber
\end{eqnarray}
Their table~1 clarifies that the parameters in this equation are: $\left( \frac{M_{\star}}{M_{\textrm{halo}}} \right)_0 = 0.0282$, $M_1 = 10^{11.884} ~M_{\odot}$, $\beta = 1.057$, and $\gamma = 0.556$. As the dark halo of each dwarf is not observable and remains hypothetical, we are only interested in whether tides are perturbing the dwarf's stellar component \citep[which they might not be even if its dark matter halo is being stripped; see][]{Smith_2016}. For this, the Shell Theorem indicates that we only need to consider the dark matter within the dwarf's optical radius. Following \citet{Venhola_2019}, we assume that this is only 4\% of the total halo mass $-$ \citet{Diaz_2016} found this fraction to be consistent with the dark matter masses within the optical radii of S\textsuperscript{4}G galaxies \citep{Sheth_2010}. Adding the halo contribution to $M_{\star}$, the total mass of the dwarf in $\Lambda$CDM for the purposes of our analysis is therefore:
\begin{eqnarray}
	M_{\textrm{dwarf, } \Lambda\textrm{CDM}} ~=~ M_{\star} + 0.04 \, M_{\textrm{halo}} \, .
	\label{M_dwarf_rule}
\end{eqnarray}
In Section~\ref{newDMfrac}, we consider other possible choices for the fraction of the halo mass within the optical radius of a dwarf.

Equation~\ref{approx_rtid} is only a very crude estimate for the tidal radius of a dwarf. While it should capture the essential physics, we expect a more careful treatment to yield an additional factor of order unity. Numerical simulations are required to capture the details of mass loss from a dwarf undergoing tidal disruption, which is expected to substantially distort its shape. To account for this, the $\Lambda$CDM expression for $r_{\textrm{tid}}$ in equation 1 of \citet{Baumgardt_2010} includes an extra factor of $2^{-1/3}$. Taking this into consideration, we adopt the following expression for $r_{\textrm{tid}}$ in $\Lambda$CDM:
\begin{eqnarray}
	r_{\textrm{tid, } \Lambda\textrm{CDM}} ~=~ \left( \frac{G \, M_{\textrm{dwarf, } \Lambda\textrm{CDM}}}{2 \Delta g_c / \Delta R} \right)^{1/3} \, .
	\label{rtid_LCDM}
\end{eqnarray}
This is based on using their study to obtain the numerical pre-factor in Equation~\ref{approx_rtid} for circular orbits in a central potential with a flat rotation curve ($\alpha = 1$) $-$ other approaches are discussed below Equation~\ref{beta_definition}. Notice that $g_c$ itself does not directly affect the tidal radius: The cluster gravity only affects the dwarf through the tidal stress it creates on the dwarf. This is not so in the corresponding expression for MOND (Equation \ref{rtid_MOND}), which we derive in Section~\ref{MOND}.

\subsection{Galaxy-galaxy harassment}
\label{harassment}

The morphology of the Fornax Cluster dwarf galaxies can also be disrupted by gravitational interactions with individual large galaxies in the cluster. This effect is called harassment \citep{Venhola_2019}. Assuming a high relative velocity between the dwarf galaxy and the larger galaxy, we can use the impulse approximation to estimate the impact of each encounter on the internal structure of the dwarf. We then need to combine the effects of many such interactions, each time adding the squares of the velocity perturbations as these would generally be in random directions, leading to a process resembling a diffusive random walk. Equivalently, we should add the energy gained by the dwarf from each encounter, leading to the concept of a heating rate $\dot{E}$ \citep[equation 8.52 of][]{Binney_Tremaine_2008}. The disruption time-scale $t_{d, \Lambda\textrm{CDM}}$ is the time-scale over which putting energy into the dwarf at the presently calculated $\dot{E}$ would cause it to become unbound given its present gravitational binding energy per unit dwarf mass of
\begin{eqnarray}
    \lvert E \rvert ~=~ \frac{G M_{\textrm{dwarf, } \Lambda\textrm{CDM}}}{2 r_{h,\textrm{dwarf}}} \, ,
    \label{Binding_energy}
\end{eqnarray}
where $r_{h,\textrm{dwarf}}$ is the half-mass radius of the dwarf. Since only the baryons are visible, we again restrict our attention to the baryonic component of each dwarf, so $r_{h,\textrm{dwarf}}$ refers to only its visible component and $M_{\textrm{dwarf, } \Lambda\textrm{CDM}}$ is again found using Equation~\ref{M_dwarf_rule}. Dividing the magnitude of the binding energy by the heating rate gives the disruption time-scale \citep[equation 8.54 of][]{Binney_Tremaine_2008}:
\begin{eqnarray}
	t_{d, \Lambda\textrm{CDM}} \equiv \frac{\lvert E \rvert}{\dot{E}} = \frac{0.043}{W_p} \frac{\sqrt{2} \sigma M_{\textrm{dwarf,} \Lambda\textrm{CDM}} r_{h,p,{\Lambda\textrm{CDM}}}^2}{G M_{p, \Lambda\textrm{CDM}}^2 n_p r_{h,\textrm{dwarf}}^3} \, .
	\label{td_LCDM}
\end{eqnarray}
The `p' subscript denotes the massive galaxy (perturber), while `dwarf' refers to the dwarf galaxy that is being perturbed. $W_p$ is a factor accounting for the shape of the perturber galaxy's mass distribution. We choose $W_p = 1$ as an intermediate value between that of the Plummer and Hernquist models \citep[chapter 8.2 of][]{Binney_Tremaine_2008}. $n_p$ is the number density of perturbers, which \citet{Venhola_2019} estimated to be $25 \, \textrm{Mpc}^{-3}$ by counting 48 large galaxies inside the virial volume of the Fornax Cluster ($R_{\textrm{vir}} = 0.77$~Mpc). Its 1D velocity dispersion is $\sigma = 370$~km/s \citep{Drinkwater_2001}, with the extra factor of $\sqrt{2}$ accounting for the fact that we need to consider the dwarf-perturber relative velocity. $M_{p, \Lambda\textrm{CDM}}$ and $r_{h,p,{\Lambda\textrm{CDM}}}$ are the perturber galaxy's mass and half-mass radius, respectively. Note that we use $r_h$ for the deprojected half-mass radius of the baryonic component. $r_h$ does not include the dark matter halo unless we explicitly say so and label it accordingly as $r_{h, \Lambda\textrm{CDM}}$. \citet{Venhola_2019} use $r_h$ for the radius containing half of the total mass including dark matter, so our notation is different in this respect.

To obtain $r_{h,\textrm{dwarf}}$ from the projected effective radius $r_e$ containing half of the dwarf's total stellar mass, we use equation B3 of \citet{Wolf_2010}, though a good approximation is that $r_{h,\textrm{dwarf}} \approx \left( 4/3 \right) r_e$. Our adopted $M_{p,*} = 10^{10}~M_{\odot}$ is the median stellar mass of the large galaxies catalogued in table~C1 of \citet{Iodice_2019} and in the FCC. In the $\Lambda$CDM case, the contribution of the dark halo should be added to this mass. Unlike with the dwarf galaxies, the full extent of the dark halo is considered for the large galaxies because these are expected to be quite robust to cluster tides, so the full halo mass should be considered when estimating the perturbation to a passing dwarf. \citet{Venhola_2019} found $M_{p, \Lambda\textrm{CDM}} = 10^{11.6} ~M_{\odot}$ following this procedure, which we also verified.

Using a single $M_p$ value for all perturbers gives only an approximate estimate of the heating rate. A more accurate calculation should use the power-law distribution of all the galaxies and make predictions based on that, but this would be extremely difficult. Moreover, the other simplifications assumed throughout the whole calculation of $t_d$ have a larger impact on the result than taking into account the right distribution of perturbing galaxy masses. Fortunately, we will see that $t_d$ greatly exceeds a Hubble time, a conclusion which should remain valid even with small adjustments to the calculation. In particular, we will show that considering the mass spectrum of perturbers should affect the estimated heating rate by only a small factor such that $t_d$ remains very long (Section~\ref{tidal_sus_Fornax}).

The $r_{h,p}$ value of the large galaxies is also obtained from the median of all the documented large galaxies (perturbers) in the cluster, yielding $r_{h,p} = 4$~kpc based on the luminous matter. This is applicable to MOND, but in the $\Lambda$CDM case, the $r_{h,p}$ of the large galaxies should account for half of the perturber's total mass, not only the stellar mass given in the catalogues. This is because the gravitational effect of the dark matter halo also contributes to perturb the stellar content of a passing dwarf. To find out the relation between $r_{h,p}$ and $r_{h,p, \Lambda\textrm{CDM}}$, \citet{Venhola_2019} looked into the Illustris cosmological simulations \citep{Pillepich_2018} to infer the relation between these two quantities in simulated large galaxies in a galaxy group with a similar mass to the Fornax Cluster, yielding $r_{h,p, \Lambda\textrm{CDM}}/r_{h, p} = 3.6$. Therefore, the half-mass radius of the perturbers in $\Lambda$CDM is taken to be $r_{h,p, \Lambda\textrm{CDM}} = 14.4$~kpc.

To summarize, the disruption time-scale in $\Lambda$CDM can be found by directly applying Equation \ref{td_LCDM} once we include the contribution of the dark matter halo to $M_{\textrm{dwarf}}$, $M_p$, and $r_{h,p}$. In Section~\ref{Harassment_MOND}, we describe how to obtain the corresponding disruption time-scale expression in MOND.

\subsection{Generalization to MOND}
\label{MOND}

The MOND model proposes that Newtonian gravity breaks down in the limit of low accelerations such that the actual gravitational field $g$ is related to the Newtonian field $g_{_N}$ according to $g = \sqrt{a_{_0} g_{_N}}$. Milgrom's constant $a_{_0} = 1.2 \times 10^{-10}$~m/s$^2$ is a new fundamental acceleration scale added by MOND. Its value has been empirically determined by matching observed galaxy rotation curves \citep{Begeman_1991, Gentile_2011, McGaugh_Lelli_2016}, which MOND does extremely well \citep{Famaey_McGaugh_2012, Lelli_2017, Li_2018}. Due to the very small numerical value of $a_{_0}$ \citep[which may be related to the quantum vacuum; see][]{Milgrom_1999, Senay_2021}, the behaviour of gravity has never been directly tested in the deep-MOND regime ($g \ll a_{_0}$). Indeed, Solar system tests are typically only sensitive to the behaviour of gravity in the regime where $g$ exceeds $a_{_0}$ by many orders of magnitude \citep[though for a proposed Solar system test in the MOND regime, see][]{Penner_2020}.

For an isolated spherically symmetric problem, the expression for the MOND gravitational field $g$ as a function of the Newtonian field $g_{_N}$ can be written as
\begin{eqnarray}
	g ~=~ g_{_N} \nu \left(g_{_N}\right) \, ,
	\label{g_g_N}
\end{eqnarray}
where $\nu$ is the interpolating function with argument $g_{_N}$. To satisfy Solar system constraints and the observed flat rotation curves in the outskirts of galaxies, this function must have the following asymptotic limits:
\begin{eqnarray}
	\nu \to \begin{cases}
	1 \, , & \textrm{if} ~g_{_N} \gg a_{_0} \, , \\
	\sqrt{\frac{a_{_0}}{g_{_N}}} \, , & \textrm{if} ~g_{_N} \ll a_{_0} \, .
	\end{cases}
	\label{nu_cases}
\end{eqnarray}
The first case is the Newtonian regime in which $\nu = 1$ and $g = g_{_N}$ to a very good approximation. In the MOND regime, $g = \sqrt{a_{_0} g_{_N}}$. This causes the gravity from an isolated point mass $M$ to decline as $1/r$ beyond its MOND radius $r_{_{\textrm{MOND}}} \equiv \sqrt{GM/a_{_0}}$, which is necessary to explain the rotation curve data using only the luminous matter. Several forms of the MOND interpolating function have been proposed \citep{Kent_1987, Hees_2014, Hees_2016, McGaugh_Lelli_2016}. Among these, the simple interpolating function \citep{Famaey_Binney_2005} seems to work better with recent observations \citep{Iocco_Bertone_2015, Banik_2018_Centauri, Chae_2018}. Therefore, we will use the simple interpolating function:
\begin{eqnarray}
	\nu \left( g_{_N} \right) ~=~ \frac{1}{2} + \sqrt{\frac{1}{4} + \frac{a_{_0}}{g_{_N}}} \, .
	\label{simple_interpolating}
\end{eqnarray}

It is well known that although MOND is capable of fitting the rotation curves of galaxies without dark matter \citep[see the review by][]{Famaey_McGaugh_2012}, it cannot fit the temperature and density profiles of galaxy clusters using only their visible mass $-$ MOND still needs an additional contribution to the gravitational field \citep{Sanders_1999, Aguirre_2001}. The central galaxy of the Fornax Cluster (NGC 1399) is no exception \citep{Samurovic_2016_Fornax}. To solve this discrepancy and to account for other observations hinting at the presence of collisionless matter in galaxy clusters \citep[most famously in the Bullet cluster;][]{Clowe_2006}, it has been proposed that MOND should be supplemented by sterile neutrinos with a rest energy of 11~eV, a paradigm known as the neutrino hot dark matter ($\nu$HDM) cosmological model \citep{Angus_2009}. $\nu$HDM can fit observations of virialized galaxy clusters using the MOND gravity of their directly detected baryons plus the sterile neutrinos \citep{Angus_2010}. It can also fit the power spectrum of anisotropies in the cosmic microwave background (CMB) because the typical gravitational field at the epoch of recombination was $\approx 20 \, a_{_0}$ and the cosmic expansion history would be standard. Neutrino free streaming reduces the power on small scales compared to $\Lambda$CDM, but this is consistent with CMB observations provided the rest energy of the neutrinos exceeds 10~eV \citep[see section 6.4.3 of][]{Planck_2016}. The gravitational fields from density perturbations would enter the MOND regime only when the redshift $\la 50$, before which the MOND corrections to General Relativity should be small \citep[for a more detailed explanation of this model, see][]{Haslbauer_2020}. $\nu$HDM relies on the existence of eV-scale sterile neutrinos, but these are also hinted at by several terrestrial experiments \citep[for a recent review, see][]{Berryman_2022}.

Equation \ref{simple_interpolating} shows that unlike Newtonian gravity, MOND is a non-linear theory of gravity. A physical consequence of this non-linearity is the so-called external field effect \citep[EFE;][]{Milgrom_1986}. This implies that the internal gravity of a system can be weakened by a constant gravitational field from its external environment even if this is completely uniform, violating the strong equivalence principle. The reason is that the MOND boost to the Newtonian gravity is approximately given by $\nu$, which is damped due to the external field. In MOND, the EFE explains why some galaxies like NGC 1052-DF2 have a very low observed velocity dispersion \citep{Van_Dokkum_2018, Famaey_2018, Kroupa_2018_DF2, Haghi_2019_DF2}, even though other galaxies like DF44 with similar properties but in a more isolated environment have a much higher velocity dispersion \citep{Van_Dokkum_2019, Bilek_2019, Haghi_2019_DF44}.\footnote{In a conventional gravity context, the very low observed velocity dispersion of NGC 1052-DF2 implies a lack of dark matter, which however is not easily explained in $\Lambda$CDM \citep{Haslbauer_2019_DF2, Moreno_2022_DF2}.} Strong evidence for the EFE has recently been obtained based on the outer rotation curves of galaxies showing a declining trend if the galaxy experiences a significant EFE, while galaxies in more isolated environments have flat outer rotation curves \citep{Haghi_2016, Chae_2020_EFE, Chae_2021}. For a discussion of observational evidence relating to the EFE, we refer the reader to section~3.3 of \citet{Banik_Zhao_2022}.

The EFE is also important to Fornax Cluster dwarfs because their low surface brightness implies rather little self-gravity, allowing the gravitational field of the cluster to dominate over that of the dwarf. As a result, the dwarf is in the quasi-Newtonian (QN) regime where its internal dynamics are similar to a Newtonian dwarf but with a renormalized gravitational constant $G_{\textrm{eff}} > G$. We need to determine $G_{\textrm{eff}}$ from the cluster gravitational field $g_c$. We do this by writing Equation \ref{simple_interpolating} in the inverse form:
\begin{eqnarray}
	g_{_N} ~&=&~ g \mu \left( g \right) \, , \text{ where} \\
	\mu \left( g \right) ~&=&~  \frac{g}{g + a_{_0}}  \, .
    \label{g_N_g}
\end{eqnarray}
As the cluster gravity is dominant over the self-gravity of the dwarf, we can set $g = g_c$, with $g_c$ obtained from observations as described in Section~\ref{cluster_tides}. Since the Newtonian gravity of the cluster is directly proportional to the Newtonian gravitational constant ($g_{c,N} \propto G$), the effective gravity of a dwarf in the cluster will be directly proportional to an analogous constant parameter $G_{\textrm{eff}}$ defined such that $g_c = \left( G_{\textrm{eff}}/G \right) g_{c,N}$. From Equation \ref{g_N_g}, we infer $G_{\textrm{eff}}$ to be:
\begin{eqnarray}
	 G_{\textrm{eff}} ~=~ \left( \frac{a_{_0} + g_c}{g_c} \right) G \, .
	\label{G_eff}
\end{eqnarray}
Note that replacing $G \to G_{\textrm{eff}}$ can only be applied if the dwarf’s self-gravity is dominated by the external field of the cluster, so that the combined gravitational field of the dwarf and cluster will remain approximately constant with increasing distance with respect to the dwarf's centre.

\subsubsection{Tidal radius}
\label{Tidal_radius_MOND}

At the tidal radius of a dwarf, the difference in cluster gravity across the dwarf is comparable to its self-gravity. Therefore, the total cluster gravity $g_c$ dominates over the dwarf's self-gravity. Thus, the MOND tidal radius of any dwarf is necessarily in the EFE-dominated/QN regime where its dynamics are approximately Newtonian but with $G \to G_{\textrm{eff}}$. Substituting this into Equation \ref{approx_rtid} gives an approximate expression for the MOND tidal radius. Accounting for additional details like the non-spherical nature of the point mass potential in the QN regime \citep[discussed further in section~2.4 of][]{Banik_Zhao_2022}, the MOND tidal radius can be expressed as \citep[equations~26 and 36 of][]{Zhao_2006}:
\begin{eqnarray}
	\label{rtid_MOND}
	&& r_{\textrm{tid, MOND}} \\
	&& = \frac{2}{3} \left. \sqrt{ \frac{\partial \ln g}{\partial \ln g_{_N}}} \right|_{g = g_c} \left[ \left( \frac{2 - \alpha}{3 - \alpha} \right) \frac{G_{\textrm{eff}} \, M_{\textrm{dwarf}}}{\Delta g_c / \Delta R} \right]^{1/3}, \nonumber
\end{eqnarray}
where the factor of order unity is the MOND Roche lobe scaling factor accounting for such subtleties. Note that we have generalized their equation 26 to write the result in terms of $G_{\textrm{eff}}$ and the tidal stress. The parameter $\alpha \equiv 2 + \frac{d\ln g_c}{d\ln r}$ has the same meaning as in Equation~\ref{M_cluster}, so its value remains 1.1. For the case of a dwarf orbiting a point mass in the deep-MOND limit ($\alpha = 1$), the numerical factors combine to give $2^{1/6}/3$, matching equation~44 of \citet{Zhao_2005}.\footnote{Equation~\ref{rtid_MOND} is the extent of the Roche Lobe in the tangential direction within the orbital plane. The extent along the orbital pole is similar, and in both cases is smaller than the extent along the radial direction \citep[see section~4.2 of][]{Zhao_2006}.}

\subsubsection{Galaxy-galaxy harassment}
\label{Harassment_MOND}

When a dwarf interacts with a massive galaxy in the Fornax Cluster environment, we need to consider both the gravity from the elliptical and the background EFE due to the cluster potential. As in Section~\ref{harassment}, we estimate the perturbation to the dwarf by assuming it is a collection of test particles that receive some impulse $\bm{u}$ from the elliptical, with the heating rate of the dwarf proportional to the square of $\lvert \Delta \bm{u} \rvert$, the spread in $\bm{u}$ across the dwarf. Once it has moved away from the elliptical, the binding energy of the dwarf is given by Equation \ref{Binding_energy} but with $G \to G_{\textrm{eff}}$ as discussed above. The main difficulty lies in estimating the energy gained by the dwarf due to interactions with impact parameter $b$, which for a high-velocity encounter is approximately the same as the closest approach distance between the dwarf and the elliptical.

We need to consider encounters in two different regimes:
\begin{enumerate}
    \item The QN regime in which $g_c \ll a_{_0}$ dominates over gravity from the elliptical; and
    \item The isolated deep-MOND (IDM) regime in which the gravity from the elliptical dominates over $g_c$ but is still much weaker than $a_{_0}$.
\end{enumerate}
We do not need to consider the Newtonian regime because the perturbers have a radius that is numerically similar to their MOND radius for the parameters given in Section~\ref{harassment}. This is not unique to the Fornax Cluster: Elliptical galaxies generally have a size similar to their MOND radius \citep{Sanders_2000}. This is because if the initial radius was much smaller and the system is nearly isothermal, then a significant proportion of the mass in the outskirts would be moving faster than the Newtonian escape velocity, causing the system to expand to its MOND radius \citep{Milgrom_1984, Milgrom_2021_polytropes}.

The QN and IDM regimes are separated by encounters with $b = r_{_{\textrm{EFE}}}$, the distance from the elliptical beyond which the cluster gravity dominates.
\begin{eqnarray}
    r_{_{\textrm{EFE}}} ~=~ \sqrt{\frac{GM_p}{g_{c,N}}} \, ,
\end{eqnarray}
where $g_{c,N} \equiv g_c \mu \left( g_c \right)$ is the Newtonian gravity of the cluster at the location of the dwarf-elliptical encounter. These encounters would generally not occur when the dwarf is at the pericentre of its orbit around the Fornax Cluster. However, encounters at this point would be more damaging because the dwarf's self-gravity would be weaker. We therefore assume that the encounters with ellipticals take place at a typical distance from the cluster of $R_{\textrm{enc}} = 0.5 \, R$, which is slightly more than the pericentre distance of $0.29 \, R$ (Appendix \ref{Rper}) but less than the present distance.

We will first consider the heating rate $\dot{E}_{\textrm{{QN}}}$ from encounters in the QN regime before turning to the heating rate $\dot{E}_{\textrm{{IDM}}}$ from encounters in the IDM regime. The total heating rate is then
\begin{eqnarray}
	\label{E_dot_MOND}
	\dot{E}_{\textrm{MOND}} ~\equiv~ \dot{E}_{\textrm{Newt}} \times \textrm{CF} ~=~ \dot{E}_{\textrm{{QN}}} + \dot{E}_{\textrm{IDM}} \, ,
\end{eqnarray}
where CF is the correction factor that needs to be applied to the Newtonian $\dot{E}$ to make it MONDian. Our approach is to assume a sharp transition between the QN and IDM regimes such that the EFE is completely dominant in the former and completely negligible in the latter. This approximate approach should be accurate to within a factor of order unity, which we will argue later is sufficient for our purposes.

In all regimes, the heating rate due to encounters with an impact parameter in the range $b \pm db/2$ ($db \ll b$) is $\dot{E}_b = \dot{C} \, \langle \Delta \widetilde{E} \rangle$, where $\dot{C} \propto b \, db$ is the average rate of such encounters and $\langle \Delta \widetilde{E} \rangle$ is the average energy gain of the dwarf per unit mass due to each such encounter. Since accelerating the dwarf as a whole does not alter its internal structure, we only need to consider the variation in the impulse $\bm{u}$ across the dwarf, so $\langle \Delta \widetilde{E} \rangle \propto {\lvert \Delta \bm{u} \rvert}^2$. In Newtonian dynamics, the magnitude of the impulse on a passing test particle is $u \propto 1/b$, so $\lvert \Delta \bm{u} \rvert \propto 1/b^2$ \citep[equation 8.41 of][]{Binney_Tremaine_2008} and $\langle \Delta \widetilde{E} \rangle \propto 1/b^4$. This explains the $1/b^3$ scaling in the integrand in equation 8.53 of \citet{Binney_Tremaine_2008}, which states that the Newtonian heating rate per unit dwarf mass is:
\begin{eqnarray}
	\dot{E}_{\textrm{Newt}} ~=~ \underbrace{\frac{14}{3} \sqrt{2\mathrm{\pi}} \frac{G^2 M^2_p n_p r^2_{h,\textrm{dwarf}}}{\sqrt{2}\sigma}}_A \int^{\infty}_{r_{h,p}} \frac{db}{b^3} ~=~ \frac{A}{2 r^2_{h,p}} \, ,
	\label{E_dot_Newt}
\end{eqnarray}
where $A$ is a constant.

We are now in a position to MONDify this result for the QN regime. Both the dwarf's self-gravity and the elliptical's gravity on the dwarf are similar to the Newtonian result but with $G \to G_{\textrm{eff}}$. The heating rate in the QN regime is thus similar to Equation \ref{E_dot_Newt}, but using $G_{\textrm{eff}}$ instead of $G$ in the calculation of the normalization constant. To distinguish this result from the Newtonian case, we call the QN normalization constant $A' = A \left(G_{\textrm{eff}}/G \right)^2$. Since by definition the QN regime involves only those encounters with $b > r_{_{\textrm{EFE}}}$, the total heating rate from encounters in this regime is
\begin{eqnarray}
	\dot{E}_{\textrm{QN}} ~=~ A' \int^{\infty}_{r_{_{\textrm{EFE}}}} \frac{db}{b^3} ~=~ \frac{A'}{2 r^2_{\textrm{EFE}}} \, .
	\label{E_dot_QN}
\end{eqnarray}

In the IDM regime, the scalings are different because the gravity from the elliptical follows an inverse distance law. Since the interaction time-scale rises linearly with the closest approach distance, the impulse becomes independent of this ($u \propto b^0$). However, as the direction from the elliptical to the dwarf is still different for different parts of the dwarf, the variation in the impulse across it scales as $\lvert \Delta \bm{u} \rvert \propto 1/b$, implying that the energy gain per encounter scales as $\langle \Delta \widetilde{E} \rangle \propto {\lvert \Delta \bm{u} \rvert}^2 \propto 1/b^2$. Since the encounter rate again behaves as $\dot{C} \propto b \, db$ due to the geometry being the same in both models, we obtain:
\begin{eqnarray}
	\label{E_dot_IDM}
	\dot{E}_{\textrm{IDM}} ~&=&~ \frac{A'}{r^2_{\textrm{EFE}}} \int^{r_{_{\textrm{EFE}}}}_{r_{_{\textrm{MOND}}}} \frac{db}{b} \\
	~&=&~ \frac{A'}{r^2_{\textrm{EFE}}} \ln \left(\frac{r_{_{\textrm{EFE}}}}{r_{_{\textrm{MOND}}}}\right) \, .
\end{eqnarray}
The normalization of the integrand ensures continuity of the specific heating rate per unit $b$ between the QN and IDM regimes.

Inserting our results for $\dot{E}_{\textrm{QN}}$ and $\dot{E}_{\textrm{IDM}}$ into Equation \ref{E_dot_MOND} and noting that $A' = A \left( G_{\textrm{eff}}/G \right)^2$, we obtain that
\begin{eqnarray}
	\textrm{CF} ~=~ \left[1 + \ln \left(\frac{a_{_0}}{g_{c,N}}\right)\right] \left(\frac{G_{\textrm{eff}} \, r_{\textrm{h,p}}}{G \, r_{_{\textrm{EFE}}}}\right)^2 \, .
	\label{corr_fact}
\end{eqnarray}
Since $t_d \equiv \lvert E \rvert/\dot{E}$ and the MONDian binding energy of the dwarf exceeds the Newtonian result (Equation \ref{Binding_energy}) by a factor of ($G_{\textrm{eff}}/G$), the effect of the MOND corrections to Newtonian gravity amount to multiplying the Newtonian $t_d$ (Equation \ref{td_LCDM}) by a factor of $\textrm{CF}^{-1} \left( G_{\textrm{eff}}/G \right)$.
\begin{eqnarray}
	\label{td_MOND}
	t_{d, \textrm{MOND}} ~&\equiv&~ \frac{\lvert E \rvert_{\textrm{MOND}}}{\dot{E}_{\textrm{MOND}}} \\
	&=& \frac{0.043}{W_{p}} \frac{\sqrt{2} \sigma \, M_{\textrm{dwarf}} \, r_{_{\textrm{EFE}}}^2}{G_{\textrm{eff}} \, M_p^2 \, n_p \, r_{h,\textrm{dwarf}}^3 \left[1 + \ln \left(\frac{a_{_0}}{g_{c,N}}\right)\right]} \, . \nonumber
\end{eqnarray}
We assume $W_p = 1$ as in the Newtonian case. Our derivation assumes that $g_{c,N} \ll a_{_0}$, which is valid in the Fornax Cluster. In general, we recommend that the logarithmic term be omitted if $g_{c,N} > a_{_0}$.

\section{Tidal susceptibility}
\label{tidal_sus}

Now that we have defined the main effects which can disturb the structure of a dwarf in a galaxy cluster, we estimate the susceptibility of a dwarf to these effects in both $\Lambda$CDM and MOND. To quantify the disturbance caused by tides from the global cluster potential, we define the tidal susceptibility as the ratio between the half-mass radius $r_h$ and the tidal radius $r_{\textrm{tid}}$ of a dwarf:
\begin{eqnarray}
	\eta_{\textrm{rtid}} ~\equiv~ \frac{r_h}{r_{\textrm{tid}}} \, .
	\label{eta_rtid}
\end{eqnarray}
From the definition of $r_{\textrm{tid}}$ in both $\Lambda$CDM (Equation \ref{rtid_LCDM}) and MOND (Equation \ref{rtid_MOND}), we have that $r_{\textrm{tid}} \propto M^{1/3}$. This implies that:
\begin{eqnarray}
    \eta_{\textrm{rtid}} ~\propto~ \frac{r_h}{M^{1/3}} ~\propto~ \rho^{-1/3} \, .
\end{eqnarray}
Therefore, only the density $\rho$ of the dwarf is relevant to its tidal susceptibility in both $\Lambda$CDM and MOND.

If a dwarf has strong self-gravity (e.g. due to being surrounded by a dark matter halo or being in the deep-MOND regime), then the point at which the tidal force of the cluster will start to dominate over the self-gravity of the dwarf will be far from the centre of the dwarf. Therefore, the dwarf's $r_{\textrm{tid}}$ will be large and its tidal susceptibility will be low. Such a dwarf should be little disturbed by the cluster tides. If instead the dwarf has only weak self-gravity (e.g. because it is a TDG with little dark matter or because it is a MONDian dwarf but the EFE from the cluster is very significant), then the point at which the tidal force of the cluster will start to dominate over the self-gravity of the dwarf will be close to the dwarf's centre. Its $r_{\textrm{tid}}$ will then be small and its tidal susceptibility high. Such a dwarf would be significantly disturbed by tides. In the extreme case that $r_{\textrm{tid}} \ll r_h$ ($\eta_{\textrm{rtid}} \gg 1$), the dwarf will be destroyed within a few dynamical times. As a result, we need to consider the maximum value of $\eta_{\textrm{rtid}}$ attained throughout the trajectory, i.e., we need to evaluate $\eta_{\textrm{rtid}}$ at pericentre.

If the disturbance is caused by interaction with massive galaxies (harassment), we define the tidal susceptibility as the ratio between the age of the elliptical galaxies in the Fornax Cluster \citep[$t_{\textrm{Fornax}} \approx 10$~Gyr;][]{Rakos_2001} and the disruption time-scale $t_d$ of the dwarf, which we assume to typically be about as old as the cluster itself.
\begin{eqnarray}
	\eta_{\textrm{har}} ~\equiv~ \frac{t_{\textrm{Fornax}}}{t_d} \, .
	\label{eta_har}
\end{eqnarray}
According to this definition, if $t_{\textrm{Fornax}} \ll t_d$ for a dwarf, then it will hardly be susceptible to the effect of galaxy-galaxy harassment. If instead $t_{\textrm{Fornax}} \gg t_d$ for a dwarf, then we expect that it will be significantly disturbed due to this process.

Although our definitions for $\eta_{\textrm{rtid}}$ and $\eta_{\textrm{har}}$ differ somewhat because the former is a ratio of radii while the latter is a ratio of time-scales, both definitions share the feature that low values of $\eta$ indicate that a dwarf should be little affected by the process under consideration. In principle, there should not be any dwarf galaxies for which either $\eta \gg 1$. It is possible to have $\eta$ slightly above 1 due to projection effects and other subtleties like the time required to achieve destruction, which can be significant for $\eta_{\textrm{rtid}}$ as multiple pericentre passages may be required and the orbital period can be long (Section~\ref{cluster_tides}). However, we should very seriously doubt the validity of any theory which tells us that a significant fraction of the dwarf galaxies in a galaxy cluster have $\eta_{\textrm{rtid}} \gg 1$ or $\eta_{\textrm{har}} \gg 1$. It is harder to falsify a theory in the opposite limit where it yields very low values for both measures of $\eta$ for all the dwarfs in a galaxy cluster. In this case, we could gain evidence against the theory if there is strong evidence that the dwarf galaxy population has been significantly affected by tides. In this project, we apply these considerations to the dwarf galaxy population in the Fornax Cluster.

\subsection{Tidal susceptibility of the Fornax dwarfs}
\label{tidal_sus_Fornax}

\begin{figure*} 
	\includegraphics[width = 8.5cm]{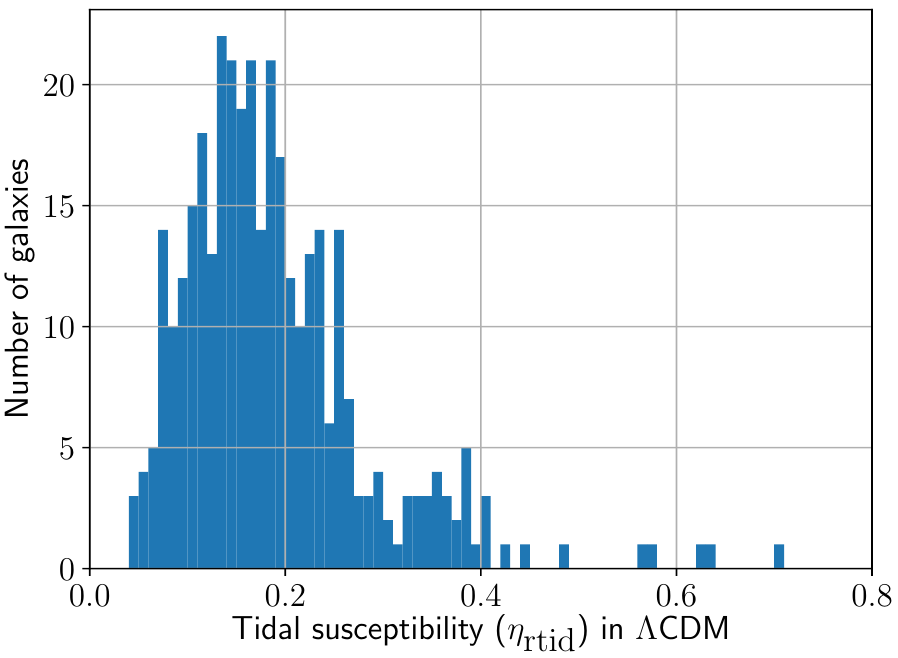} \hspace{0.07cm}
	\includegraphics[width = 8.5cm]{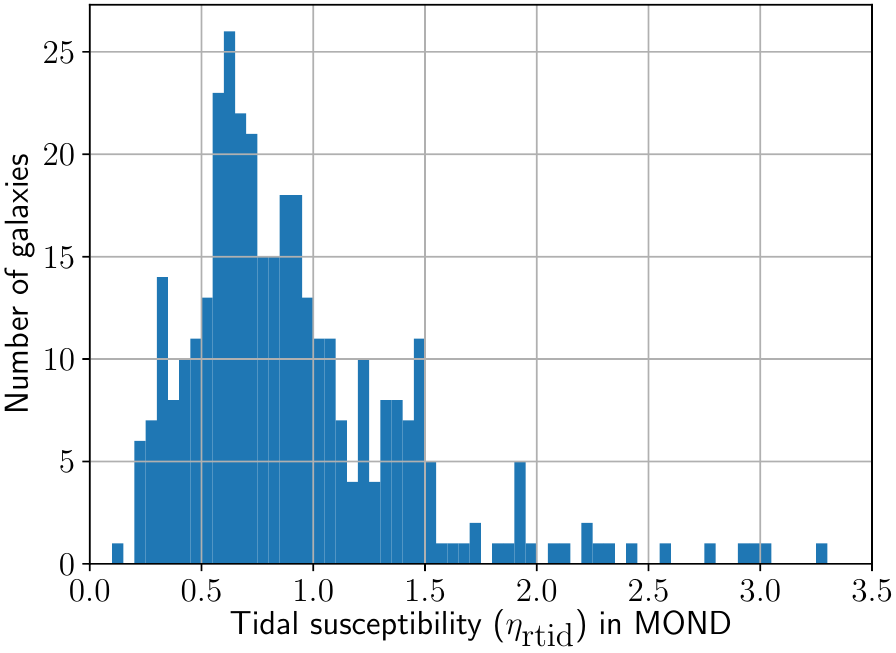} \\
	\includegraphics[width = 8.5cm]{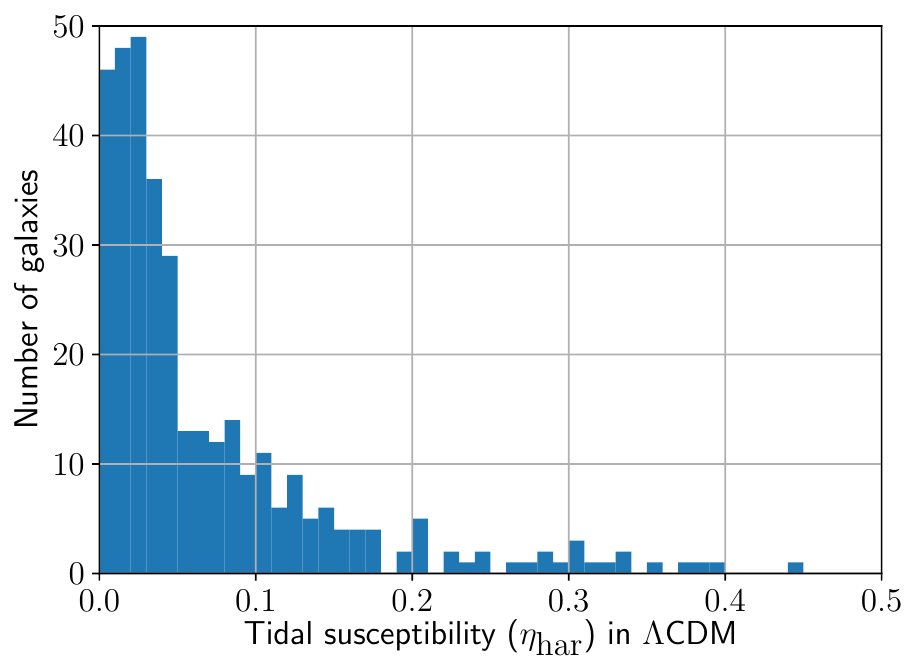} \hspace{0.07cm}
	\includegraphics[width = 8.5cm]{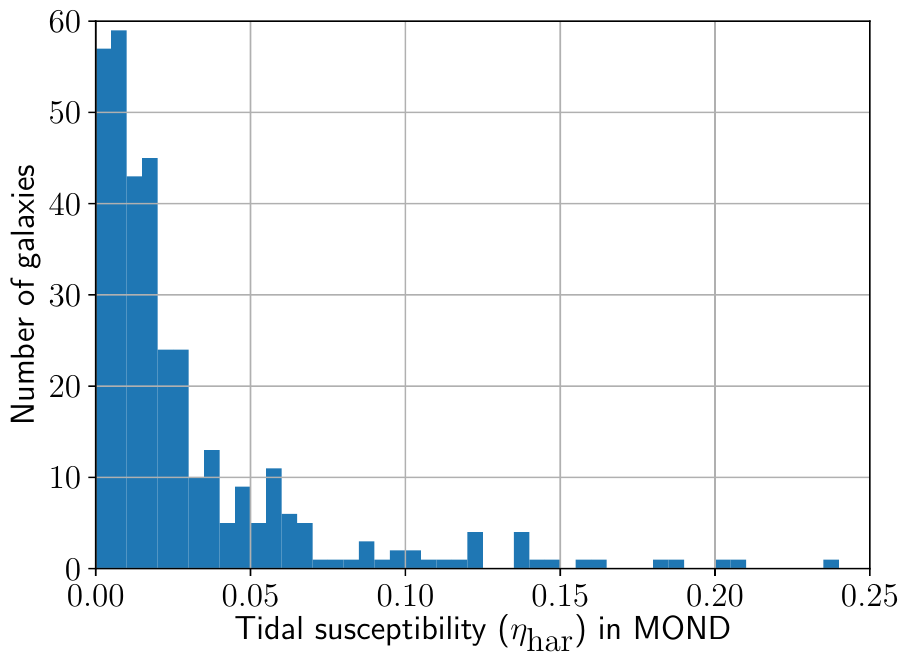}
	\caption{Histogram of the tidal susceptibility values of Fornax Cluster dwarfs in $\Lambda$CDM (left column) and MOND (right column) to tides from the overall cluster potential (top row) and harassment by interactions with individual massive galaxies (bottom row). The bin widths are: 0.01 (top left), 0.05 (top right), 0.01 (bottom left), and 0.005 (bottom right). Notice the different $\eta_{\textrm{rtid}}$ scales for $\Lambda$CDM and MOND. In both theories, we typically have $\eta_{\textrm{har}} \ll \eta_{\textrm{rtid}}$.}
	\label{fig:hist_tidal_sus}
\end{figure*}

Our first quantitative result is the susceptibility of dwarfs in the FDS catalogue to cluster tides, which we calculate in $\Lambda$CDM and MOND using Equations \ref{rtid_LCDM} and \ref{rtid_MOND}, respectively. We show the results as histograms in the top row of Fig.~\ref{fig:hist_tidal_sus}, with $\Lambda$CDM shown on the left and MOND on the right. The $\eta_{\textrm{rtid}}$ values are $\approx 5\times$ higher in MOND than in $\Lambda$CDM. Since an isolated dwarf has a similar amount of self-gravity in both frameworks by construction, the difference in $\eta_{\textrm{rtid}}$ values is primarily caused by the EFE weakening the self-gravity of a MONDian dwarf as it approaches the cluster centre (Section~\ref{Introduction}). This effect does not exist for a $\Lambda$CDM dwarf, which would retain the same dark matter fraction within its baryonic extent throughout its trajectory.

The bottom row of Fig.~\ref{fig:hist_tidal_sus} shows the susceptibility of FDS dwarfs to galaxy-galaxy harassment according to $\Lambda$CDM (Equation~\ref{td_LCDM}) and MOND (Equation~\ref{td_MOND}). In both theories, the histogram of $\eta_{\textrm{har}}$ peaks at very low values such that $\eta_{\textrm{har}} \ll \eta_{\textrm{rtid}}$ and $\eta_{\textrm{har}} \ll 1$. Therefore, both frameworks predict that the FDS dwarfs should be little affected by interactions with massive elliptical galaxies in the Fornax Cluster.

This implies at face value that in $\Lambda$CDM, the observed signs of tidal disturbance \citep[section 7.4 of][]{Venhola_2022} cannot be assigned to either cluster tides or to harassment. Since we explore the impact of cluster tides more carefully later in this contribution, we briefly reconsider our calculation of $\eta_{\textrm{har}}$. As explained in Section~\ref{harassment}, one simplifying assumption we made is that there are 48 equal mass and equal size perturbers within the 0.77~Mpc virial radius of the Fornax Cluster. However, the heating rate due to any individual perturber scales as $\dot{E} \propto \left( M_p/r_{h,p} \right)^2$ (Equation~\ref{E_dot_Newt}). We can use this to find the ratio $\dot{E}/\dot{E}_{\textrm{fid}}$ between the heating rate due to individual perturbers and the assumed heating rate $\dot{E}_{\textrm{fid}}$ for an `average' perturber with $M_{\star} = 10^{10} \, M_{\odot}$ and $r_{h,p} = 4$~kpc, taking into account that the actual mass and size are larger in $\Lambda$CDM and assuming a de Vaucouleurs profile for the stars \citep{De_Vaucouleurs_1948}. We obtain that in descending order of $M_{\star}$, the ratio $\dot{E}/\dot{E}_{\textrm{fid}}$ for the perturbers listed in table~C1 of \citet{Iodice_2019} is 14.7 (FCC 219), 42.7 (FCC 167), 10.3 (FCC 184), 4.76 (FCC 161), 5.41 (FCC 147), 11.8 (FCC 170), 1.06 (FCC 276), 1.93 (FCC 179), and 0.13 (FCC 312). Other perturbers have $M_{\star} < 10^{10} \, M_{\odot}$, so we assume their contribution to the heating rate is small. Adding up the above ratios and averaging over 48 perturbers (many of which are too low in mass to appreciably harass Fornax dwarfs), we get that $\dot{E}/\dot{E}_{\textrm{fid}}$ is on average 1.9. Therefore, using a more accurate treatment of the heating rate would not change our conclusion that the FDS dwarfs are not really susceptible to galaxy-galaxy harassment: Doubling all the $\eta_{\textrm{har}}$ values would still lead to its distribution having a mode $<0.1$ and all the dwarfs having $\eta_{\textrm{har}} < 1$.

Moreover, using $t_{\textrm{Fornax}}$ as the time-scale for interactions is an optimistic assumption $-$ dwarfs in $\Lambda$CDM may have been accreted by the cluster long after they formed, while in MOND they could be TDGs that formed more recently \citep{Renaud_2016}. This implies that the dwarfs would not have experienced that many encounters with elliptical galaxies, which themselves might only have been accreted $\ll 10$~Gyr ago. As an example, we may consider the case of FCC 219 $\equiv$ NGC 1404, the most massive perturber listed in table~C1 of \citet{Iodice_2019} in terms of $M_{\star}$. Its radial velocity exceeds that of the brightest cluster galaxy NGC 1399 by 522~km/s, but modelling indicates that the relative velocity could be higher still as most of it should lie within the sky plane \citep{Machacek_2005}. Moreover, NGC 1404 appears to lie in front of the Fornax Cluster: Its heliocentric distance is only $18.7 \pm 0.3$~Mpc \citep{Hoyt_2021}, whereas the distance to NGC 1399 is $20.0 \pm 0.3$~Mpc \citep{Blakeslee_2009}. Detailed modelling in a $\Lambda$CDM context indicates that although NGC 1404 is not on a first infall, it has likely spent $\la 3$~Gyr within the cluster \citep{Shearman_2018}. During this time, the high relative velocity would have reduced the heating rate on any dwarf galaxy that it came near (Equation~\ref{E_dot_Newt}). It is therefore clear that $\eta_{\textrm{har}}$ is overestimated by assuming that both all the dwarfs and all 48 massive ellipticals were in the virial volume of the Fornax Cluster over the last 10~Gyr.

Based on this, we will neglect the role of harassment in what follows and focus on cluster tides.\footnote{This is consistent with the previous $\Lambda$CDM result that harassment is not very significant for dwarfs in a Virgo-like cluster \citep{Smith_2015}.} Thus, $\eta$ will be used to mean $\eta_{\textrm{rtid}}$ unless stated otherwise. An important example of this is our discussion of Newtonian TDGs that are purely baryonic, where $\eta_{\textrm{har}}$ plays an important role (Appendix~\ref{tidal_sus_newton}).

\subsection{Testing the effect of cluster tides on Fornax dwarfs}
\label{surf_dens_dwarfs}

\begin{figure}
	\centering
	\includegraphics[width = 8.5cm]{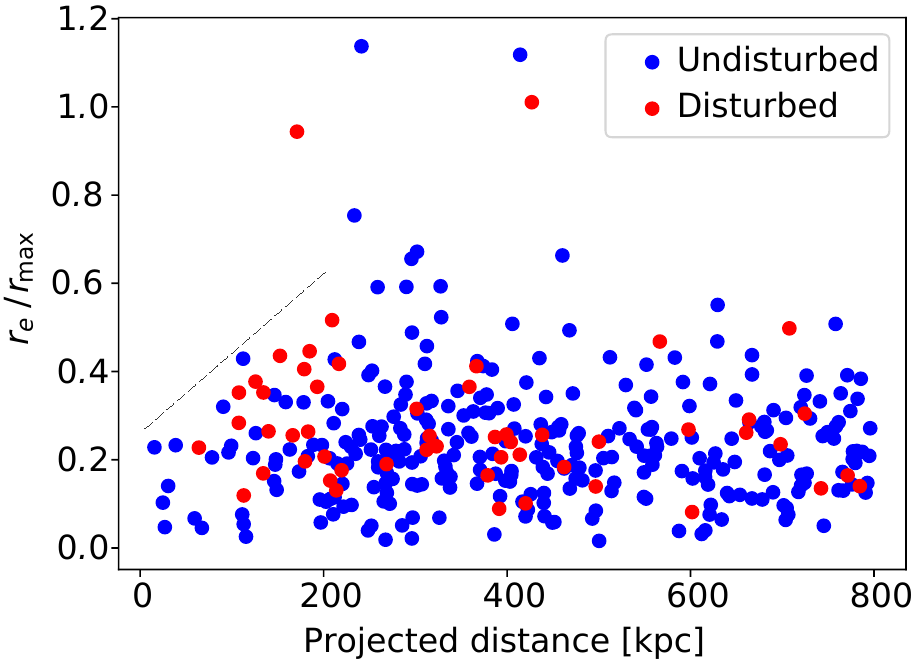}
	\caption{Distribution of the projected distances of Fornax Cluster dwarfs against $r_e/r_{\textrm{max}}$, where $r_e$ is the projected half-light radius and $r_{\textrm{max}}$ is the maximum $r_e$ at fixed $M_{\star}$ that the dwarf can have to remain detectable given the surface brightness limit of the survey of $27.8 \, \textrm{mag} \, \textrm{arcsec}^{-2}$ in the $r'$ band. Dwarfs visually classified as `undisturbed' are shown in blue, while those classified as `disturbed' are shown in red. Notice the lack of low surface brightness dwarfs near the cluster centre. We have emphasized this by drawing a dashed grey line for illustrative purposes, which we interpret as a tidal edge. This interpretation is bolstered by the lack of dwarfs above this line and the high proportion of disturbed dwarfs just below this line.}
	\label{fig:tid_edge}
\end{figure}

A significant fraction of the FDS dwarfs appear disturbed in a manual visual classification \citep[Fig.~\ref{fig:tid_morph_class}; see also][]{Venhola_2022}. To check if cluster tides are truly the main mechanism responsible for the apparent disturbance of the Fornax dwarfs $-$ as our results in Fig.~\ref{fig:hist_tidal_sus} seem to suggest $-$ in Fig.~\ref{fig:tid_edge} we plot the projected distance of the selected Fornax dwarfs against the ratio between their effective radius $r_e$ and $r_{\textrm{max}}$, where $r_{\textrm{max}}$ is the maximum $r_e$ that the dwarf could have to remain detectable given its $M_{\star}$ and the FDS detection limit of $27.8 \, \textrm{mag} \, \textrm{arcsec}^{-2}$. Dwarfs with larger size at fixed stellar mass $-$ i.e., lower surface brightness dwarfs $-$ are more susceptible to tides and will be more easily destroyed, especially near the cluster centre where the tides are stronger. In Fig.~\ref{fig:tid_edge}, we can see a deficit of low surface brightness dwarfs near the cluster centre. The absence of dwarfs in this region of the parameter space cannot be explained by the survey detection limit as we find an increasing number of dwarfs with the same or lower surface brightness at larger $R_{\textrm{sky}}$, e.g. if we consider a horizontal line at $r_e/r_{\textrm{max}} = 0.4$. This tendency is highlighted in Fig.~\ref{fig:tid_edge} using a sloped dotted line that appears to be a tidal edge. Further from the cluster, its tides become weaker, so it is quite possible that dwarfs in this region are not much affected by tides.

Additional evidence for the importance of tides towards the cluster centre comes from the colours of the dots in Fig.~\ref{fig:tid_edge}, which indicate whether the dwarf visually appears disturbed (red) or undisturbed (blue). Just below the claimed tidal edge, we would expect that the dwarfs are much more likely to appear disturbed as they should be close to the threshold of being destroyed altogether. This is indeed apparent: The proportion of disturbed galaxies is much higher in this part of the parameter space.\footnote{This is not expected if the disturbances are due to harassment because dwarfs subject to this would be well mixed throughout the cluster \citep{Smith_2015}.}

\begin{figure}
	\centering
	\includegraphics[width = 8.5cm]{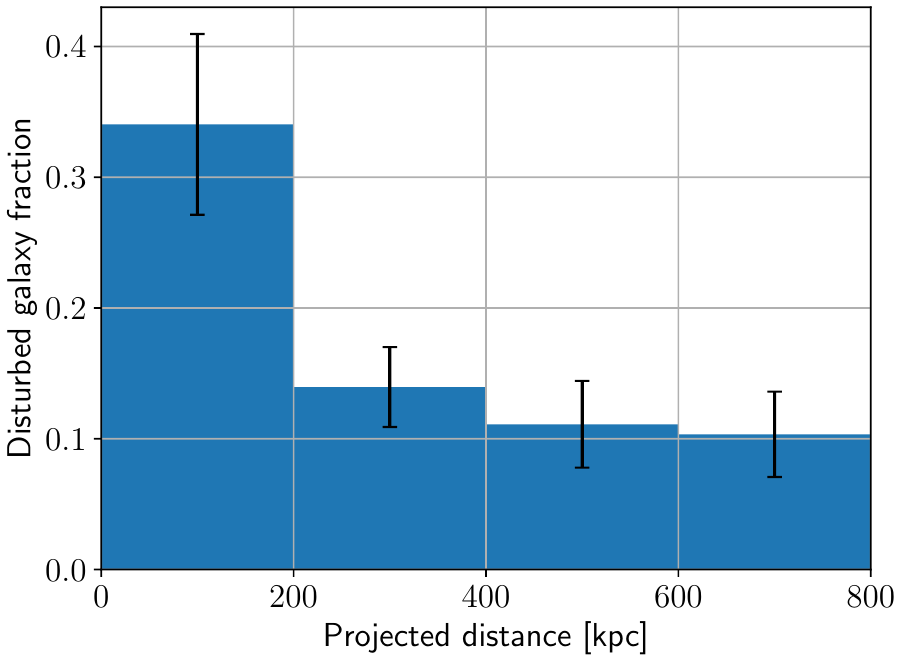}
	\caption{The proportion of Fornax dwarfs that appear disturbed in different projected separation bins of width 200~kpc. The error bars show the binomial uncertainty assuming the likelihood of appearing disturbed is the same as the proportion of disturbed dwarfs.\textsuperscript{\ref{Uncertainty_fixed_P}}}
	\label{fig:hist_disturb}
\end{figure}

To emphasize this trend further, we use Fig.~\ref{fig:hist_disturb} to show the observed fraction of disturbed dwarfs ($f_d$) in different $R_{\textrm{sky}}$ bins. This is found as $f_d = S/T$, with the uncertainties calculated using binomial statistics as $\sqrt{S \left( T - S \right)/T^3}$, where $T$ is the number of galaxies in each $R_{\textrm{sky}}$ bin and $S \leq T$ is the number of these galaxies which appear disturbed.\footnote{\label{Uncertainty_fixed_P}This is based on the binomial uncertainty in $S$ assuming that the probability of a galaxy appearing disturbed in each $R_{\textrm{sky}}$ bin is $f_d = S/T$. In reality, $f_d$ is not precisely constrained by the observations $-$ we handle this complexity later (Equation~\ref{Bernoulli_mean_stdev}).} As expected from our previous results, $f_d$ is very high in the central 200 kpc of the Fornax Cluster. Although $f_d$ is very low further out, it is still non-zero and remains so out to the largest distances covered by our dataset. We attribute this to the complexities of visually assessing whether a dwarf is tidally disturbed: If a dwarf appears asymmetric due to observational difficulties or due to a dense star cluster on one side, this could lead to a false positive. It is also possible that the dwarf is genuinely disturbed due to a recent close encounter with a massive galaxy in the cluster, which could happen even in the cluster outskirts. When we construct a detailed model of the Fornax Cluster dwarf galaxy population in Section~\ref{subsubsec:dwarf_disturbance}, we will need to allow a non-zero likelihood that a dwarf appears disturbed even if it is unaffected by cluster tides.

\subsection{Correlating tidal susceptibility with the observed level of disturbance}
\label{comparison_disturbance}

Having obtained the tidal susceptibility $\eta$ of each Fornax dwarf in our sample (Section~\ref{tidal_sus_Fornax}), we can compare this to its visual level of disturbance. We do so using the proportion of dwarfs classified as disturbed in each $\eta$ bin, which is similar to the analysis shown in Fig.~\ref{fig:hist_disturb} but binning in $\eta$ instead of $R_{\textrm{sky}}$. We consider each $\eta$ bin as an experiment with $T$ trials (dwarfs) out of which $S$ are `successes' (disturbed-looking dwarfs). We then use binomial statistics to infer the probability distribution of the disturbed fraction $f_d$ assuming a uniform prior over the range $0-1$ and applying Bayes' Theorem. The mean and standard deviation of $f_d$ are:
\begin{eqnarray}
    \label{Bernoulli_mean_stdev}
    \textrm{mean} &=& \frac{S + 1}{T + 2} \, , \\
	\textrm{standard deviation} &=& \frac{1}{T + 2}\sqrt{\frac{ \left( S + 1 \right) \left( T - S + 1 \right)}{ \left( T + 3 \right)}} \, . \nonumber
\end{eqnarray}
For the extreme case $S = T = 0$, we expect that the probability distribution of $f_d$ is uniform over the range $0-1$ as there is no data. In this case, we recover the standard result that the mean of this distribution is $1/2$ and its variance is $1/12$.

\begin{figure}
	\centering
	\includegraphics[width = 8.5cm]{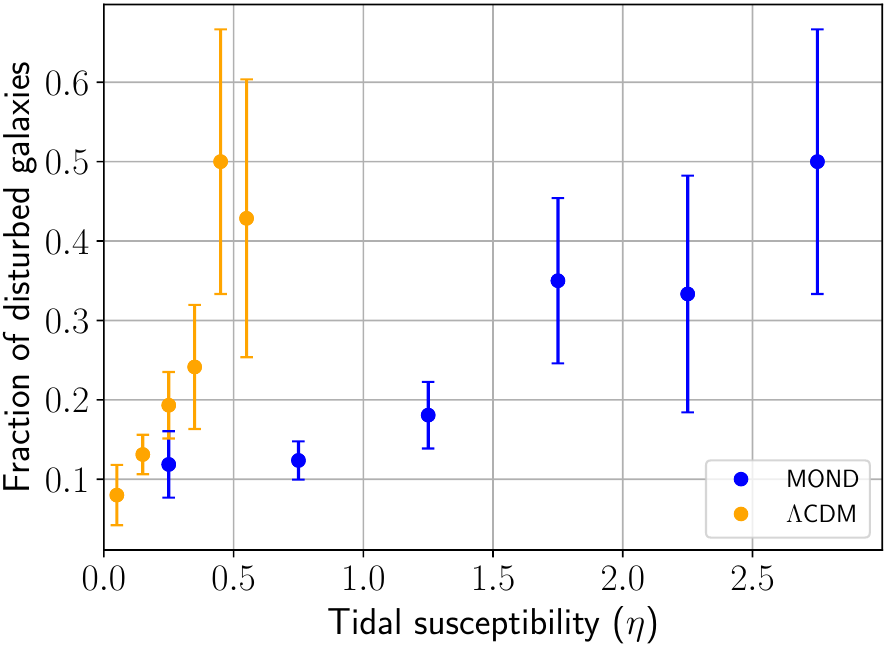}
	\caption{The likelihood that a dwarf appears disturbed in each tidal susceptibility bin in $\Lambda$CDM (orange) and MOND (blue). The value and uncertainty are calculated using binomial statistics (Equation \ref{Bernoulli_mean_stdev}) and plotted at the centre of each bin. The bin width is 0.5 for MOND and 0.1 for $\Lambda$CDM. In both cases, the last bin also includes all dwarfs with higher $\eta$. Notice that the likelihood of a dwarf appearing disturbed rises with $\eta$. The higher uncertainties at high $\eta$ are due to a small sample size (see the $\eta$ distribution in Fig.~\ref{fig:hist_tidal_sus}).}
	\label{fig:tid_sus_obs}
\end{figure}

We use Fig.~\ref{fig:tid_sus_obs} to plot the mean and standard deviation obtained in this way against the central $\eta$ value for the bin under consideration. In both $\Lambda$CDM and MOND, a clear trend is apparent whereby dwarfs with higher $\eta$ are more likely to appear disturbed. We quantify this by dividing the FDS sample into two subsamples where $\eta$ is below or above some threshold $\eta_t$, thereby assuming only a monotonic relation between $f_d$ and $\eta$ that is not necessarily linear. Appendix~\ref{Binomial_significance} explains how we obtain the likelihood that the same $f_d$ can explain the number of disturbed dwarfs and the total number of dwarfs in both subsamples given binomial uncertainties. Using this method, we find that the `signal' is maximized in $\Lambda$CDM if we use $\eta_t = 0.36$, in which case the null hypothesis of $f_d$ being the same in both subsamples can be rejected at a significance of $P = 4.1 \times 10^{-3}$ ($2.87\sigma$). If instead we use MOND, the optimal $\eta_t = 0.85$ and the significance rises to $P = 4.4 \times 10^{-4}$ ($3.52\sigma$).\footnote{Section~\ref{discussion} provides a more rigorous quantification of how confident we can be that $f_d$ rises with $\eta$.} Though both theories imply that $f_d$ is higher in the high $\eta$ subsample, $f_d$ starts rising at a much lower value of $\eta$ in $\Lambda$CDM than in MOND, as clearly shown by the optimal $\eta_t$ values. We may expect that dwarfs start to look disturbed when their half-mass radius is about the same as their tidal radius, so $f_d$ should start rising only when $\eta \ga 0.5$. This is not the case in $\Lambda$CDM, which implies that dwarfs are more likely to be classified as disturbed once their $\eta \ga 0.1-0.2$. A dwarf with such a low $\eta$ should be little affected by tides, indicating a problem for this framework. In the MOND case, we see that dwarfs start being classified as disturbed more often once their $\eta \ga 1-1.5$, which is much more plausible physically.

Another important aspect is the overall distribution of $\eta$, whose decline towards the highest bin is responsible for a larger uncertainty in the probability of appearing disturbed. The distribution of $\eta$ is shown explicitly in the top row of Fig.~\ref{fig:hist_tidal_sus}. There are no $\Lambda$CDM dwarfs with $\eta > 0.7$, even though a dwarf with $\eta = 0.7$ should still be tidally stable. In MOND, the maximum $\eta \approx 3$, though there are very few dwarfs with $\eta > 1.5$. The high calculated $\eta$ for these dwarfs could indicate that they lie very close to the cluster centre in projection but not in reality. To handle such projection effects and other uncertainties like the unknown orbital eccentricity distribution of the dwarfs, we next construct a test mass simulation of the dwarf galaxy population in the Fornax Cluster.

\section{Test mass simulation of the Fornax Cluster}
\label{test_mass}

In order to quantify the aforementioned trends and thereby obtain the range of values that the minimum $\eta$ required for disturbance and the $\eta$ required for destruction can have to be consistent with observations $-$ both in $\Lambda$CDM and in MOND $-$ we need to construct a forward time-evolution model of the Fornax Cluster. With this forward model, we can also account for projection effects that can make dwarfs appear closer to the cluster centre than they actually are. In this section, we describe the set-up of the simulated Fornax system with test masses, as well as the methods that we use to quantify the properties of the Fornax dwarfs and their orbits. Here we focus only on those dwarfs classified as `non-nucleated' as this type of galaxy is more numerous than the `nucleated' type. Moreover, having the same deprojection method (Appendix~\ref{deproj}) for all dwarfs will simplify the analysis. Removing the nucleated dwarfs from the sample leaves us with 279 dwarfs.

\subsection{Orbit integration}
\label{orbit_integration}

The first step in building a simulation of test masses orbiting in the observed cluster potential is to generate a grid of orbits for a wide range of semi-major axis ($R_i$) and eccentricity ($e$) values, with the integrations started at $R = R_i$. The initial radii have a range of values from 15~kpc to 2015~kpc, while the eccentricities cover the full range of values for an ellipse ($0 < e < 1$). The grid is divided into $100 \times 100$ cells. Initially, we assign the test mass a mass and half-mass radius which are typical for a Fornax dwarf ($M_{\textrm{dwarf}} = 3.16 \times 10^7~M_{\odot}$ and $r_h = 0.84$~kpc), but these values are not relevant as the results will be rescaled later according to the distribution of dwarf densities in the system (Section~\ref{subsubsec:dwarf_density}).

We initialize the simulated dwarfs for every possible combination of $R_i$ and $e$ as described below. We start the simulation at the semi-major axis of the orbit, where the velocity $v$ satisfies
\begin{eqnarray}
	v ~=~ v_c ~=~ \sqrt{-\bm{r} \cdot \bm{g}} \, .
	\label{v_i}
\end{eqnarray}
As discussed in section~2.3.1 of \citet{Banik_2018_Centauri}, the eccentricity $e$ is defined such that
\begin{eqnarray}
	e ~\equiv~ \lvert \widehat{\bm{r}} \cdot \widehat{\bm{v}} \rvert \, ,
	\label{e_i}
\end{eqnarray}
where $r = R_i \widehat{\bm{r}}$ and $\bm{g} = -g_c \widehat{\bm{r}}$, with $\widehat{\bm{v}}$ indicating the unit vector parallel to any vector $\bm{v}$ of length $v$. The modulus is not required in our case because we start with the dwarf going away from the cluster if $e > 0$. Using Cartesian coordinates, we define the initial positions and velocities of the orbit as:
\begin{eqnarray}
	x ~&=&~ R_i \, , \\
	y ~&=&~ 0 \, , \\ \vspace{10pt}
	v_x ~&=&~ ve \, , \\
	v_y ~&=&~ v\sqrt{1 - e^2} \, .
\end{eqnarray}
Equation \ref{v_i} defines $v$ and Equation \ref{e_i} sets the component of $\bm{v}$ along the radial direction. $v_y$ is the remaining tangential velocity.

In order to obtain the positions and velocities of the simulated dwarf at each point of the orbit, we implement a fourth-order Runge-Kutta integrator in 2D. To ensure that the time-step we use for each iteration is computationally efficient but also small enough to yield accurate results, we use an adaptive time-step that depends on the dynamical time-scale at the instantaneous orbital radius $R$:
\begin{eqnarray}
    dt ~=~ 0.01 \sqrt{\frac{R}{g_c}} \, .
\end{eqnarray}
We evolve the system for $t_{\textrm{Fornax}} = 10$~Gyr, the estimated age of the system \citep{Rakos_2001}. At each time-step, we calculate the tidal radius of the simulated dwarf at its current position and, by comparing this with the half-mass radius, we obtain its instantaneous tidal susceptibility $\eta$. We record the $e$ value of each simulated orbit and its final $R$, the distance with respect to the cluster centre at which we should be seeing the dwarf today. We also record two $\eta$ values in each orbit simulation: the maximum $\eta$ over the whole simulation ($\eta_{\textrm{max}}$), and the maximum $\eta$ in the last 2~Gyr ($\eta_{\textrm{max, recent}}$). We use $\eta_{\textrm{max}}$ to decide whether the dwarf is destroyed and should be removed from our statistical analysis. If not, then $\eta_{\textrm{max, recent}}$ is used to set the likelihood that the dwarf appears disturbed. This is because we expect a dwarf to return to a nearly undisturbed appearance if it experiences only low $\eta$ values along its orbit for over 2~Gyr, provided $\eta$ is never so high as to destroy the dwarf.

\subsection{Assigning probabilities to the orbits}
\label{orbits_probability}

The orbital and internal properties of the Fornax dwarfs (e.g., the radial profile of the orbits, the distribution of their eccentricities, the likelihood of appearing perturbed) follow certain probability distributions. Because of this, we assign probabilities to each of our simulated orbits by fitting them to a few crucial observed properties (next subsection) in order to make our simulated system as similar as possible to the observed Fornax dwarf galaxy system. The parameters governing these probability distributions are described below.

\subsubsection{Number density of dwarfs}
\label{subsubsec:dwarf_number_density}

The number density $n$ of dwarfs is assumed to be a function only of the distance $R$ from the cluster centre. It is related to the radial probability distribution $P_r$ as: $n \propto P_r/R^2$. We assume that $P_r$ is described by a double power-law:
\begin{eqnarray}
	P_r ~=~ R^2 \left( R + r_{\textrm{core}} \right)^{\textrm{Slope}_{P_r}} \, ,
	\label{P_r}
\end{eqnarray}
where $r_{\textrm{core}}$ is the radius of the constant density central region of the Fornax Cluster and $\textrm{Slope}_{P_r}$ is the power-law slope of the radial profile in the cluster outskirts. To obtain a convergent number of dwarfs, $\textrm{Slope}_{P_r} < -3$.

\subsubsection{The eccentricity distribution}
\label{subsubsec:dwarf_eccentricity}

For the probability distribution of the orbital eccentricities, we assume a linear function as in \citet{Banik_2018_Centauri}:
\begin{eqnarray}
	P_e ~=~ 1 + \textrm{Slope}_{P_e} \left( e - \frac{1}{2} \right) \, ,
	\label{P_e}
\end{eqnarray}
where $\textrm{Slope}_{P_e}$ is the slope of the eccentricity probability distribution.

\subsubsection{Distribution of dwarf densities}
\label{subsubsec:dwarf_density}

The tidal susceptibility of a dwarf depends on both its mass and its radius, which in general differ from the values assumed in our test mass simulation. As discussed below Equation~\ref{eta_rtid}, the mass and radius of a dwarf affect its tidal susceptibility only to the extent that they affect its density $\rho$. Therefore, the $\eta$ values that we recorded in Section~\ref{orbit_integration} should be multiplied by a density-related factor accounting for the difference between the intended density $\rho$ and the fixed value $\rho_0$ assumed in that section. We therefore set
\begin{eqnarray}
    \eta_{\textrm{max}} ~&=&~ \eta_{\textrm{max, 0}} \left( \frac{\rho}{\rho_0} \right)^{-1/3} \, , \\
    \eta_{\textrm{max, recent}} ~&=&~ \eta_{\textrm{max, recent, 0}} \left( \frac{\rho}{\rho_0} \right)^{-1/3} \, ,
\end{eqnarray}
where the `0' subscript denotes values obtained in Section~\ref{orbit_integration}. The $-1/3$ exponent comes from the fact that $\eta \propto {r_h}/M^{1/3}$ in both theories.

\begin{figure}
	\centering
	\includegraphics[width = 8.5cm]{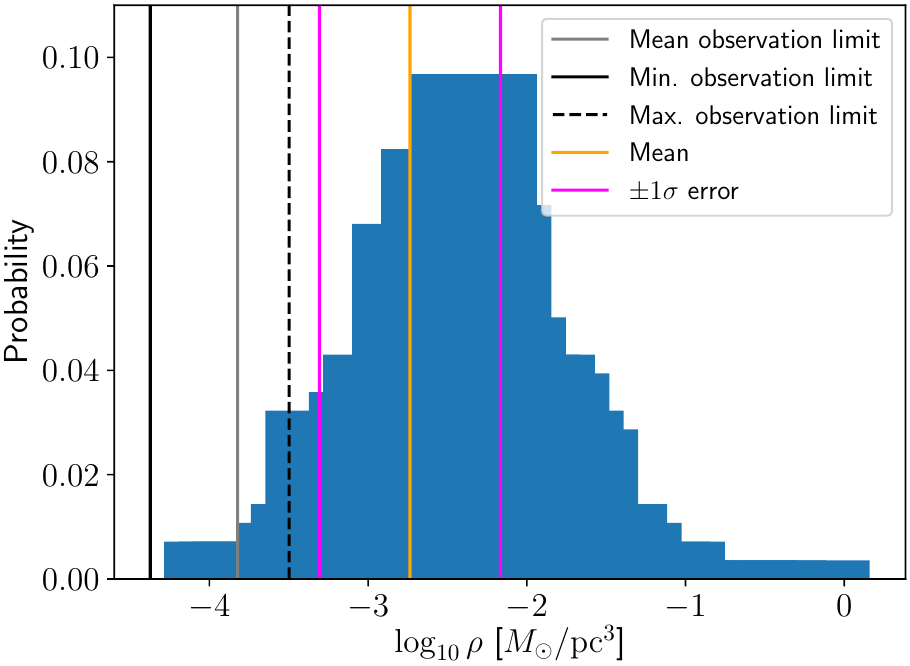}
	\caption{The distribution of each dwarf galaxy's mean baryonic density $\rho$ within its half-mass radius. The orange vertical line shows the sample mean, while the magenta lines offset by $\pm 0.57$~dex show the standard deviation around it. The grey line shows the density of a dwarf corresponding to the observational surface density detection limit of $\Sigma_{\textrm{min}} = 0.26~M_{\odot}/\textrm{pc}^2$ assuming the mean $M/L_{r'} = 1.10$ and mean $\rho / \Sigma = 0.59 \, \textrm{kpc}^{-1}$. If instead we add (subtract) the standard deviation in the $M/L_{r'}$ ratios, we have that $\Sigma_{\textrm{min}} = 0.35 \left( 0.17 \right) \, M_{\odot}/\textrm{pc}^2$. From these and by adding (subtracting) the standard deviation in the $\rho/\Sigma$ ratios, we obtain the $\rho$ value given by the dashed (solid) black line.}
	\label{fig:hist_dens}
\end{figure}

The density $\rho$ of each Fornax dwarf within its $r_h$ can be inferred from the data in the FDS catalogue using $\rho = 3 M_{\star}/ \left( 8 \mathrm{\pi} r_h^3\right)$. Fig.~\ref{fig:hist_dens} shows a histogram of the so-obtained densities of these dwarfs, from which it can be seen that the FDS distribution of $\log_{10} \, \rho$ follows a Gaussian distribution with mean $-2.74$ in units of $M_{\odot}/\textrm{pc}^3$. Therefore, when we assign a density to each of the simulated dwarfs obtained in Section~\ref{orbit_integration}, we associate a probability to this density according to a log-normal distribution. This is assumed to be independent of $R_i$ since the central region of a cluster should be able to accrete dwarfs that formed further out, leading to mixing of dwarfs that formed in different positions within the cluster.

\begin{figure}
	\centering
	\includegraphics[width = 8.5cm]{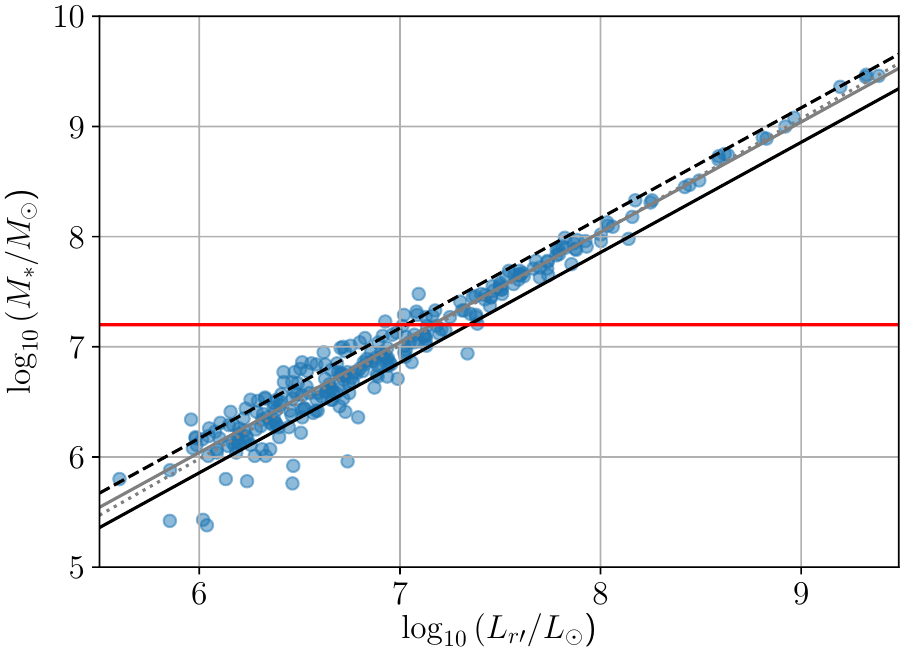}
	\caption{The relation between mass and luminosity in the $r'$-band for the non-nucleated sample of dwarfs in the FDS catalogue, with both shown on a $\log_{10}$ scale. The assumed $M/L$ ratios are based on empirical relations with the colour (Section~\ref{Fornax}). The dotted grey line shows the linear regression, while the solid grey line shows our adopted fit assuming a slope of 1. The dashed (solid) black line shows one standard deviation above (below) the mean $M_{\star}/L_{r'}$. The horizontal red line at $10^{7.2} \, M_{\odot}$ shows the stellar mass below which core formation is inefficient in $\Lambda$CDM (Section~\ref{discussion}).}
	\label{fig:M_L}
\end{figure}

\begin{figure}
	\centering
	\includegraphics[width = 8.5cm]{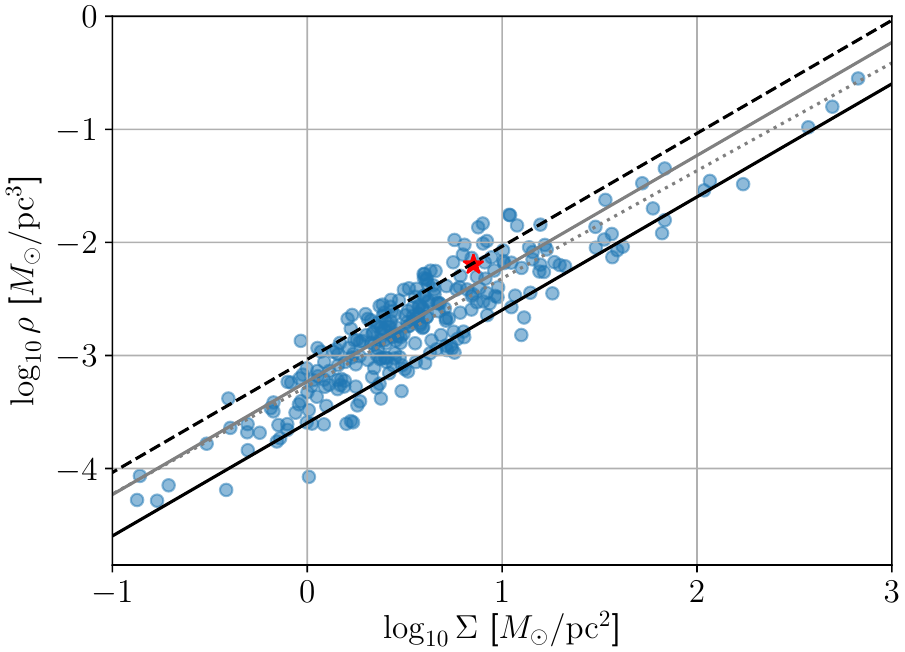}
	\caption{The relation between the average 3D mass density of baryons within their half-mass radius and their surface mass density within their projected half-light radius for our galaxy sample. The dotted grey line shows the linear regression, while the solid grey line shows our adopted fit assuming a slope of 1. The dashed (solid) black line shows one standard deviation above (below) the mean $\rho/\Sigma$. The red star shows the values for the dwarf galaxy used in our \textit{N}-body simulations (Section~\ref{Nbody_sim}).}
	\label{fig:surfdens_voldens}
\end{figure}

In order to set the lowest density that can be assigned to a dwarf in a way that is consistent with the observational constraints of the FDS, we check down to which surface brightness $\mu$ dwarfs can be detected in this survey. The limiting $\mu$ is given by the $1\sigma$ signal-to-noise threshold per pixel, which in the FDS is $27.8 \, \textrm{mag} \, \textrm{arcsec}^{-2}$ in the red band \citep[section~4.1 of][]{Venhola_2018}. To infer the corresponding $\rho$, we first convert this $\mu$ value to astronomical units ($L_{\odot}/\textrm{pc}^2$):
\begin{eqnarray}
	\log_{10} \, \mu \left[ L_{\odot}/\textrm{pc}^2 \right] = \frac{\mu \left[ \textrm{mag} \, \textrm{arcsec}^{-2} \right] - 21.57 - \textrm{Mag}_{\odot}}{-2.5} \, ,
\end{eqnarray}
where $\textrm{Mag}_{\odot} = 4.65$ is the absolute magnitude of the Sun in the red band \citep[table 3 of][]{Willmer_2018}. This gives $\mu_{\textrm{min}} = 0.23 \, L_{\odot}/\textrm{pc}^2$. We then use the mass-luminosity relation (solid grey line in Fig.~\ref{fig:M_L}) to obtain that $M/L_{r'} = 1.10 \pm 0.38 \, M_{\odot}/L_{\odot, r'}$. From this we can convert $\mu_{\textrm{min}}$ to a surface density $\Sigma_{\textrm{min}}$ with some error due to the scatter in $M/L_{r'}$, yielding $\Sigma_{\textrm{min}} = 0.26 \pm 0.09 \, M_{\odot}/\textrm{pc}^2$. Finally, we can convert this $\Sigma_{\textrm{min}}$ to a threshold density $\rho_t$ by plotting the surface density of the Fornax dwarfs against their volume density and doing a linear regression (Fig.~\ref{fig:surfdens_voldens}). Since the slope is very close to 1, we fix it to 1 for simplicity, leading to a fixed ratio of $\rho/\Sigma = 0.59 \pm 0.33 \, \textrm{kpc}^{-1}$. The limiting $\rho$ of the Fornax survey that we obtain with this method is $\rho_t = 1.51^{+1.67}_{-1.09} \times 10^{-4} \, M_{\odot}/\textrm{pc}^3$ considering the $1\sigma$ lower and upper limits to both $M/L_{r'}$ and $\rho/\Sigma$. From Fig.~\ref{fig:hist_dens}, we can see that the distribution of dwarfs is only included in its entirety if we take the lower limit and thus adopt a threshold of $\rho_t = \rho_{\textrm{min}} = 4.2 \times 10^{-5} \, M_{\odot}/\textrm{pc}^3$. Given that the images of the dwarfs have been carefully analysed by observers and labelled as `unclear' whenever the image was not clear enough, we assume that all the considered dwarfs were observed without difficulty by the FDS. Therefore, we consider that a reasonable lower limit for the density distribution in our statistical analysis should encompass all the dwarfs in the dataset, so we take $\rho_t = \rho_{\textrm{min}} = 4.2 \times 10^{-5} M_{\odot}/\textrm{pc}^3$ (black line in Fig.~\ref{fig:hist_dens}) as our nominal lower limit to the density distribution. This choice of $\rho_t$ is 0.09~dex below $\rho_{\textrm{min, FDS}}$, the lowest $\rho$ of any considered dwarf in the FDS. If instead we had assumed that $\rho_t = \rho_{\textrm{mean}} = 1.51 \times 10^{-4} M_{\odot}/\textrm{pc}^3$ (grey line in Fig.~\ref{fig:hist_dens}), we would have needed to discard 7 of the observed dwarfs in the FDS. These and other choices for $\rho_t$ are discussed in Section~\ref{discussion}.

In the $\Lambda$CDM case, we need to include the halo mass within the baryonic extent of each dwarf (Equation~\ref{M_dwarf_rule}), leading to higher volume densities. This causes a steeper slope and a larger amount of scatter in the mass-luminosity relation, making it difficult to follow the above-mentioned method. To keep the procedure similar, we set $\rho_{\textrm{min}}$ to a value 0.09~dex below $\rho_{\textrm{min, FDS}}$ as this is the gap assumed for MOND. The steps involved with this model are shown in Appendix~\ref{dwarf_dens_LCDM}.

\subsubsection{Disturbance to the dwarf structure}
\label{subsubsec:dwarf_disturbance}

Assuming that tides are the main cause of the apparent disturbance to the structure of many Fornax dwarfs, we expect the probability of a dwarf appearing perturbed to grow with its tidal susceptibility. We assume a linear relation between $\eta$ and the probability of disturbance ($P_{\textrm{dist}}$) with slope:
\begin{eqnarray}
	\textrm{Slope}_{P_{\textrm{dist}}} ~=~ \frac{P_{\textrm{dist, ceiling}} - P_{\textrm{dist, floor}}}{\eta_{\textrm{destr}} - \eta_{\textrm{min, dist}}} \, ,
\end{eqnarray}
where $\eta_{\textrm{min, dist}}$ is the lowest $\eta$ value at which the dwarf is disturbed by tides, $\eta_{\textrm{destr}}$ is the $\eta$ value at which the dwarf is destroyed (the algorithm rejects all simulated orbits in which $\eta_{\textrm{max}}$ surpasses this value), $P_{\textrm{dist, ceiling}}$ is the probability for a dwarf to appear disturbed right before it gets destroyed at $\eta = \eta_{\textrm{destr}}$, and $P_{\textrm{dist, floor}}$ is the minimum probability for a dwarf to appear disturbed if $\eta_{\textrm{max, recent}} < \eta_{\textrm{min, dist}}$. We allow $P_{\textrm{dist, floor}} > 0$ to capture the possibility that a dwarf appears disturbed for reasons unrelated to cluster tides, e.g. asymmetric star formation. Similarly, we expect that $P_{\textrm{dist, ceiling}} < 1$ because a significantly perturbed dwarf might be elongated along the line of sight and thus appear circular. For a dwarf with $\eta_{\textrm{max, recent}} \geq \eta_{\textrm{min, dist}}$, the probability of disturbance is:
\begin{eqnarray}
	P_{\textrm{dist}} ~=~ P_{\textrm{dist, floor}} + \textrm{Slope}_{P_{\textrm{dist}}} \left( \eta_{\textrm{max, recent}} - \eta_{\textrm{min, dist}} \right).
\end{eqnarray}

\subsection{Comparison with observations}
\label{comparison_observations}

The observed parameters of the Fornax dwarfs that we aim to reproduce in our simulation are:
\begin{enumerate}
    \item The distribution of sky-projected distances ($R_{\textrm{sky}}$) to the cluster centre;
    \item The distribution of apparent $\eta$ values at pericentre ($\eta_{\textrm{obs}}$); and
    \item The disturbed fraction of dwarfs as a function of $\eta_{\textrm{obs}}$.
\end{enumerate}
Because these quantities are projected or depend on the deprojection method, we need to obtain the $R_{\textrm{sky}}$ values of our simulated dwarfs and then deproject them using the same method that we use for the observed dwarfs. To obtain the $R_{\textrm{sky}}$ values for each 3D distance $R$ of the simulated dwarf, we consider the view if it is observed from all possible angles $0^{\circ} \leq \theta \leq 90^{\circ}$ in steps of $1^{\circ}$, where $\theta$ is the angle between $\bm{R}$ and the line of sight. The projected distance is given by
\begin{eqnarray}
	R_{\textrm{sky}} ~=~ R \sin \theta \, .
	\label{R_sky}
\end{eqnarray}
Each value of $\theta$ is statistically weighted by the difference in $\cos \theta$ across the corresponding bin. We then apply the deprojection method described in Appendix \ref{deproj} and obtain the corresponding distance at pericentre (Appendix~\ref{Rper}). With this, we can calculate $R_{\textrm{tid}}$ and $\eta$ at pericentre in a similar way to that in which we obtain these parameters for the observed dwarfs. We name the new $\eta$ parameter that we obtain with this method $\eta_{\textrm{obs}}$. Therefore, the simulated quantities that we compare to the previously mentioned observables are: $R_{\textrm{sky}}$, the distribution of $\eta_{\textrm{obs}}$, and the probability of disturbance at each $\eta_{\textrm{obs}}$.

To do the comparison, we start by dividing the range of $R_{\textrm{sky}}$ and $\eta_{\textrm{obs}}$ into several bins. We then classify the observed dwarfs into these bins according to their values of projected distance or estimated $\eta$ at pericentre. To obtain the probability for a dwarf to have a projected distance or $\eta_{\textrm{obs}}$ which falls in the range of values delimited by each of these bins, we count the number of dwarfs in each bin and compare it to the total number of dwarfs. To obtain the probability of disturbance, we count the number of dwarfs classified as disturbed in each $\eta_{\textrm{obs}}$ bin and compare it to the total number of dwarfs in that bin.

For the simulated sample (i.e., the dwarfs generated for all possible combinations of $R_i$, $e$, $\rho$, and $\theta$), we consider the same bins as for the observed sample. For each bin, we add the probability that each simulated dwarf has $R_{\textrm{sky}}$ or $\eta_{\textrm{obs}}$ values that fall in the range given by the bin. We then normalize this by the sum of all the probabilities in all bins. For the probability of disturbance, we apply an additional factor of $P_{\textrm{dist}}$ to the likelihood of each $\left( R_i, e, \rho, \theta \right)$ combination and add this to the appropriate $\eta_{\textrm{obs}}$ bin. We then divide this sum by the probability of $\eta_{\textrm{obs}}$ falling in that bin (i.e., without considering $P_{\textrm{dist}}$).

To quantify how closely the properties of the simulated sample of dwarfs resemble the properties of the observed FDS dwarfs in terms of each of the above-mentioned observables, we use the binomial probability
\begin{eqnarray}
	P_{\textrm{x}} ~=~ \prod_{\textrm{Bins}} \frac{T!}{ \left( T - S \right)! S!} p^S \left( 1 - p \right)^{T - S} \, ,
	\label{binomial}
\end{eqnarray}
where $T$ is the total number of observed dwarfs, $S$ is the number of observed dwarfs in a bin, $p$ is the simulated probability that a dwarf is in that bin, and the `x' subscript refers to the observable under consideration. If this is the disturbed fraction, $T$ is the total number of observed dwarfs in a particular $\eta_{\textrm{obs}}$ bin, $S$ is the observed number of disturbed dwarfs in that bin, and $p$ is the probability given by the simulation that a dwarf in that bin is disturbed. The total probability is given by multiplying all probabilities for all the bins and all the observables:
\begin{eqnarray}
    P_{\textrm{total}} ~=~ P_{R_{\textrm{sky}}} P_{\eta_{\textrm{obs}}} P_{\left.\textrm{perturbed}\right|\eta_{\textrm{obs}}} \, .
    \label{P_tot}
\end{eqnarray}

In order to maximize this $P_{\textrm{total}}$, we leave as free parameters: $r_{\textrm{core}}$, $\textrm{Slope}_{P_r}$, $\textrm{Slope}_{P_e}$, $\eta_{\textrm{min, dist}}$, $\eta_{\textrm{destr}}$, $P_{\textrm{dist, floor}}$, and $P_{\textrm{dist, ceiling}}$. We explore this set of parameter values using the Markov Chain Monte Carlo (MCMC) method discussed below.

\subsubsection{MCMC analysis}
\label{MCMC}

The MCMC method generates a sequence of parameter values in such a way that their frequency distribution matches the posterior inference on the model parameters. The basic idea is to start with some initial guess for the parameters with likelihood $P_{\rm{total}}$ and generate a proposal by adding Gaussian random perturbations to the parameters, leading to a likelihood of $P_{\rm{next}}$ with the revised parameters. The proposal is accepted if $P_{\rm{next}} > P_{\rm{total}}$ or if a random number drawn uniformly from the range $\left( 0-1 \right)$ is $<P_{\rm{next}}/P_{\rm{total}}$. If the proposal is rejected, the parameter perturbations are not applied but the previous parameters must be recorded once more.

\begin{table}
	\centering
	\caption{Priors for the free parameters in our model of the Fornax Cluster dwarf galaxy population.}
	\begin{tabular}{ccc}
	\hline
	Parameter & Minimum & Maximum \\ \hline
	$\textrm{Slope}_{P_r}$ & $-9$ & $-3$ \\
	$\textrm{Slope}_{P_e}$ & $-2$ & 2 \\
	$r_{\textrm{core}}$/Mpc & 0.01 & 3 \\
	$P_{\textrm{dist, floor}}$ & 0 & 1 \\
	$P_{\textrm{dist, ceiling}}$ & 0 & 1 \\
	$\eta_{\textrm{min, dist}}$ & 0 & 5 \\
	$\eta_{\textrm{destr}}$ & $\eta_{\textrm{min, dist}}$ & 5 \\ \hline
	\end{tabular}
	\label{tab:priors}
\end{table}

We run a total of $10^5$ trials in each chain and check that the acceptance fraction is close to 0.234, the optimal acceptance rate for an efficient MCMC algorithm \citep*{Gelman_1997}. This is achieved by rerunning the chain a few times to determine the optimal step sizes for the parameter perturbations. To ensure that the algorithm chooses physically reasonable parameter values, we impose the priors listed in Table~\ref{tab:priors}. If the algorithm chooses a value for any of these parameters outside the specified range, it is asked to draw another proposal, but this does not count as a new MCMC trial. We let the algorithm consider a sufficiently large number of proposals at each stage in the chain that we are sure to obtain a physically plausible proposal for the parameter combination to try next, even if this is rejected because it fits the observations poorly.

To prevent the MCMC algorithm from starting with a set of values which is too far away from the optimal set, we first fit the simulation's free parameters to the observations using a gradient ascent algorithm \citep{Fletcher_1963}. This maximizes $P_{\rm{total}}$ by increasing or decreasing the step size according to how much $P_{\textrm{total}}$ increased or decreased with respect to the previous set of parameter values that it tested. This is done until the step size becomes very small, indicating that the algorithm cannot increase $P_{\textrm{total}}$ any more. Then the algorithm converges and returns the optimal set of parameter values.


\section{Results of the statistical analysis}
\label{Results}

We present our best-fitting model in each theory (Section~\ref{sec:best_model}) before discussing the parameter uncertainties obtained with the MCMC method (Section~\ref{sec:uncertainties}).

\subsection{The best-fitting model}
\label{sec:best_model}

\begin{table}
	\centering
	\caption{The parameters of our best-fitting model in each theory, obtained with the gradient ascent method (columns $2-3$) and based on $10^5$ MCMC trials (columns $4-5$). The last row shows the likelihood of the model (Equations~\ref{binomial} and \ref{P_tot}).}
	\begin{tabular}{ccccc}
	\hline
	& \multicolumn{2}{c}{Gradient ascent} & \multicolumn{2}{c}{MCMC} \\
	Parameter & $\Lambda$CDM & MOND & $\Lambda$CDM & MOND \\ \hline
	$\textrm{Slope}_{P_r}$ & $-3.77$ & $-3.67$ & $-5.85$ & $-4.55$ \\ 
	$\textrm{Slope}_{P_e}$ & $-1.55$ & 0.34 & $-1.98$ & $-1.70$ \\
	$r_{\textrm{core}}$ & 0.62 & 0.65 & 1.35 & 0.90 \\
	$P_{\textrm{dist, floor}}$ & 0.09 & 0.04 & 0.10 & 0.02 \\
	$P_{\textrm{dist, ceiling}}$ & 0.65 & 0.76 & 0.54 & 0.53 \\
	$\eta_{\textrm{min, dist}}$ & 0.11 & 0.24 & 0.12 & 0.10 \\
	$\eta_{\textrm{destr}}$ & 0.24 & 1.88 & 0.23 & 1.24 \\ \hline
	$\log_{10} P_{\textrm{total}}$ & $-30.69$ & $-32.46$ & $-30.53$ & $-32.25$ \\ \hline
	\end{tabular}
	\label{tab:best_fit}
\end{table}

The optimal set of parameters found by the gradient ascent algorithm are given in columns 2 and 3 of Table \ref{tab:best_fit} for $\Lambda$CDM and MOND, respectively. These are the initial values at which we start the MCMC chains. Due to the use of $10^5$ trials, the MCMC method provides a set of parameter values (a model) that fits the observations slightly better (higher $P_{\textrm{total}}$ in Equation~\ref{P_tot}) than we achieved with gradient ascent. The best-fit parameter values in the MCMC chain are also given in Table~\ref{tab:best_fit} (columns $4-5$) along with the goodness of fit to the observations (last row). In this regard, there is little difference between the theories, though the optimal parameters are rather different. We will return to this later (Section~\ref{discussion}).

\begin{figure*} 
	\includegraphics[width = 8.5cm]{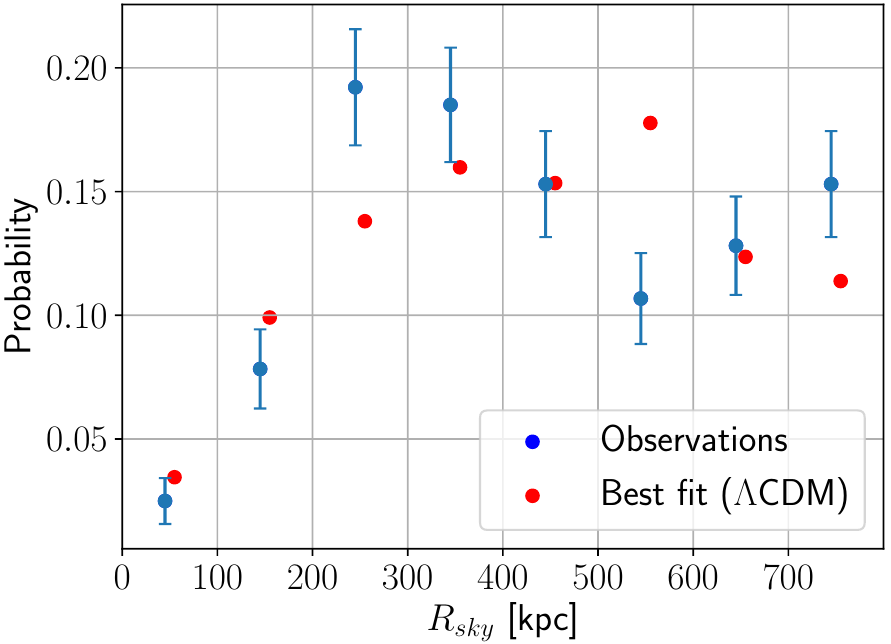} \hspace{0.07cm}
	\includegraphics[width = 8.5cm]{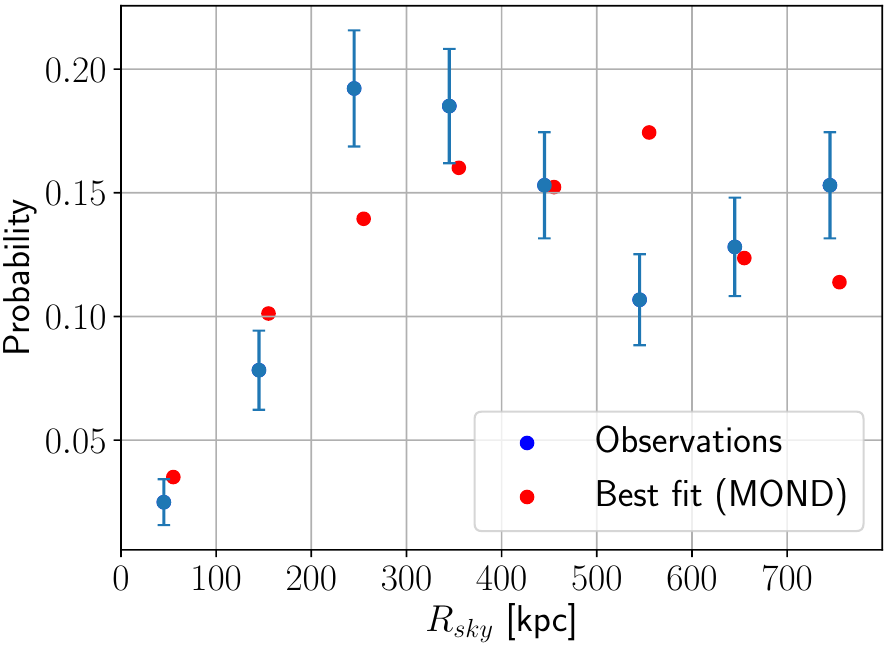} \\
	\includegraphics[width = 8.5cm]{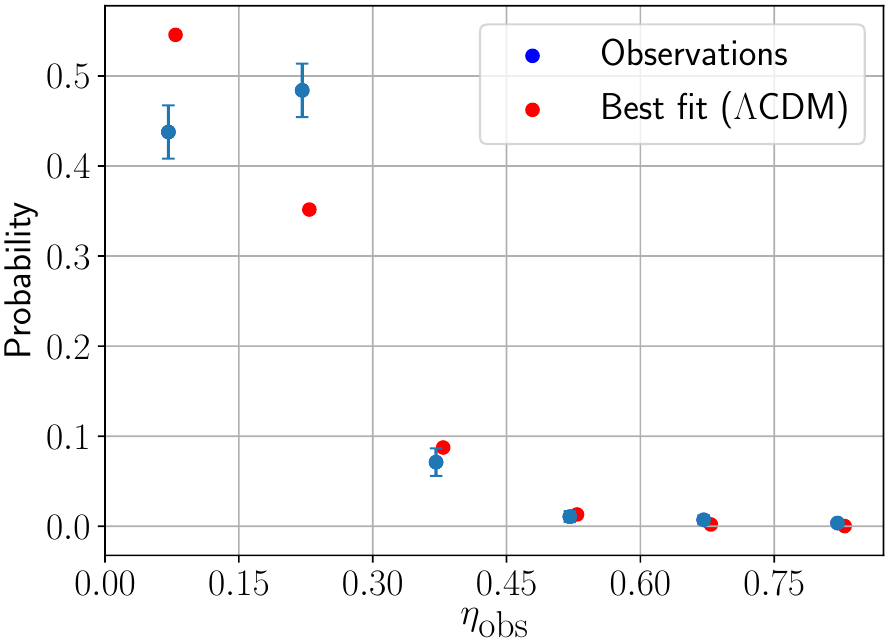} \hspace{0.07cm}
	\includegraphics[width = 8.5cm]{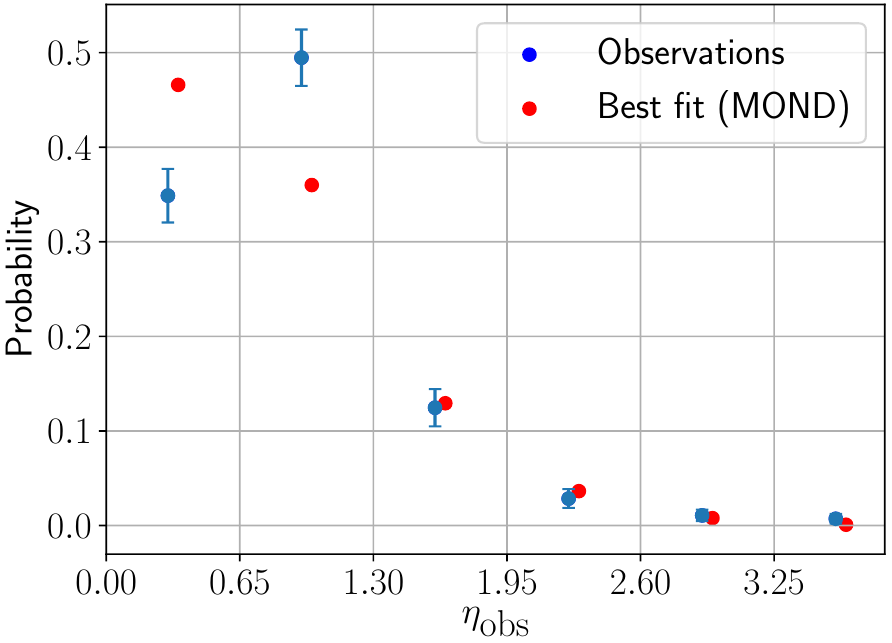} \\
	\includegraphics[width = 8.5cm]{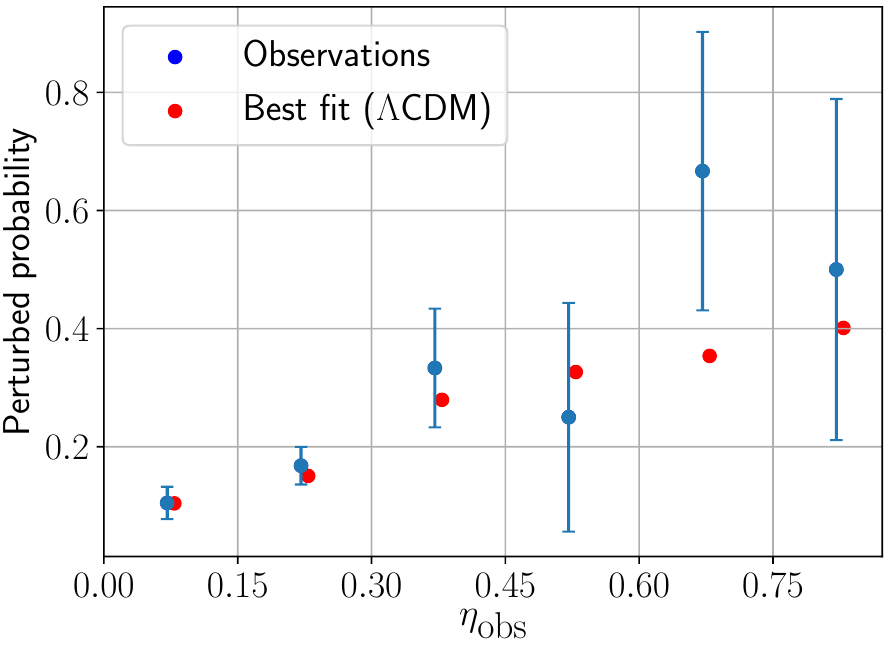}
	\hspace{0.07cm}
	\includegraphics[width = 8.5cm]{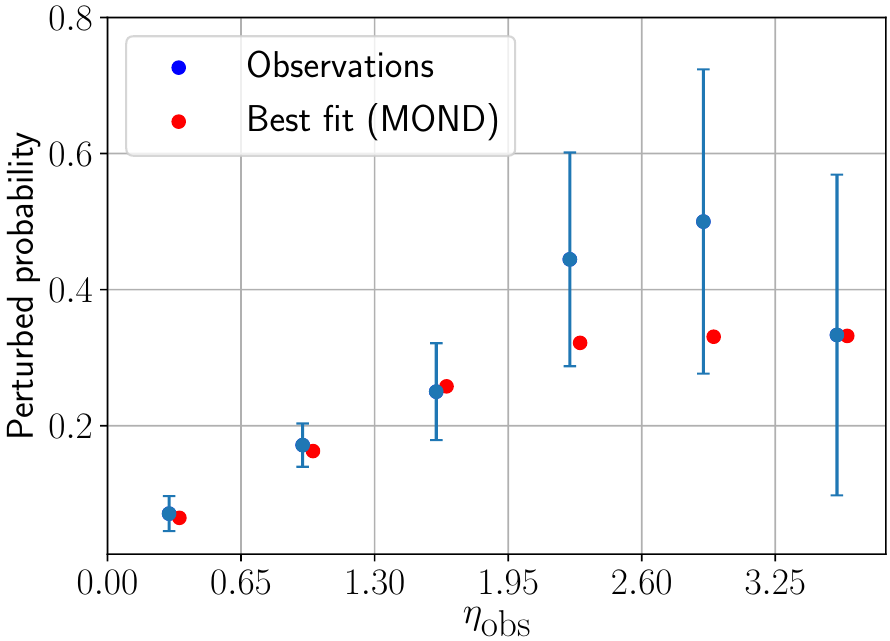}
	\caption{Comparison between observations and the best-fitting simulation in $\Lambda$CDM (left column) and MOND (right column) in terms of the distribution of projected separation $R_{\textrm{sky}}$ from the cluster (first row), tidal susceptibility $\eta_{\textrm{obs}}$ (second row), and likelihood that a dwarf appears disturbed as a function of $\eta_{\textrm{obs}}$ (third row). The observations (blue points with error bars) and the best-fitting simulation in each theory (red points) are plotted at the centre of each bin, but dithered slightly along the $x$-axis for clarity in case the model works well. The bin width in $R_{\textrm{sky}}$ is 100~kpc in both theories. For $\eta_{\textrm{obs}}$, the bin width is 0.15 in $\Lambda$CDM and 0.65 in MOND.}
	\label{fig:prob_dist}
\end{figure*}

Using these parameters, Fig.~\ref{fig:prob_dist} shows the simulated and observed probability distributions of $R_{\textrm{sky}}$, $\eta_{\textrm{obs}}$, and disturbed fraction vs. $\eta_{\textrm{obs}}$, revealing a good overall fit to the observations in both theories.\footnote{The low values of $P_{\textrm{total}}$ arise due to the large sample size.} In particular, the rising likelihood of a dwarf appearing disturbed as a function of $\eta_{\textrm{obs}}$ is nicely reproduced by the best-fitting models.

\subsection{Parameter uncertainties}
\label{sec:uncertainties}

\renewcommand{\arraystretch}{1.2}
\begin{table}
	\centering
    \caption{The most likely value and $1\sigma$ confidence interval of each model parameter in our test mass simulation of the Fornax Cluster dwarf galaxy population, based on $10^5$ MCMC trials.}
	\begin{tabular}{ccc}
	\hline
	Parameter & $\Lambda$CDM & MOND \\ \hline
	$\textrm{Slope}_{P_r}$ & $-7.43_{-0.99}^{+2.24}$ & $-7.58_{-0.88}^{+2.18}$ \\
	$\textrm{Slope}_{P_e}$ & $-1.65_{-0.30}^{+1.80}$ & $0.75_{-1.22}^{+1.20}$ \\
	$r_{\textrm{core}}$ & $2.00_{-0.98}^{+0.34}$ & $2.02_{-0.88}^{+0.52}$ \\
	$P_{\textrm{dist, floor}}$ & $0.10_{-0.03}^{+0.03}$ & $0.07_{-0.03}^{+0.04}$ \\
	$P_{\textrm{dist, ceiling}}$ & $0.49_{-0.15}^{+0.30}$ & $0.79_{-0.20}^{+0.15}$ \\
	$\eta_{\textrm{min, dist}}$ & $0.11_{-0.06}^{+0.05}$ & $0.24_{-0.19}^{+0.24}$ \\
	$\eta_{\textrm{destr}}$ & $0.25_{-0.03}^{+0.07}$ & $1.88_{-0.53}^{+0.85}$ \\ \hline
	\end{tabular}
    \label{tab:MCMC_paramvalues}
\end{table}
\renewcommand{\arraystretch}{1}

To fit the test mass simulation of the Fornax dwarf galaxy system to its observed properties, we require several free parameters in the model (Section~\ref{test_mass}). Having discussed the values of these parameters in the most likely model (Table~\ref{tab:best_fit}), we now find the most likely value of each parameter and its uncertainty. This is somewhat different because instead of considering the most likely model, we use the MCMC chain to obtain the posterior inference on each model parameter, which we then characterize using its mode and $1\sigma$ confidence interval. The results are shown in Table~\ref{tab:MCMC_paramvalues}.

\begin{figure*}
	\includegraphics[scale = 1.5]{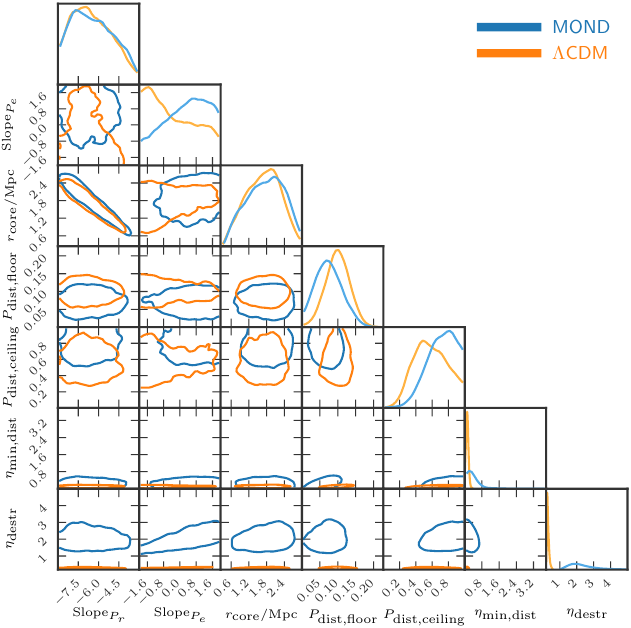}
	\caption{The $1\sigma$ confidence regions for the parameters in our model of the Fornax Cluster dwarf galaxy population using $\Lambda$CDM (orange) and MOND (blue), based on the priors listed in Table~\ref{tab:priors}. The top panel in each column shows the inference on a single parameter, while the other panels show the $1\sigma$ confidence region for a pair of parameters. The results shown in this `triangle plot' are based on $10^5$ MCMC trials (Section~\ref{MCMC}). All the triangle plots shown in this contribution were generated using the \textsc{pygtc} package \citep{Bocquet_2016}.}
	\label{fig:triang_MONDLCDM}
\end{figure*}

We also use Fig.~\ref{fig:triang_MONDLCDM} to show the results of the MCMC analysis by plotting the probability distribution of each parameter and showing contour plots for all possible parameter pairs. The parameters $\textrm{Slope}_{P_r}$, $r_{\textrm{core}}$, $P_{\textrm{dist, floor}}$, and $P_{\textrm{dist, ceiling}}$ cover a similar range of values in both theories. This is to be expected because the distribution of dwarfs in the Fornax Cluster is known observationally such that $\textrm{Slope}_{P_r}$ and $r_{\textrm{core}}$ are not strong tests of the gravity law, while $P_{\textrm{dist, floor}}$ and $P_{\textrm{dist, ceiling}}$ are set by the proportion of dwarfs in different $\eta_{\textrm{obs}}$ bins that appear disturbed (Fig.~\ref{fig:tid_edge}). Unlike these four parameters, $\textrm{Slope}_{P_e}$, $\eta_{\textrm{min, dist}}$, and $\eta_{\textrm{destr}}$ cover very different ranges in these two models. As discussed below, these are the parameters which can help us discern between $\Lambda$CDM and MOND, allowing us to assess which model performs better when compared to observations.

The inference on $\textrm{Slope}_{P_e}$ (shown in the top panel of column~2 of Fig.~\ref{fig:triang_MONDLCDM}) peaks close to the minimum allowed value of $-2$ in $\Lambda$CDM. The opposite happens in MOND, where the peak is close to 1. Negative slopes in Equation \ref{P_e} assign higher probabilities to nearly circular orbits. However, according to \citet{Ambartsumian_1937}, we expect the eccentricity distribution to be thermal and thus have $\textrm{Slope}_{P_e} \approx 2$ \citep[for a derivation, see section~4.2 of][]{Kroupa_2008}. In this regard, MOND performs better than $\Lambda$CDM.


\begin{figure}
	\centering
	\includegraphics[width = 8.5cm]{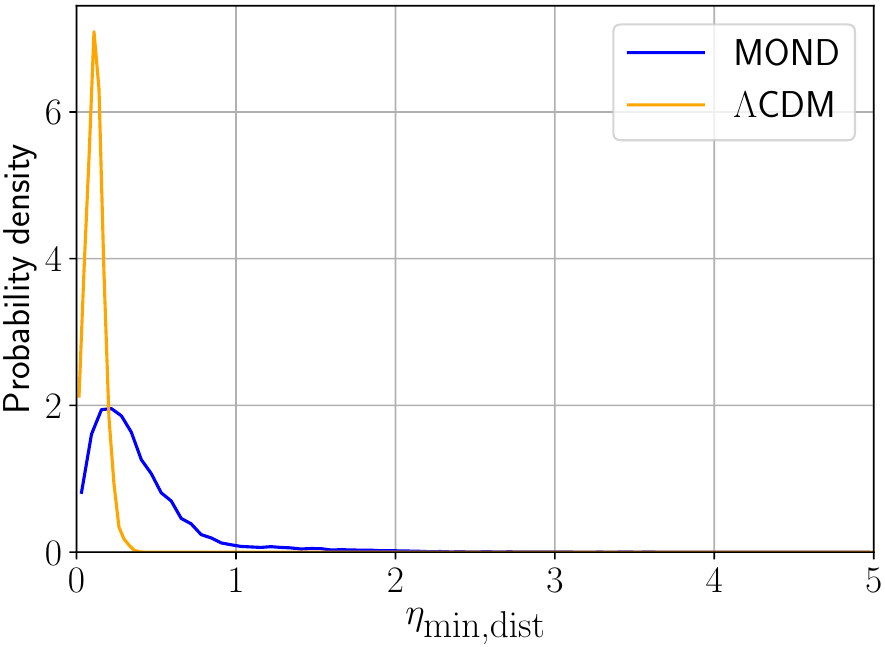}
	\caption{The probability distribution of the tidal susceptibility above which a dwarf is more likely to appear disturbed (Section~\ref{subsubsec:dwarf_disturbance}). Notice that the MCMC analysis prefers significantly higher values for MOND (blue) than for $\Lambda$CDM (orange).}
	\label{fig:Min_eta_dist}
\end{figure}

\begin{figure}
	\centering
	\includegraphics[width = 8.5cm]{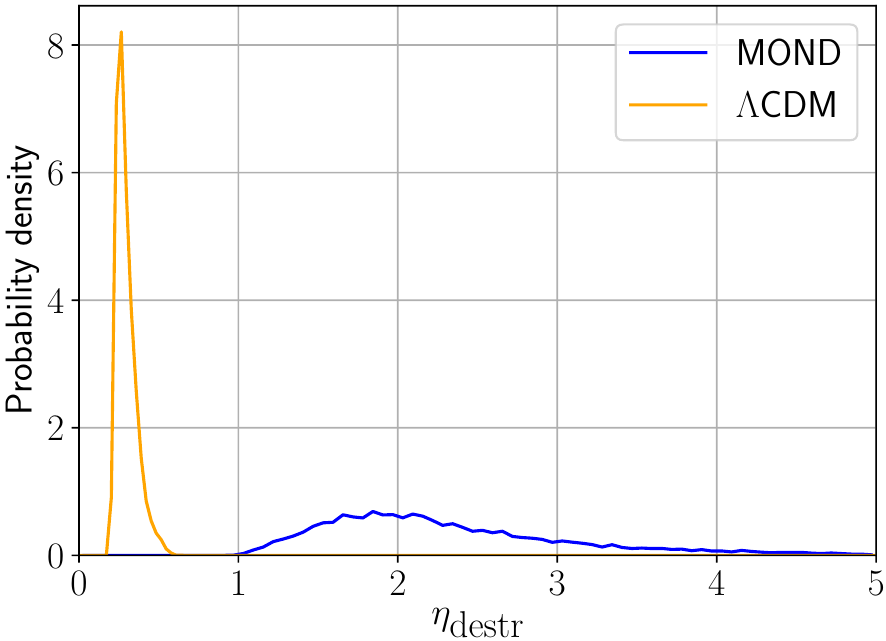}
	\caption{The probability distribution of the tidal susceptibility at which a dwarf is destroyed (Section~\ref{subsubsec:dwarf_disturbance}) according to $\Lambda$CDM (orange) and MOND (blue).}
	\label{fig:eta_destr}
\end{figure}

The major differences between $\Lambda$CDM and MOND are in the parameters $\eta_{\textrm{min, dist}}$ and $\eta_{\textrm{destr}}$, whose posterior inferences are shown in detail in Figs.~\ref{fig:Min_eta_dist} and \ref{fig:eta_destr} due to their importance to our argument. The low values in $\Lambda$CDM arise because dwarfs have quite strong self-gravity by virtue of being embedded in a dominant dark matter halo throughout their trajectory. This makes them less susceptible to the effect of tides (stronger self-gravity raises $r_{\textrm{tid}}$ and thus reduces $\eta$; see Equation~\ref{eta_rtid}). As a result, the algorithm needs to set $\eta_{\textrm{min, dist}}$ and $\eta_{\textrm{destr}}$ to very low values in order to match the observed fact that many dwarfs are morphologically disturbed and we do not observe dwarfs beyond a certain limiting $\eta$. MOND also boosts the baryonic self-gravity of a dwarf, but this boost is damped due to the EFE of the cluster's gravitational field. This effect gets stronger as dwarfs approach the pericentre of their orbits, to the point that dwarfs which are sufficiently close to the cluster centre can become almost Newtonian despite a very low internal acceleration. Because of this, MONDian dwarfs are significantly more susceptible to tides than their $\Lambda$CDM counterparts. This causes the algorithm to choose significantly higher $\eta_{\textrm{min, dist}}$ and $\eta_{\textrm{destr}}$ values in the MOND case.

\textit{N}-body simulations of dwarf galaxies show that $\eta_{\textrm{destr}}$ should be $\approx 1$ in $\Lambda$CDM \citep{Penyarrubia_2009, Van_den_Bosch_2018}. However, fitting the observations with our MCMC method gives a much lower value of $\eta_{\textrm{destr}} = 0.25^{+0.07}_{-0.03}$. This implies an important discrepancy between model expectations in $\Lambda$CDM and actual observations of dwarf galaxies in the Fornax Cluster.

Turning to MOND, comparing the $\eta_{\textrm{destr}}$ value inferred from observations with that obtained using simulations is not so straightforward given that the best available \textit{N}-body simulations studying the resilience of Milgromian dwarf galaxies to tides is by now very old and poorly suited to the present study \citep{Brada_2000_tides}. Because of this, we perform our own \textit{N}-body simulations of a typical Fornax Cluster dwarf galaxy, as described next.

\section{\textit{N}-body simulations of a Fornax dwarf}
\label{Nbody_sim}


As the last part of this project, we conduct our own \textit{N}-body simulations of a typical Fornax dwarf to find out the expected $\eta_{\textrm{destr}}$ in MOND. The motivation is that while the analytic formula for the tidal radius (Equation~\ref{rtid_MOND}) should capture the scalings with the relevant variables like the tidal stress and the EFE, there could be a constant numerical pre-factor that arises from a detailed simulation. We investigated this using the Milgromian \textit{N}-body code \textsc{phantom of ramses} (\textsc{por}) developed in Bonn by \citet{Lughausen_2015}, who adapted it from the Newtonian \textit{N}-body code \textsc{ramses} \citep{Teyssier_2002}. As a result, \textsc{por} inherits many features of \textsc{ramses}, including the adaptive mesh refinement technique to better resolve denser regions. \textsc{por} can work with both particle and gas dynamics. It is suited for simulations of isolated galaxies \citep{Banik_2020_M33, Roshan_2021_disc_stability, Banik_2022_fake_inclination}, interacting galaxies \citep{Renaud_2016, Thomas_2017, Thomas_2018, Bilek_2018, Banik_2022_satellite_plane}, galaxy formation \citep{Wittenburg_2020}, and even for cosmological structure formation (N. Wittenburg et al., in preparation). The main difference between \textsc{por} and \textsc{ramses} is the fact that \textsc{por} solves the ordinary Poisson equation twice, with $\bm{g}_{_N}$ found using standard techniques in the first stage and the following equation solved in the second stage to implement the MOND corrections:
\begin{eqnarray}
    \nabla \cdot \bm{g} ~=~ \nabla \cdot \left( \nu \bm{g}_{_N} \right) \, ,
\end{eqnarray}
where $\nu$ was defined in Equation~\ref{simple_interpolating}. The boundary condition for the Milgromian potential $\Phi$ is:
\begin{eqnarray}
    \Phi ~=~ \sqrt{GMa_{_0}} \ln r \, ,
\end{eqnarray}
where $M$ is the total mass in the simulation volume and $r$ is the distance from the barycentre in the simulation unit of length, the choice of which has no bearing on the result.

Since Fornax Cluster dwarfs are expected to contain little gas (Section~\ref{effects_gravi}), we can simplify the set-up greatly by using the `particle-only' version of the \textsc{por} code. In particular, we use the `staticparts' patch \citep[described in section~4.1 of][]{Nagesh_2021} which allows the use of particles that provide gravity but do not move if their mass exceeds a user-defined threshold. This is helpful because we treat the cluster gravity as sourced by a point mass fixed at the origin, with the dwarf at three possible initial distances $R_i$. To ensure the gravity on the dwarf is the same as in the Fornax Cluster, we use Equation \ref{M_cluster} to obtain $g_{c}$ and then obtain the corresponding $g_{c,N}$ with the simple interpolating function in the inverse form (Equation~\ref{g_N_g}), from which we get the central mass:
\begin{eqnarray}
    M_c ~\equiv~ \frac{g_{c,N}R_i^2}{G} \, .
\end{eqnarray}
The different MOND dynamical cluster masses obtained in this way are: $M_c = 2.18 \times 10^{12}~M_{\odot}$ at 150~kpc, $M_c = 2.89 \times 10^{12}~M_{\odot}$ at 300~kpc, and $M_c = 3.31 \times 10^{12}~M_{\odot}$ at 450~kpc. We use $7-13$ refinement levels and set the box length to $6 \, R_i$ as the apocentre could be at almost $2 \, R_i$.

For the dwarf, we use a half-mass radius of $r_h = 0.84$~kpc and a total mass of $M_{\textrm{dwarf}} = 3.16 \times 10^7~M_{\odot}$ represented by $10^5$ particles, making the mass resolution $316 \, M_{\odot}$. These are typical parameters for a dwarf in the Fornax Cluster (see the red star in Fig.~\ref{fig:surfdens_voldens}). Setting the velocity dispersion $\sigma$ is non-trivial because we need to account for the cluster EFE when we initiate the simulation. We do this by using the Fornax dwarf templates kindly provided by Prof. Xufen Wu, who used a similar method to that described in section~3.3 of \citet{Haghi_2019_DF2} to generate these templates. The idea is to take a Newtonian template and then enhance the velocities by the factor needed to ensure virial equilibrium given the enhanced gravity \citep{Wu_2013}.

To set up the dwarf, we apply a Galilean transformation to the template whereby the Cartesian positions of all particles are boosted by ($x_0 = R_i$, $y_0 = 0$, $z_0 = 0$) and the velocities are boosted depending on the circular velocity at $R_i$ and the orbital eccentricity $e$, as described in Section~\ref{orbit_integration}. We start the simulation with the dwarf at the semi-major axis of its orbit and receding from the cluster. We then evolve the system until shortly after the dwarf reaches apocentre for the second time so that there is ample time to assess the impact of the pericentre passage. The code generates an output of the mass, position, and velocity of every particle every 20~Myr, allowing us to analyse the structure of the dwarf and find out if it has been destroyed.

Our main objective is to find the threshold value of $\eta$ at pericentre beyond which the dwarf gets destroyed in the simulation. This requires us to perform multiple simulations with different eccentricities in order to obtain different $\eta$ values at pericentre. To guide our choice of parameters, we use a simple MOND Runge-Kutta orbit integrator of a point mass orbited by a test particle in 2D. This is also very helpful when deciding the appropriate duration for each simulation, which we keep fixed for models with the same $R_i$.

\subsection{Analysis}

We extract the particle positions $\bm{r}_i$, velocities $\bm{v}_i$, and masses $m_i$ using \textsc{extract\_por} \citep{Nagesh_2021}, with the index $i$ used in what follows to distinguish the particles. To assess if a dwarf has been destroyed, we infer three properties of the dwarf from the output at each snapshot: its half-mass radius, velocity dispersion, and aspect ratio. Unlike in Newtonian gravity, the time-varying EFE implies that these quantities are expected to vary around the orbit even if the dwarf is completely tidally stable ($\eta \ll 1$), perhaps most famously for the velocity dispersion \citep{Kroupa_2018_DF2}. To assess tidal stability, we check whether the dwarf responds adiabatically to the time-varying EFE. Tidal stability requires the dwarf to recover the initial values for these parameters after the pericentre passage, at least by the time of the next apocentre. If this is not the case, then the dwarf is either destroyed or unstable, in which case several pericentre passages may be required to destroy the dwarf. However, it is beyond the scope of this project to simulate multiple pericentre passages.


\subsubsection{Finding the barycentre}

We apply an iterative outlier rejection scheme to accurately obtain the barycentre position $\overline{\bm{r}}$ and velocity $\overline{\bm{v}}$ based on the positions and velocities of the particles. In the first iteration, we consider all the particles and calculate
\begin{eqnarray}
    \overline{\bm{r}} ~\equiv~ \frac{\sum_i m_i \bm{r}_i}{M} \, , \quad M \equiv \sum_i m_i \, .
\end{eqnarray}
We use a similar definition for $\overline{\bm{v}}$. The barycentre position and velocity are then used to find the root mean square (rms) dispersion in position and velocity.
\begin{eqnarray}
    r^2_{\textrm{rms}} ~\equiv~ \frac{\sum_i m_i {\lvert \bm{r}_i - \overline{\bm{r}} \rvert}^2}{M} \, ,
    \label{r_rms}
\end{eqnarray}
with a similar definition used for $v_{\textrm{rms}}$, which we call $\sigma$ for consistency with other workers. This lets us define a $\chi^2$ statistic for each particle based on its position.
\begin{eqnarray}
    \chi^2_{\textrm{pos}} ~\equiv~ \left( \frac{\left| \bm{r}_i - \overline{\bm{r}} \right|}{r_{\textrm{rms}}} \right)^2 \, ,
\end{eqnarray}
with a similar definition used for $\chi^2_{\textrm{vel}}$ based on the velocity.

In the second iteration, we repeat the above steps for only those particles whose $\chi^2_{\textrm{pos}}$ and $\chi^2_{\textrm{vel}}$ are both below 25, which changes the calculated quantities. In subsequent iterations, we expect to have pinned down the barycentre more precisely, so we use the stricter condition that
\begin{eqnarray}
    \chi^2_{\textrm{pos}} + \chi^2_{\textrm{vel}} ~<~ \chi^2_{\textrm{max}} \, ,
\end{eqnarray}
where $\chi^2_{\textrm{max}} = 11.83$ is set so that the likelihood of the $\chi^2$ statistic for two degrees of freedom exceeding $\chi^2_{\textrm{max}}$ is the same as the likelihood of a Gaussian random variable deviating from its mean value by $\geq 3\sigma$. Our procedure can thus be thought of as $3\sigma$ outlier rejection.

We consider the algorithm to have converged once the difference in $\overline{\bm{r}}$ and $\overline{\bm{v}}$ between successive iterations is so small that
\begin{eqnarray}
    \frac{\left| \Delta \overline{\bm{r}} \right|^2}{r^2_{\textrm{rms}}} + \frac{\left| \Delta \overline{\bm{v}} \right|^2}{\sigma^2} ~<~ 10^{-5} \, ,
\end{eqnarray}
with the additional requirement that the number of `accepted' particles deviates from that in the previous iteration by no more than the Poisson uncertainty. In the analyses described below, we will only consider those particles which are accepted on the final iteration.


\subsubsection{Velocity dispersion}

The velocity dispersion $\sigma$ is already available as part of our $3\sigma$ outlier rejection system for finding the barycentre of the dwarf. This 3D $\sigma$ is found by applying Equation~\ref{r_rms} but using velocities rather than positions. If the dwarf were isolated and unaffected by tides, equation 14 of \citet{Milgrom_1994_virial} tells us to expect that
\begin{eqnarray}
    \sigma ~=~ \left( \frac{4}{9} GMa_{_0} \right)^{1/4} \, .
    \label{v_MOND_iso}
\end{eqnarray}
This assumes dynamical equilibrium and the deep-MOND limit, but does not make any assumptions concerning whether the orbits are mostly radial or tangential. If the system is not spherically symmetric, the velocity dispersion would not be the same along every direction, but the bulk 3D velocity dispersion above would still hold. Another important caveat is that the system should consist only of particles with $m_i \ll M$.

\subsubsection{Half-mass radius}

To obtain the half-mass radius $r_h$, we order the particles in ascending order of their distance to the above-determined dwarf barycentre $\overline{\bm{r}}$. We then find the index $p$ such that
\begin{eqnarray}
	\sum_{i=1}^p m_i ~=~ \frac{M}{2} \, ,
\end{eqnarray}
with the total mass $M$ of all accepted particles in general being slightly below the initial mass of the dwarf. By definition, $r_h$ is the distance of particle $p$ from the dwarf's barycentre.
\begin{eqnarray}
	r_h ~\equiv~ \lvert \bm{r}_p - \overline{\bm{r}} \rvert \, .
\end{eqnarray}

\subsubsection{Aspect ratio}

To quantify the shape of the simulated dwarf, we obtain its inertia tensor
\begin{eqnarray}
    \matr{I}_{jk} ~\equiv~ \sum_i m_i \left( \bm{r} - \overline{\bm{r}} \right)_j \left( \bm{r} - \overline{\bm{r}} \right)_k \, ,
\end{eqnarray}
where the spatial indices $j$ and $k$ take values in the range $1-3$ because there are three dimensions. We then find the eigenvalues of $\matr{I}$. The aspect ratio of the dwarf is defined as
\begin{eqnarray}
    \textrm{aspect ratio} ~\equiv~ \sqrt{\frac{\lambda_{\textrm{min}}}{\lambda_{\textrm{max}}}} \, ,
    \label{Aspect_ratio_def}
\end{eqnarray}
where $\lambda_{\textrm{min}}$ ($\lambda_{\textrm{max}}$) is the smallest (largest) eigenvalue.

\subsection{Results}

\begin{figure*} 
	\includegraphics[width = 5.7cm]{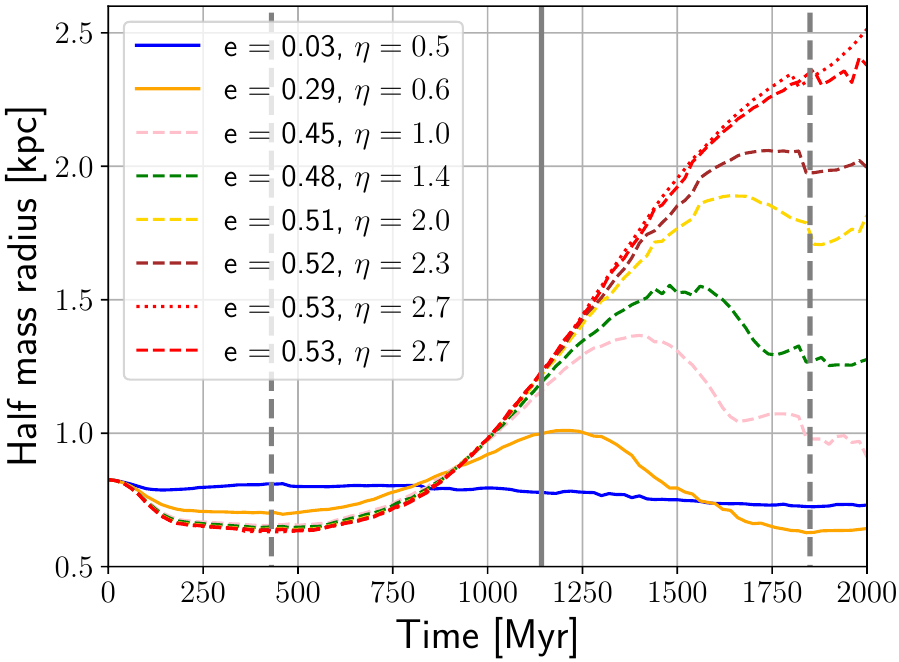} \hspace{0.07cm}
	\includegraphics[width = 5.7cm]{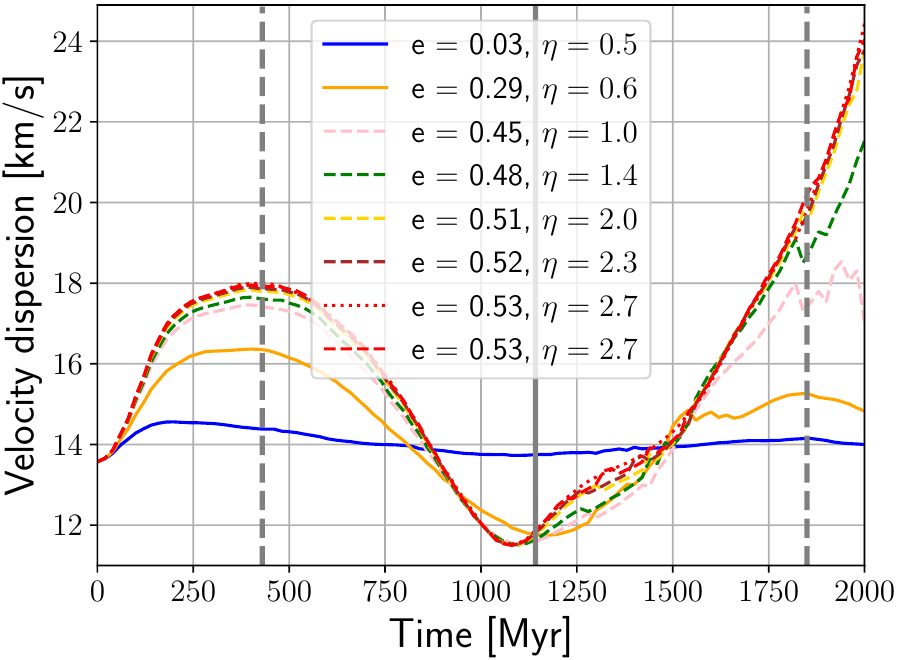} \hspace{0.07cm}
	\includegraphics[width = 5.7cm]{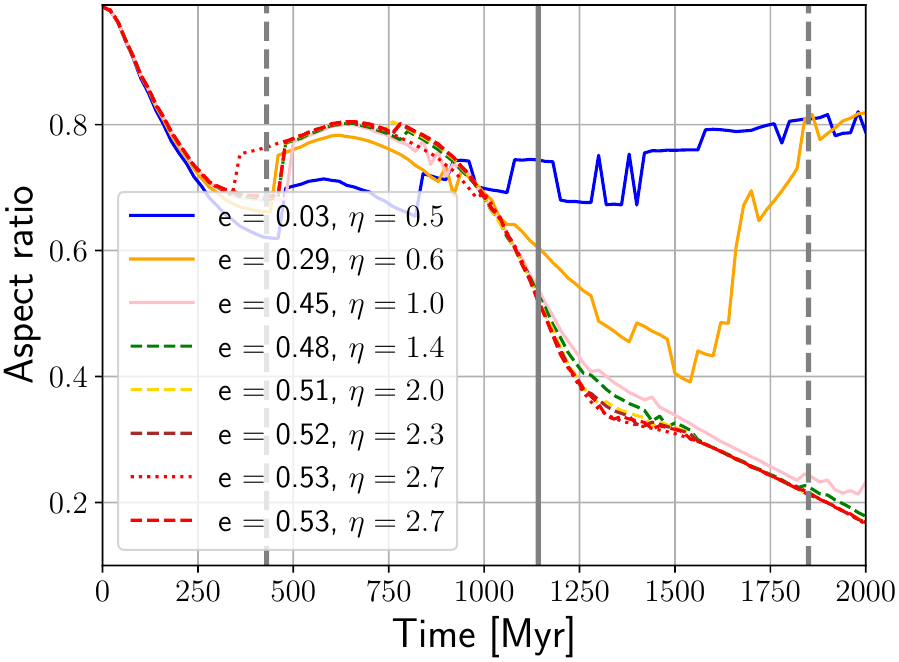} \\
	\includegraphics[width = 5.7cm]{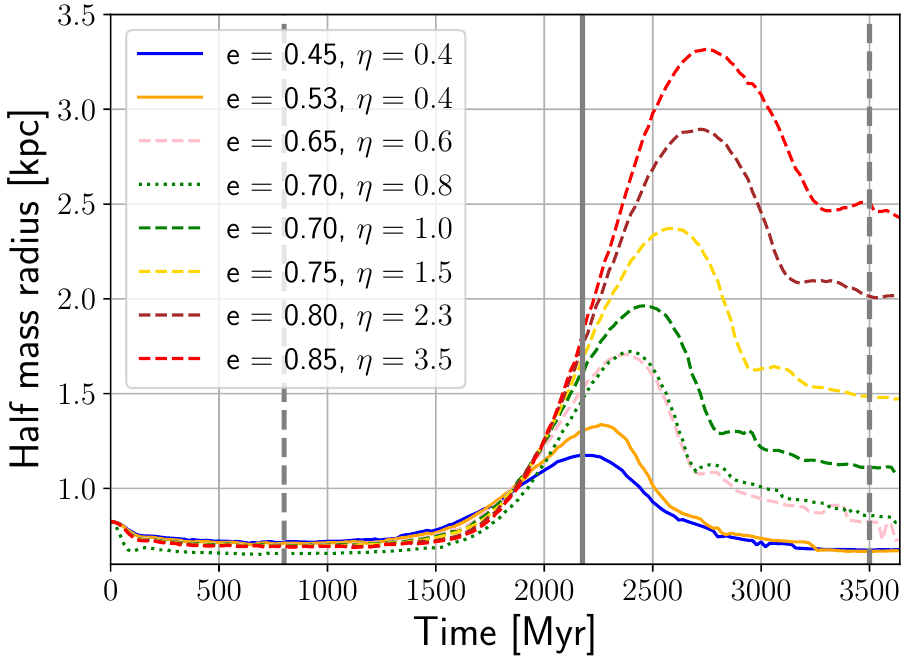} \hspace{0.07cm}	
	\includegraphics[width = 5.7cm]{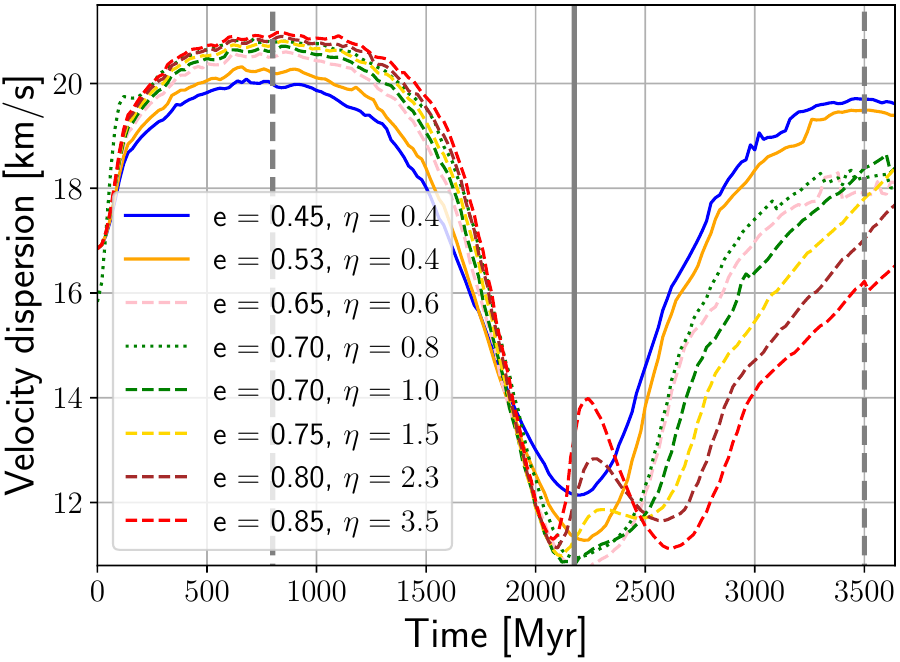} \hspace{0.07cm}
	\includegraphics[width = 5.7cm]{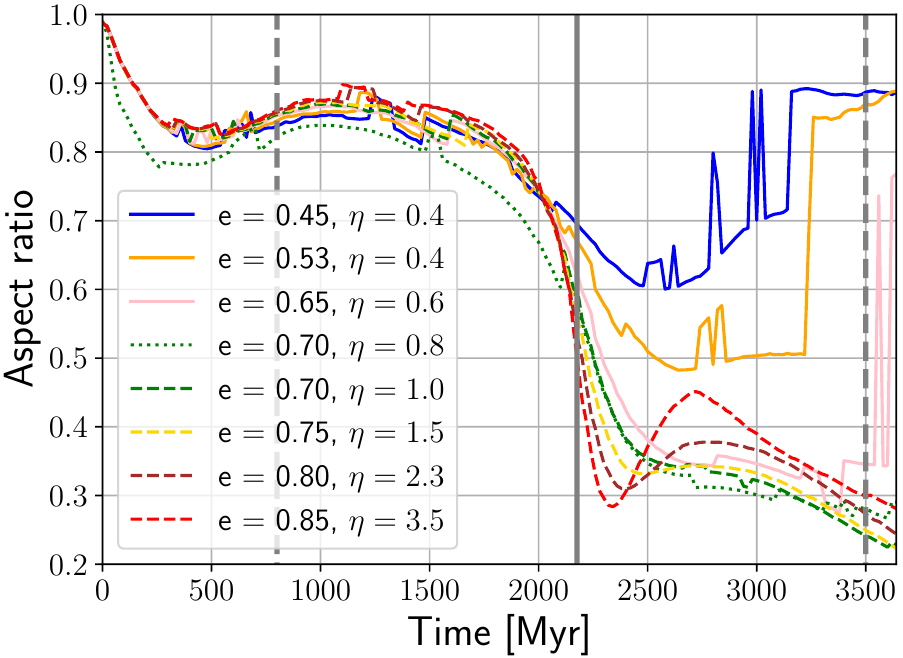} \\
	\includegraphics[width = 5.7cm]{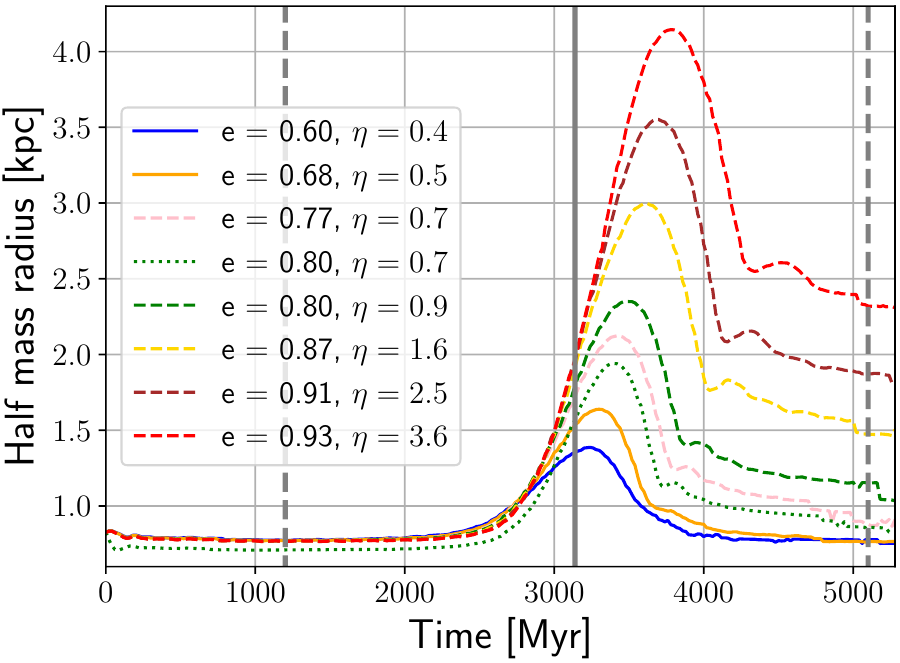} \hspace{0.07cm}
	\includegraphics[width = 5.7cm]{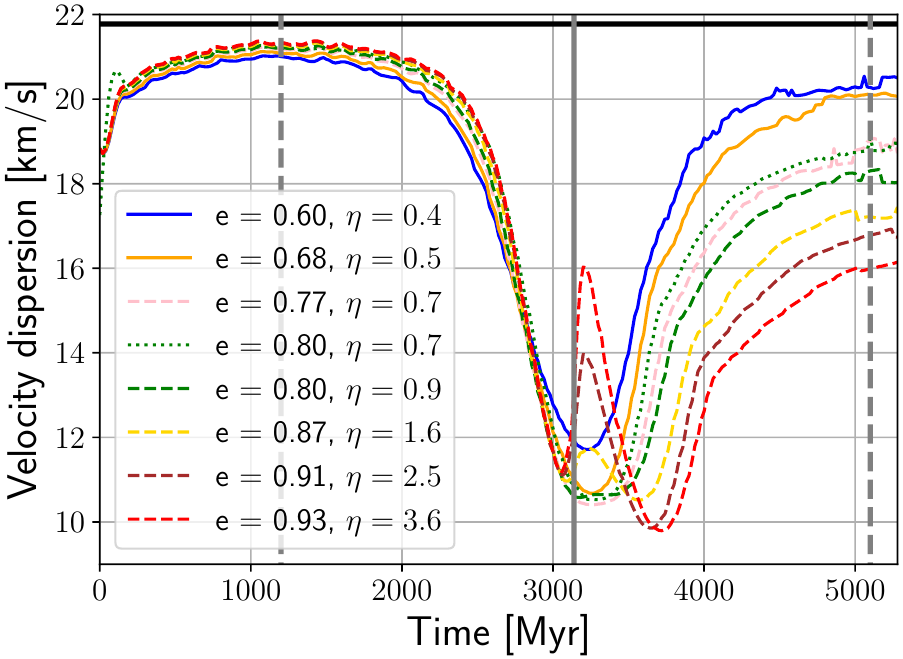} \hspace{0.07cm}
	\includegraphics[width = 5.7cm]{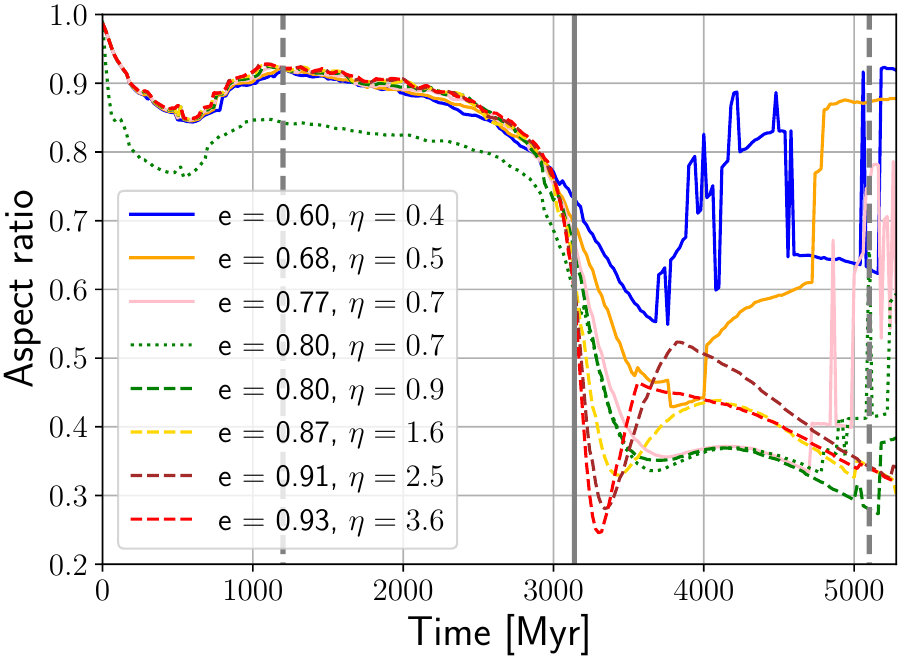}
	\caption{Evolution of the half-mass radius (first column), 3D velocity dispersion (second column), and aspect ratio (third column) of the simulated dwarfs over time starting from an initial distance of $R_i=150$~kpc (first row), $R_i=300$~kpc (second row), and $R_i=450$~kpc (third row). In each panel, the different curves show different orbital eccentricities as indicated in the legend, which gives their corresponding $\eta$ values at pericentre (solid grey line) based on the EFE and tidal stress there but with the mass and half-mass radius found at the next apocentre (see the text). The mass at that time is similar to the initial value. The vertical dashed grey lines represent the first and second apocentre of the orbit. The solid (dashed) coloured lines represent those dwarfs which do (do not) recover their initial properties. The dotted lines that repeat one of the eccentricities in each panel correspond to a higher resolution simulation ($8 \times 10^5$ particles), indicating that resolution hardly affects our results. The horizontal black line in the lower middle panel represents the expected velocity dispersion of the dwarf in the isolated deep-MOND limit (Equation \ref{v_MOND_iso}).}
	\label{fig:dwarf_simulations}
\end{figure*}

The results of our \textsc{por} simulations are shown in Fig.~\ref{fig:dwarf_simulations}. Unlike in the Newtonian case, even dwarfs with a very low tidal susceptibility exhibit significant variations in their properties due to the time-varying EFE. We can see that in the cases with low $e$, the dwarf manages to recover the properties it had before pericentre. However, in the cases with higher $e$, these properties do not regain their initial values, indicating that the dwarf is tidally unstable.\footnote{This seems to be the case for the MW satellite Crater II \citep{Torrealba_2016}, whose low surface brightness, small pericentre \citep{Hefan_Li_2021}, and low velocity dispersion for $\Lambda$CDM \citep{Caldwell_2017} suggest that it is the remnant of an originally smaller object that got severely disrupted by tides during its perigalacticon passage \citep{Borukhovetskaya_2022, Errani_2022}. It is also expected to be tidally unstable in MOND \citep[see section~3.3 of][]{Banik_Zhao_2022}.} This was expected because dwarfs with more eccentric orbits have closer pericentre passages and thus higher $\eta$ values at pericentre.

To assess whether a dwarf is destroyed in the simulation, the criterion that we apply is to consider destroyed those dwarfs which have a higher $r_h$ at the second apocentre than at pericentre. Since the dwarf is likely to expand even further as it heads towards its next pericentre, this implies that the dwarf has been too destabilized by tides to contract back to its size at its first pericentre passage. As a result, the dwarf would have an even higher tidal susceptibility at subsequent pericentres. This makes it very likely that the dwarf would not be able to survive multiple pericentre passages. On the other hand, if a dwarf that experiences a pericentre passage has a smaller $r_h$ at the subsequent apocentre and is contracting further, then it may well get back to its size at first pericentre by the time it reaches its second pericentre. This should allow it to survive multiple pericentre passages, which in the Fornax Cluster case should allow survival over a Hubble time.

To fairly compare our \textit{N}-body results with our MCMC analysis, we should consider how observers calculate $\eta_{\textrm{obs}}$. The $r_h$ entering into Equation~\ref{eta_rtid} is the observed size, so ideally we would calculate $\eta$ at pericentre using the EFE and tidal stress there but using the presently observed size. As a proxy for this, we use the size at apocentre since this is the orbital phase at which we are most likely to observe the dwarf. Physically, the tidal stability of a dwarf depends on the ratio between its size and tidal radius at pericentre. Using the ratio between the tidal radius at pericentre and the half-mass radius at apocentre may seem somewhat counter-intuitive. However, the $\eta_{\textrm{destr}}$ values obtained in this way are much more comparable to those obtained from our MCMC analysis of the Fornax Cluster for the reasons discussed above. In what follows, we will use $\eta$ to mean the value calculated in this way, though Table~\ref{tab:e_eta} also shows results based on the size at pericentre.

\begin{table}
	\centering
	\caption{Summary of our MOND \textit{N}-body simulation results for a Fornax dwarf with an initial distance of $R_i = 150$~kpc and different orbital eccentricities (first column). The tidal susceptibility is calculated assuming the EFE and tidal stress at pericentre but using the half-mass radius of the dwarf at pericentre (second column) or at the subsequent apocentre (third column), which we argue in the text is more comparable to our MCMC results. The fourth column gives our assessment of the simulation based on the top left panel of Fig.~\ref{fig:dwarf_simulations}.}
	\begin{tabular}{cccc}
	\hline
	& \multicolumn{2}{c}{$\eta$ using $r_h$ at $\ldots$} & \\
	$e$ & Pericentre & Apocentre & Outcome \\ \hline
	0.03 & 0.6 & 0.5 & Stable \\
	0.29 & 0.9 & 0.6 & Stable \\
	0.45 & 1.2 & 1.0 & Stable \\
	0.48 & 1.3 & 1.4 & Marginal \\
	0.51 & 1.4 & 2.0 & Unstable \\
	0.52 & 1.4 & 2.3 & Unstable \\
	0.53 & 1.4 & 2.7 & Destroyed \\ \hline
	\end{tabular}
	\label{tab:e_eta}
\end{table}

To constrain $\eta_{\textrm{destr}}$, we focus mainly on models with $R_i = 150$~kpc as dwarfs with a larger semi-major axis would typically be observed much further out than the region contributing to the apparent tidal edge in Fig.~\ref{fig:tid_edge}, especially if the eccentricity is significant. The results of these models are summarized in Table~\ref{tab:e_eta}. The models with $\eta \leq 1.0$ respond adiabatically. We choose $\eta_{\textrm{destr}} =  1.4$ as the lowest value at which a dwarf can get destroyed in MOND since dwarfs with this $\eta$ still seem to be marginally capable of contracting their $r_h$ back to their pericentre value by the time they reach apocentre.\footnote{This certainly appears to be the case for the $\eta = 1.5$ model with $R_i = 300$~kpc.} For the upper limit to $\eta_{\textrm{destr}}$ at pericentre, we choose a value of 2.0 because for this $\eta$, the dwarfs in our simulations are clearly larger at apocentre than at pericentre and are still expanding at the end of the simulation, indicating irreversible behaviour. We therefore infer that $\eta_{\textrm{destr}} =  1.70 \pm 0.30$ if $r_h$ is measured at the second apocentre. If instead we obtain $r_h$ at pericentre, then $\eta_{\textrm{destr}}$ has a slightly lower value of $1.35 \pm 0.05$.

As expected, $\eta_{\textrm{destr}}$ is of order unity because the main physics should be captured by analytic arguments \citep{Zhao_2005, Zhao_2006}. Our numerical results suggest that it would be more accurate to drop the factor of $\frac{2}{3}$ in Equation~\ref{rtid_MOND}, which would also reconcile the numerical pre-factor with that in the Newtonian tidal radius formula (Equation~\ref{rtid_LCDM}) for the case $\alpha = 1$ and $g \gg a_{_0}$. This seems to indicate that we should identify the tidal radius with the distance to the L1 Lagrange point in the derivation of \citet{Zhao_2006} $-$ their equation~36 introduces a factor of $\frac{2}{3}$ in the Newtonian limit because the Roche Lobe extends to a shorter distance in the two non-radial directions than in the radial direction by about this factor. However, it could be that for somewhat eccentric orbits, the Roche Lobe's extent along the radial direction is the limiting factor to the dwarf's size.\footnote{Without the $\frac{2}{3}$ factor in Equation~\ref{rtid_MOND}, the tidal susceptibility threshold is $\eta_{\textrm{destr}} = 1.13 \pm 0.20$ when using $r_h$ at apocentre and $\eta_{\textrm{destr}} = 0.90 \pm 0.03$ when using $r_h$ at pericentre. Note that the MOND tidal susceptibilities of FDS dwarfs would also be reduced by a factor of $\frac{2}{3}$ in this case, which would affect the inferred $\eta_{\textrm{destr}}$ posterior.}

Our simulations also show that the higher the initial distance to the cluster, the more resilient the dwarf is to the effect of cluster tides. This is because a more eccentric orbit implies a shorter amount of time spent near pericentre, so the dwarf is exposed to a high $\eta$ value for only a very brief period, allowing it to recover. Therefore, we would probably still be able to observe dwarfs which have $\eta = 2.4$ (or higher) at pericentre if these have sufficiently large apocentric distances. Given that in our analysis we considered dwarfs up to 800~kpc from the cluster centre, it is likely that there are several dwarfs in our sample which experienced a somewhat higher $\eta$ at some point in their past $-$ but for a sufficiently brief period that the dwarf remained intact. This is fairly consistent with the results of our MCMC analysis, which found that $\eta_{\textrm{destr}} = 1.88_{-0.53}^{+0.85}$.

The observed shape of a dwarf is one of the indicators for whether it has been perturbed. Therefore, to estimate the $\eta$ at which simulated dwarfs should start appearing morphologically disturbed, we look at the evolution of their aspect ratio (Equation~\ref{Aspect_ratio_def}). We need to bear in mind that even a uniform external field can cause a MONDian dwarf to become deformed because the potential of a point mass is not spherical once the EFE is considered \citep{Banik_2018_EFE}. \textit{N}-body simulations of dwarfs experiencing the EFE but not tides explicitly show that this process can yield axis ratios of $\approx 0.7$ \citep{Wu_2017}. This is very much in line with our lowest eccentricity orbit with $R_i = 150$~kpc, so the mild degree of flattening evident here is not necessarily indicative of tidal effects. We find that models with $R_i = 150$~kpc start to acquire significantly elongated morphologies throughout most of their trajectories only when $\eta \ga 0.6$ (see column~3 in Fig.~\ref{fig:dwarf_simulations}). Therefore, we take $\eta_{\textrm{min, dist}} \approx 0.6$. This is slightly higher than what our MCMC analysis requires ($\eta_{\textrm{min, dist}} = 0.24_{-0.19}^{+0.24}$). One possible explanation is that dwarfs with higher $R_i$ start acquiring elongated morphologies at lower $\eta$.


To check if increasing the resolution would affect our results, we perform a high-resolution rerun of one of our models for each $R_i$. This is shown using the dotted line in each panel of Fig.~\ref{fig:dwarf_simulations}. The only resolution-related effect which we can observe is that the half-mass radius of a distant dwarf expands less than at lower resolution. Because of this, we obtain slightly lower pericentric $\eta$ values for the same orbit with higher resolution. However, the evolution of the dwarf properties as a function of $\eta$ at pericentre remains almost the same as for the low-resolution model. Therefore, our conclusions should barely be affected by the resolution of the simulation.

\section{Discussion}
\label{discussion}


Observations of Fornax Cluster dwarf galaxies show that some of them present a detectable level of disturbance in their morphology. Among the environmental effects inside a galaxy cluster that could be causing this disturbance, we found that gravitational tides from the cluster are the most likely cause (Section~\ref{effects_gravi}). The condition for a dwarf galaxy in a galaxy cluster to be tidally stable is approximately the same as the requirement that the dwarf's density exceed the average density of the cluster interior to the dwarf's orbit (Equation~\ref{approx_rtid}).\footnote{The tidal stress $\Delta g_c/\Delta r$ is related to the cluster mass profile $M_c \left( <R \right)$ by $\Delta g_c/\Delta r = GM_c\left( 2 - \alpha \right)/R^3$, from which it follows that $r_{\textrm{tid}}^3/M_{\textrm{dwarf}} \approx R^3/M_c$. Thus, a dwarf with $r_h \approx r_{\textrm{tid}}$ has $M_{\textrm{dwarf}}/r_h^3 \approx M_c/R^3$.} This should be the case for a $\Lambda$CDM dwarf in a cluster because we expect the dwarf to be dominated by dark matter and to have formed much earlier than the cluster, at which time the cosmic mean density was higher. Therefore, in this paradigm, the dwarf galaxies in the Fornax Cluster should be little affected by the tides it raises. This is indeed what our calculations show (Fig.~\ref{fig:hist_tidal_sus}).

In MOND, the enhancement to the Newtonian gravity of an isolated dwarf is similar to that provided by the dark matter halo in $\Lambda$CDM. However, MONDian dwarfs in a galaxy cluster are also affected by the resulting EFE, which weakens their self-gravity. As a result, they are more susceptible to tides than dwarfs in $\Lambda$CDM, which has no EFE due to the strong equivalence principle. Therefore, observations of Fornax dwarfs can be used to compare which of the two models performs better.

To check if tides might be important in the Fornax Cluster, we plotted the projected separation ($R_{\textrm{sky}}$) of each FDS dwarf against a measure of its surface brightness (Fig.~\ref{fig:tid_edge}). This revealed a lack of low surface brightness dwarfs in the central $\approx 200$~kpc even though such dwarfs are evident further out, indicating that selection effects are not responsible for the tentative tidal edge marked on this figure as a grey line. Just below this, the proportion of apparently disturbed dwarfs is also much higher than elsewhere in the cluster (see Fig.~\ref{fig:hist_disturb}). We quantified this trend by plotting the disturbed fraction as a function of the tidal susceptibility $\eta$ of each dwarf (Equation~\ref{eta_rtid}), revealing a clear rising trend detected at $2.9\sigma$ significance in $\Lambda$CDM and $3.5\sigma$ in MOND (Fig.~\ref{fig:tid_sus_obs}). These arguments suggest that the dwarf galaxy population in the FDS catalogue has been significantly shaped by tides, as previously argued by \citet{Venhola_2022}.

\begin{figure}
	\centering
	\includegraphics[width = 8.5cm]{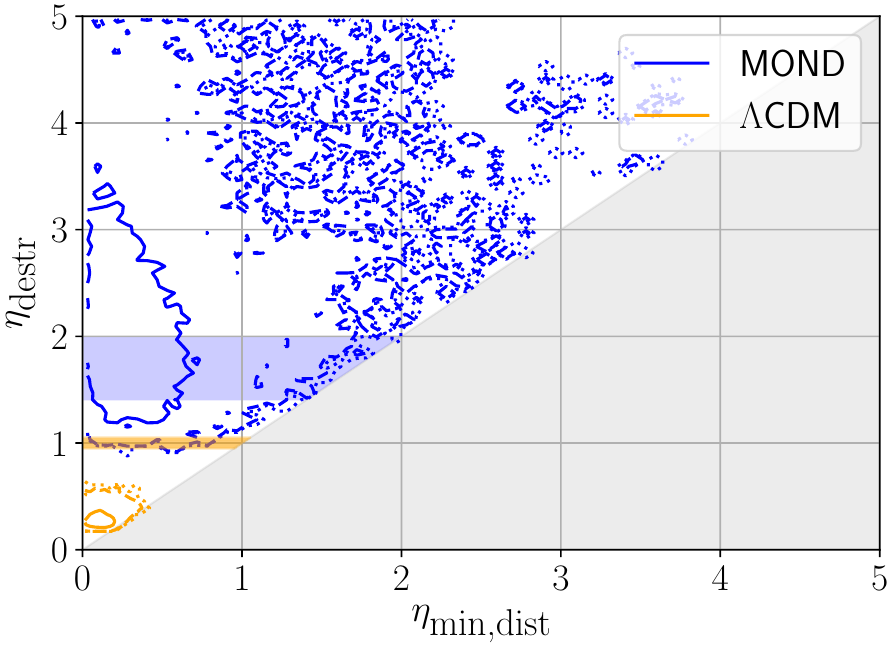}
	\caption{Joint inference on $\eta_{\textrm{min, dist}}$ and $\eta_{\textrm{destr}}$ (Section~\ref{subsubsec:dwarf_disturbance}). We show the $1\sigma$ (inner solid line), $3\sigma$ (dashed line), and $5\sigma$ (outer dotted line) confidence region for MOND (blue) and $\Lambda$CDM (orange). The thick orange line shows the $\Lambda$CDM expectation that $\eta_{\textrm{destr}} \approx 1$. For MOND, the corresponding expectation from our \textit{N}-body simulations (Section~\ref{Nbody_sim}) is that $\eta_{\textrm{destr}} = 1.70 \pm 0.30$ (horizontal blue stripe). The grey shaded region below the line of equality is not allowed by our choice of prior because it is unphysical.}
	\label{fig:eta_min_dist_destr}
\end{figure}

However, the overall distribution of $\eta$ only goes up to $\approx 0.5$ in $\Lambda$CDM (Fig.~\ref{fig:hist_tidal_sus}). We expect a dwarf to be destroyed or severely disturbed only if $\eta \approx 1$, as indicated by $\Lambda$CDM \textit{N}-body simulations \citep{Penyarrubia_2009, Van_den_Bosch_2018}. We quantified this discrepancy using our MCMC analysis, which shows that the tidal stability limit of the Fornax dwarfs should be $\eta_{\textrm{destr}} = 0.25^{+0.07}_{-0.03}$ to match observations. Therefore, $\Lambda$CDM dwarfs should be destroyed when the tidal force that they experience is $\approx 0.25^3 = 1.56 \times 10^{-2}$ times smaller than their internal gravity (tidal force/internal gravity $\approx \eta^3$). Not only is this unrealistic, but also such a low $\eta_{\textrm{destr}}$ is in $>5\sigma$ tension with the $\eta_{\textrm{destr}}$ value of 1 inferred from $\Lambda$CDM \textit{N}-body simulations (Fig.~\ref{fig:eta_min_dist_destr}). The highest $\eta_{\textrm{destr}}$ value achieved with our MCMC analysis for $\Lambda$CDM is only 0.60. This corresponds to the $4.42\sigma$ upper limit because we ran $10^5$ MCMC trials. Since the uncertainty on $\eta_{\textrm{destr}}$ towards higher values from the mode is only 0.07, it is clear that $\eta_{\textrm{destr}} = 1$ is strongly excluded by the observations if the tidal susceptibilities are calculated within the $\Lambda$CDM framework.

These calculations are based on Equation~\ref{rtid_LCDM}, which can be written in the alternative form
\begin{eqnarray}
    \frac{r_{\textrm{tid, } \Lambda\textrm{CDM}}}{R} ~=~ \left( \frac{M_{\textrm{dwarf}}}{\beta M_c \left( < R \right)} \right)^{1/3} \, , \quad \beta ~=~ 2 \left( 2 - \alpha \right) \, ,
    \label{beta_definition}
\end{eqnarray}
where $\alpha = 1.1$ (defined in Equation~\ref{M_cluster}) is the logarithmic slope of the Fornax Cluster mass profile $M_c \left( <R \right)$ based on hydrostatic equilibrium of the gas around its central galaxy \citep{Paolillo_2002}. This implies $\beta = 1.8$. Other workers use slightly different definitions for the tidal radius, which affects the results somewhat because the calculated $\eta \propto \beta^{1/3}$. For example, equation~6 of \citet{Wasserman_2018} gives $\beta = 2 - \alpha = 0.9$ for radial orbits and $\beta = 3 - \alpha = 1.9$ for circular orbits. Allowing even a modest amount of eccentricity, it is clear that $\beta$ in their tidal radius definition is smaller than our adopted 1.8, so their formula generally gives even lower $\eta$ values, worsening the problem for $\Lambda$CDM. Meanwhile, equation~3 of \citet{Penyarrubia_2009} gives $\beta = 3$, though this is for circular orbits and lacks a rigorous derivation \citep[see section~3.1 of][]{Penyarrubia_2008}. $\beta = 2$ is more appropriate to account for elongation in the potential along the radial direction \citep{Innanen_1983, Zhao_2006}. However, even if we adopt $\beta = 3$, this would only raise our calculated $\eta$ values by a factor of $\left( 3/1.8 \right)^{1/3}$, or equivalently imply that we can keep our definition but should consider dwarfs to be destroyed at $\eta_{\textrm{destr}} = \left( 1.8/3 \right)^{1/3} = 0.84$. This is still well above the value given by any of the $10^5$ trials in the MCMC analysis. A more recent detailed derivation affirms that for circular orbits, the appropriate value of $\beta = 3 - \alpha = 1.9$ in the Fornax case \citep[equation~5 of][]{Van_den_Bosch_2018}, which is very similar to our adopted value of 1.8. Although this could be somewhat higher with a lower value for $\alpha$, we can get $\beta = 3$ only for circular orbits around a point mass ($\alpha = 0$), which is not consistent with the Fornax Cluster having an extended dark matter halo. Moreover, a dwarf on an elliptical orbit is exposed to the pericentre value of $\eta$ for only a short time. We may intuitively expect that a dwarf would be disrupted only if it experiences $\eta > 1$ for a significant duration, since otherwise there is not enough time for tidal forces to disrupt the dwarf. This could explain why \citet{Van_den_Bosch_2018} found that dwarfs are actually quite robust to tides, more so than in many numerical simulations where apparent tidal destruction could be a numerical artefact \citep[see also][]{Webb_2020}. It could well be that the appropriate $\eta_{\textrm{destr}}$ is slightly above 1, as in the MOND case. Moreover, \citet{Van_den_Bosch_2018} found that galaxy-galaxy harassment is much less damaging than the tidal shock from pericentre passage. While their work addressed subhaloes in a MW-like halo and neglected hydrodynamics, it is still very useful in showing that a subhalo can resist disruption even if the energy it gains from harassment exceeds the binding energy, justifying our neglect of the harassment scenario (Section~\ref{tidal_sus_Fornax}).

In MOND, we obtained a tidal stability limit with the MCMC analysis of $\eta_{\textrm{destr}} = 1.88^{+0.85}_{-0.53}$, which is closer to the expected value of $\approx 1$ based on analytic arguments (Equation~\ref{rtid_MOND}). To check if this limit is accurate, we performed several \textit{N}-body simulations of a dwarf orbiting a central potential similar to the Fornax Cluster (Section~\ref{Nbody_sim}). These simulations suggest that cluster tides would make Fornax dwarfs appear disturbed when $\eta_{\textrm{min,dist}} \ga 0.6$ and destroy them at $\eta_{\textrm{destr}} = 1.70 \pm 0.30$, which is in good agreement with our MCMC results (see Fig.~\ref{fig:eta_min_dist_destr}).

\begin{figure}
	\centering
	\includegraphics[width = 8.5cm]{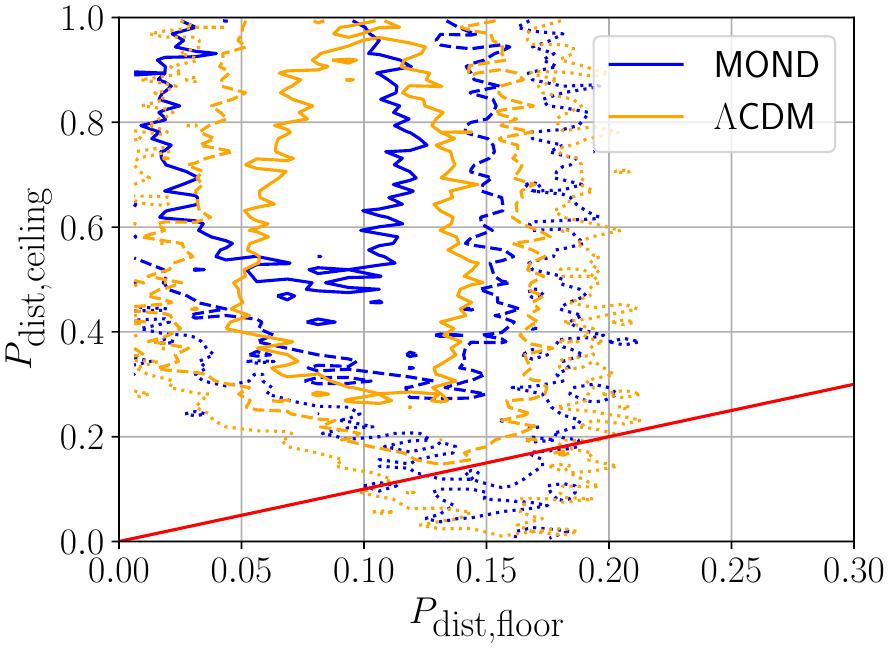}
	\caption{Joint inference on $P_{\textrm{dist, floor}}$ and $P_{\textrm{dist, ceiling}}$ (Section~\ref{subsubsec:dwarf_disturbance}). We show the $1\sigma$ (inner solid line), $2\sigma$ (dashed line), and $3\sigma$ (outer dotted line) confidence region for MOND (blue) and $\Lambda$CDM (orange). Physically, we expect to get values above the solid red line of equality ($P_{\textrm{dist, ceiling}} \geq P_{\textrm{dist, floor}}$), though this is not imposed as a prior. Even so, this is favoured by the MCMC analysis, which gives a likelihood that $P_{\textrm{dist, ceiling}} \leq P_{\textrm{dist, floor}}$ of only $3.14 \times 10^{-3}$ ($2.95\sigma$) in MOND and $6.43 \times 10^{-3}$ ($2.73\sigma$) in $\Lambda$CDM. Both theories prefer a non-zero false positive rate of $P_{\textrm{dist, floor}} \approx 0.1$, which is related to the similar fraction of dwarfs classified as disturbed in the outskirts of the Fornax Cluster where tides should be unimportant (Fig.~\ref{fig:hist_disturb}).}
	\label{fig:P_floor_ceiling}
\end{figure}

We considered several possible explanations for the discrepancy between the low tidal susceptibility values of $\Lambda$CDM dwarfs and the fact that some of the observed Fornax dwarfs appear disturbed. This could be due to the fact that cluster tides are not the main effect responsible for the observed morphological disturbances. However, there are several trends in the FDS that suggest exactly this. These trends are as follows:
\begin{enumerate}
    \item There are fewer low surface brightness dwarfs towards the centre of the cluster, where they are most susceptible to tides (Fig.~\ref{fig:tid_edge}). Since such dwarfs are detectable further out, this feature cannot be ascribed to selection effects. A related finding is that FDS dwarfs are typically larger towards the cluster centre, which could be related to tidal heating \citep[for a more detailed discussion, see section~7.4 of][]{Venhola_2022}; and
    \item The algorithm in charge of fitting the simulated Fornax system to the observations clearly noticed a rising trend between $\eta$ and the probability of disturbance ($P_{\textrm{dist}}$). This is shown by the fact that the algorithm chose $P_{\textrm{dist, ceiling}} > P_{\textrm{dist, floor}}$ with $\approx 3\sigma$ confidence in both $\Lambda$CDM and MOND (see Fig.~\ref{fig:P_floor_ceiling}), even though we did not impose this condition a priori.
\end{enumerate}
We have seen that these trends cannot be understood in $\Lambda$CDM as a direct consequence of cluster tides given the very low $\eta$ values. Moreover, the other major environmental effect that could be causing the observed disturbance (galaxy-galaxy harassment) also presents very low $\eta$ values (see Section~\ref{tidal_sus_Fornax}).

Another possibility is that our results could be affected by some of the assumptions or choices that we made during the analysis. To check if this is the case, we repeat the procedures described in Section~\ref{test_mass} but change some of the assumed conditions and/or parameters in the following ways:
\begin{enumerate}
    \item{Considering that the FDS dwarfs could have a lower dark matter fraction within their optical radius:} We consider the possibility that the dark matter fraction of the FDS dwarfs is lower than assumed in our nominal case (this is motivated in Section~\ref{newDMfrac}). Assuming that $\Lambda$CDM explains the properties of isolated dwarfs, we use the velocity dispersions of nearby isolated dwarfs to estimate their typical dark matter fraction, which returns a somewhat lower value than assumed in our nominal analysis. Substituting this fit (Equation~\ref{revised_DM_frac}) into our MCMC chain raises $\eta_{\textrm{destr}}$ slightly, but it is still only $0.33^{+0.04}_{-0.05}$. We then consider a very conservative scenario in which there is only $10\times$ as much dark matter as stars within the optical extent of each dwarf, which requires altering Equation~\ref{M_dwarf_rule} to $M_{\textrm{dwarf, } \Lambda\textrm{CDM}} = 11~M_{\star}$. For this very low dark matter fraction, we obtain that $\eta_{\textrm{destr}} = 0.54^{+0.19}_{-0.09}$, which reduces the tension between observations and $\eta_{\textrm{destr}} = 1$ (as expected from \textit{N}-body simulations) to $2.29\sigma$ (the triangle plot for this analysis is shown in Fig.~\ref{fig:triang_newDMfrac}). While this is a significant improvement with respect to the $>5\sigma$ tension in the nominal case, we see that even when considering one of the most conservative assumptions for the amount of dark matter contained within the optical radius of a dwarf, $\eta_{\textrm{destr}} \geq 1$ is still excluded at 97.8\% confidence. Moreover, we show in Section~\ref{newDMfrac} that in a recent high-resolution cosmological $\Lambda$CDM simulation, the dark matter fraction within the stellar $r_h$ of a dwarf is far higher than this at the relevant $M_{\star}$, and is actually quite close to our nominal assumption;

	\item{Changing the lower limit to the distribution of dwarf densities in the test mass simulation}: To check if the adopted detection limit to the density of the Fornax dwarfs significantly affects the results, we repeat the analysis using a density threshold $\rho_t$ that is $5\sigma$ below the mean logarithmic density. We also consider a density limit of $\rho_{\textrm{mean}}$ (grey line in Fig.~\ref{fig:hist_dens}). For reference, the nominal $\rho_t$ in MOND is $2.88\sigma$ below the mean logarithmic density, while $\rho_{\textrm{mean}}$ is $1.91\sigma$ below. The corresponding values in $\Lambda$CDM are $3.58\sigma$ and $2.56\sigma$, respectively (see Appendix~\ref{dwarf_dens_LCDM}). Fig.~\ref{fig:3dens} shows the triangle plots comparing the results obtained using these two density limits with the nominal one for $\Lambda$CDM and MOND. From these plots (described further in Appendix~\ref{sec:triang}), we can see that choosing a lower $\rho_t$ worsens the tension for $\Lambda$CDM while maintaining consistency in MOND. Using a higher $\rho_t$ helps to increase the estimated values for $\eta_{\textrm{destr}}$ in $\Lambda$CDM. However, even if we use $\rho_t = \rho_{\textrm{mean}}$, the inferred $\eta_{\textrm{destr}}$ is still significantly below the threshold of $\approx 1$ required in \textit{N}-body simulations, while the inference on $\eta_{\textrm{min, dist}}$ hardly changes. Thus, choosing even higher $\rho_t$ could perhaps help $\Lambda$CDM to reach a reasonable $\eta_{\textrm{destr}}$. However, taking such high values for $\rho_t$ would be in disagreement with observations as the whole point of $\rho_t$ is that dwarfs are not detectable if they have a lower density, but dwarfs with a lower density are clearly observed if we adopt such a high $\rho_t$;

	\item{Changing the values of the deprojection parameters (see Appendix \ref{deproj})}: The deprojection parameters in our nominal analysis were $\textrm{offset} = 0.4^{\circ}$ and $\textrm{nnuc}_{\textrm{floor}} = 1.2^{\circ}$ based on fig.~6 of \citet{Venhola_2019}. We repeat our analysis using deprojection values at the upper limit of the envelope in this figure: $\textrm{offset} = 0.5^{\circ}$ and $\textrm{nnuc}_{\textrm{floor}} = 1.5^{\circ}$. Fig.~\ref{fig:deproj} shows the triangle plots comparing the results for these two different deprojections in $\Lambda$CDM and MOND. From these plots, we can see that these two deprojections give almost the same results in either theory;

	\item{Changing the ratio between present and pericentre distances (see Appendix~\ref{Rper})}: A related change we could make is to consider altering the assumed ratio of 0.29 between the average $R$ and the pericentre distance. This is valid for a thermal eccentricity distribution with $\textrm{Slope}_{P_e} = 2$, which is expected theoretically but is the highest possible value (Equation~\ref{P_e}). With a lower $\textrm{Slope}_{P_e}$, the ratio would rise as orbits would typically be more circular, reducing the calculated tidal susceptibility at pericentre. This would worsen the problem for $\Lambda$CDM; and

	\item{Increasing the resolution}: In Section~\ref{orbit_integration}, we created a grid of $100 \times 100$ cells for different values of the orbital eccentricity ($e$) and initial distance to the cluster centre ($R_i$). We increase the resolution to $200 \times 200$ and repeat the analysis to check if this has any effect on the results. The triangle plots showing the results in $\Lambda$CDM and MOND for these two resolutions are shown in Fig.~\ref{fig:highres}. From these plots, we can see that the results are nearly identical for the high- and low-resolution cases.
\end{enumerate}

From these tests, we infer that our results are not significantly affected by modelling assumptions.

\subsection{The dark matter content of dwarf galaxies in $\Lambda$CDM}
\label{DM_content}

Our conclusion that $\Lambda$CDM is inconsistent with the FDS dwarfs relies heavily on their low values of $\eta$ in this paradigm, which in turn relies on the assumption that they should be dominated by dark matter. We therefore explore whether consistency could be gained by partially relaxing this assumption in a manner consistent with other constraints.

To try and raise $\eta$ while continuing to use Newtonian gravity, we consider the possibility that the FDS dwarfs are TDGs. Our results are presented in Appendix~\ref{tidal_sus_newton}. We see that this scenario is also not viable because the elliptical galaxies in the cluster must still contain substantial dark matter haloes, leading to highly efficient disruption of dwarfs through galaxy-galaxy harassment.

It thus seems clear that the FDS dwarfs should be primordial. In this case, we may consider whether the dark matter density in their central regions could be substantially less than assumed here, raising their tidal susceptibility within the $\Lambda$CDM framework. The transformation of central cusps in the dark matter density profile into cores is expected to be rather inefficient for dwarfs with $M_{\star} \la 10^{7.2} \, M_{\odot}$ \citep{Di_Cintio_2014, Dutton_2016, Tollet_2016}. Most FDS dwarfs have a lower $M_{\star}$ (i.e., they lie below the red line in Fig.~\ref{fig:M_L}). This makes it unlikely that baryonic feedback has substantially reduced the central dark matter density of most FDS dwarfs, especially at the low mass and low surface brightness end important to our argument about tidal stability. Adiabatic contraction could actually raise the central dark matter density \citep{Li_2022, Moreno_2022_expansion}, as could tidal stripping of the dark matter halo \citep{Penyarrubia_2008}. The colours of the FDS dwarfs also indicate that star formation stopped early, most likely due to ram pressure stripping of the gas (Section~\ref{effects_gravi}). Thus, it would only be possible for strong feedback to substantially reduce the baryonic potential depth once. This is insufficient to substantially affect the central dark matter density even in the extreme case that the entire gas disc is instantaneously removed \citep{Gnedin_2002}. Multiple bursts of star formation would be required to substantially affect the dominant dark matter halo \citep{Pontzen_2012}, but it is very unlikely that this occurred in most FDS dwarfs. Consequently, they should still have a significant amount of dark matter in their central regions, as is the case with Galactic satellites whose star formation ended early \citep{Read_2019_core}. Moreover, the low surface brightness nature of the FDS dwarfs considered here implies an atypically large size at fixed $M_{\star}$, causing the baryonic portion of the dwarf to enclose a larger amount of dark matter than for the more typical Illustris galaxies considered by \citet{Diaz_2016}.

Another way in which FDS dwarfs could lose dark matter is through interactions with a massive elliptical galaxy. This scenario has been shown to lead to a dwarf like DF2 with an unusually low dark matter content \citep{Shin_2020}. However, such examples are rare in cosmological simulations \citep{Haslbauer_2019_DF2, Moreno_2022_DF2}. In addition, the possibility that most FDS dwarfs lack dark matter altogether runs into severe difficulties based on simple analytic arguments: Newtonian TDGs would be very fragile and easily disrupted by interactions with massive cluster ellipticals, which must have substantial dark matter haloes in a $\Lambda$CDM context (Appendix~\ref{tidal_sus_newton}). MOND seems to offer the right level of tidal stability: neither too much such that all the dwarfs are completely shielded from tides and the observed signs of tidal disturbance remain unexplained, nor too little such that the dwarfs would have been destroyed by now in the harsh cluster environment studied here. The FDS dwarfs behave just as they ought to in MOND.

This conclusion is in agreement with the recent work of \citet{Keim_2022}, which used the observed tidal disturbance of the dwarf galaxies NGC 1052-DF2 and NGC 1052-DF4 to argue that they must be `dark matter free', since otherwise their dark matter halo would have shielded them from tides. Phrased in a less model-dependent way, these observations indicate much weaker self-gravity than for a typical isolated dwarf, which is a clear prediction of MOND due to the EFE \citep{Famaey_2018, Kroupa_2018_DF2, Haghi_2019_DF2}. In the more isolated galaxy DF44, the self-gravity is stronger despite a similar baryonic content \citep{Van_Dokkum_2019}, but this too is in line with MOND expectations \citep{Bilek_2019, Haghi_2019_DF44}. Strong evidence for the EFE has also been reported from the outer rotation curves of spiral galaxies, which tend to be flat for isolated galaxies but have a declining trend for galaxies in a more crowded environment \citep{Haghi_2016, Chae_2020_EFE, Chae_2021}.

Our results with the FDS are similar to those of \citet{Chilingarian_2019} and \citet{Freundlich_2021}, who also report signs of tidal disturbance in some of the dwarf galaxies in the Coma cluster. Another case in point is the recent study of the dwarf galaxy population in the Hydra I cluster, where the proximity to the cluster centre seems to be affecting the morphology of the dwarfs in a manner suggestive of tidal effects \citep[e.g., larger half-mass radii for dwarfs closer to the cluster centre;][]{La_Marca_2022}. Closer to home, the MW satellites also show signs of tidal disturbance like elliptical isophotes \citep{McGaugh_Wolf_2010}. There is a good correlation between these features and the value of $\eta$ in MOND, which moreover has a maximum value very close to 1 (see their fig.~6). However, the maximum $\eta$ in $\Lambda$CDM is $\la 0.2$, making it difficult to understand the observations in this framework.

\subsubsection{Revised dark matter fraction in $\Lambda$CDM dwarfs}
\label{newDMfrac}

Throughout our analysis, we followed the \citet{Diaz_2016} prescription that $4\%$ of the total dark matter halo of each dwarf lies within its optical radius, with the total halo mass $M_{\textrm{halo}}$ following from $M_{\star}$ through the \citet{Moster_2010} abundance matching relation. The factor of $4\%$ was obtained by fitting to the dynamically inferred dark matter masses $M_{\textrm{DM}}$ within the optical radii of S\textsuperscript{4}G galaxies, as shown in fig.~6 of \citet{Diaz_2016}. In this figure, we can see that for low-mass galaxies ($M_{\star} \la 10^9 M_{\odot}$), the $M_{\textrm{DM}}/M_{\star}$ vs. $M_{\star}$ relation seems to flatten at $M_{\textrm{DM}} \approx 10 \, M_{\star}$. However, this is unclear because S\textsuperscript{4}G has very few well-observed galaxies with such a low mass.

We can use other surveys to extend the S\textsuperscript{4}G results to even lower mass by using measurements of the baryonic properties of dSph galaxies and their line of sight velocity dispersion $\sigma_{\textrm{los}}$. The Newtonian dynamical masses of galaxies from the other surveys are found using equation~2 in \citet{Wolf_2010}:
\begin{eqnarray}
    M_{\textrm{dyn}} \left( < r_h \right) ~=~ \frac{3 r_h \langle \sigma^2_{\textrm{los}} \rangle}{G} \, ,
    \label{M_dyn}
\end{eqnarray}
where $M_{\textrm{dyn}} \left( < r_h \right)$ is the mass within the baryonic $r_h$. Note that when using this to estimate $M_{\textrm{DM}}/M_{\star}$, we account for the fact that only half the stellar mass is enclosed within $r_h$.

\begin{figure}
	\includegraphics[width = 8.5cm]{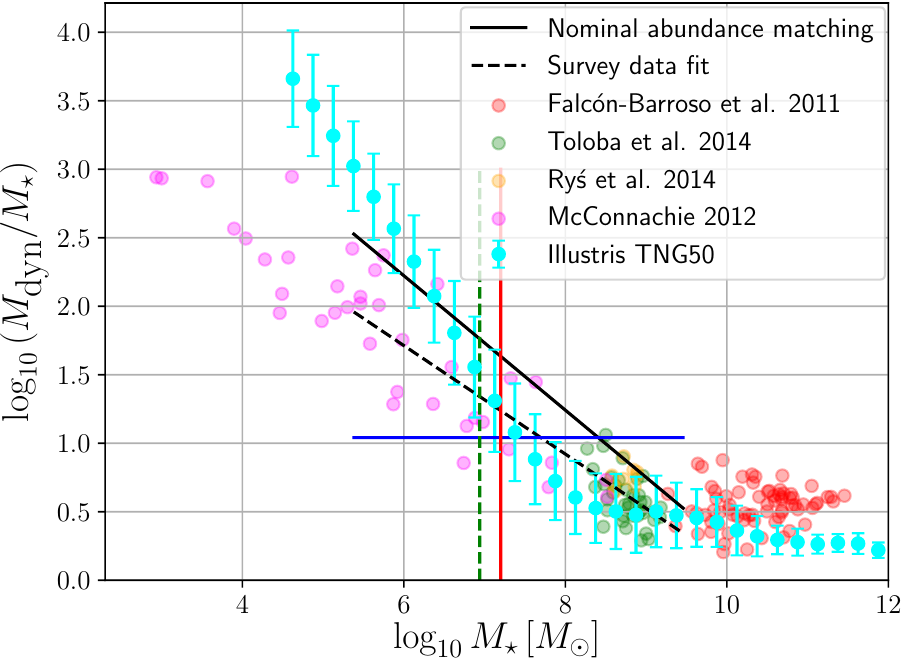}
	\caption{The relation between stellar mass and Newtonian dynamical mass (Equation~\ref{M_dyn}). The semi-transparent coloured dots represent galaxies from four different galaxy surveys as indicated in the legend \citep{Falcon_2011, McConnachie_2012, Rys_2014, Toloba_2014}. The cyan dots with error bars represent the logarithmic mean and dispersion of the total/stellar mass ratio within the stellar $r_h$ of the dwarfs in each $M_{\star}$ bin in the $\Lambda$CDM cosmological simulation Illustris TNG50 \citep{Pillepich_2018, Pillepich_2019, Nelson_2019, Nelson_2019b}. These bins have a width of 0.25~dex and cover the mass range $\log_{10} \left( M_{\star} /M_\odot \right) = 4.5 - 12$. The solid black line represents the expected trend in $\Lambda$CDM with the nominal dark matter fraction from abundance matching (Equation~\ref{M_dwarf_rule}). The dashed black line represents the fit to the dwarfs from the aforementioned galaxy surveys in the mass range covered by the FDS dwarfs. The horizontal blue line at $\log_{10} 11$ shows our conservative assumption that $M_{\textrm{DM}} = 10 \, M_{\star}$. In these three cases, the relations are only plotted over the $M_{\star}$ range of the FDS dwarfs. Their median mass is shown with the dashed green vertical line. The solid red vertical line corresponds to the stellar mass below which core formation is inefficient in $\Lambda$CDM (see the text).}
	\label{fig:M_dyn/M_star}
\end{figure}

To check the consistency between the assumed dark matter fraction and observations of isolated dwarfs, we use Fig.~\ref{fig:M_dyn/M_star} to plot $M_{\textrm{dyn}}/M_{\star}$ of the galaxies in four different galaxy surveys (semi-transparent coloured dots), assuming the \citet{Diaz_2016} result for the dark matter fraction as used in our nominal analysis (black line; Equation~\ref{M_dwarf_rule}), and assuming conservatively that $M_{\textrm{DM}} = 10 \, M_{\star}$ (blue line). We can see that it is rather unlikely that the FDS dwarfs generally have much less dark matter in their baryonic region than we assumed, since the linear regression to the survey data over the $M_{\star}$ range of the FDS dwarfs (dashed black line) is quite close to our nominal dark matter fraction. We can also use the Illustris TNG50 cosmological simulation \citep{Pillepich_2018, Pillepich_2019, Nelson_2019, Nelson_2019b} to check the dark matter fraction that we expect dwarfs to have in the $\Lambda$CDM paradigm. We do this in Fig.~\ref{fig:M_dyn/M_star}, where we show the mean and standard deviation of $M_{\textrm{DM}}/M_{\star} + 1$ within the stellar $r_h$ in $M_{\star}$ bins of width 0.25~dex (cyan dots with error bars). The trend followed by these simulated dwarfs is even steeper than that given by the observed dwarfs, though both give a similar dark matter fraction at the low-mass end crucial to our analysis (the median $M_{\star}$ of the FDS dwarfs is shown by the vertical dashed green line at $\log_{10} \left( M_{\star}/M_{\odot} \right) = 6.96$). This further supports our nominal choice for the dark matter fraction of FDS dwarfs. One reason for their high expected dark matter fraction is that the vast majority of them have too little stellar mass for efficient core formation, the threshold for which is shown by the red vertical line at $\log_{10} \left( M_{\star}/M_{\odot} \right) = 7.2$ for the reasons discussed above. All these arguments highlight that the $M_{\textrm{DM}} = 10 \, M_{\star}$ case is clearly very conservative given the steep relation followed by low-mass galaxies that we expect from abundance matching arguments, Illustris TNG50 results, and the velocity dispersions of nearby dwarfs.

To assess the sensitivity of our analysis in Section \ref{test_mass} to the assumed dark matter fraction, we repeat it with the dark matter fraction given by the linear fit \citep[equations~18 and 19 of][]{Banik_2018_escape} to the observed isolated dwarfs in Fig.~\ref{fig:M_dyn/M_star}:
\begin{eqnarray}
     \log_{10} \left( \frac{M_{\textrm{DM}}}{M_{\star}} + 1 \right) ~=~ 4.089 - 0.396~\log_{10} \left( \frac{M_{\star}}{M_{\odot}} \right) \, ,
    \label{revised_DM_frac}
\end{eqnarray}
where $M_{\textrm{DM}}/M_{\star}$ is the ratio of dark matter to stars within the stellar $r_h$. The typical dwarf densities in this case are about 0.5~dex lower than with the nominal dark matter fraction. As a result, the logarithmic mean is lower than in the nominal case by a similar amount: It is now $\log_{10} \, \rho \left( M_{\odot}/\textrm{pc}^3 \right) = -1.41$. In this case, the density threshold $\rho_t = 5.85 \times 10^{-4}~M_{\odot}/\textrm{pc}^3$ is $2.44\sigma$ below the mean. To keep our statistical analysis comparable to our nominal one, we use the same 6 bins in $\eta_{\textrm{obs}}$ as before. In this way, we obtain that Equation~\ref{revised_DM_frac} gives a slightly higher $\eta_{\textrm{destr}} = 0.33^{+0.04}_{-0.05}$. The maximum value achieved by the MCMC chain is only 0.59, which implies that the $\Lambda$CDM model is still in $>5\sigma$ tension with the expected value of 1.

For completeness, we repeat our analysis with the very conservative assumption that $M_{\textrm{DM}} = 10 \, M_{\star}$. In this case, the distribution of dwarf densities is similar to that in MOND (Fig.~\ref{fig:hist_dens}) but scaled up $11\times$. Thus, the logarithmic dispersion remains $\sigma = 0.57$~dex and the density threshold $\rho_t = 4.66 \times 10^{-4}~M_{\odot}/\textrm{pc}^3$ is still $2.88\sigma$ below the mean $\log_{10} \, \rho$, which is now $-1.69$ in these units. As expected, the $\rho_t$ value is $11\times$ higher than in the MOND model $-$ and thus much less than in our nominal $\Lambda$CDM analysis. We found that in this reduced density case, $\eta_{\textrm{destr}} = 0.54^{+0.19}_{-0.09}$ and the probability that $\eta_{\textrm{destr}} \geq 1$ is $2.23 \times 10^{-2}$ ($2.29\sigma$).

Appendix~\ref{sec:triang} shows the complete triangle plot with the distributions of the model parameters and parameter pairs for the nominal $\Lambda$CDM analysis and the two revised cases described above. There is little impact to the inferences on parameters other than $\eta_{\textrm{destr}}$, $\eta_{\textrm{min, dist}}$, and $\textrm{Slope}_{P_e}$.

Therefore, it is clear that assuming a lower dark matter fraction for the $\Lambda$CDM dwarfs helps to alleviate the tension between observations and \textit{N}-body simulations only if this fraction is reduced significantly. However, having a dark matter fraction of $M_{\textrm{DM}}/M_{\star} = 10$ within the optical radius is a very conservative assumption at odds with many other lines of evidence, including cosmological simulations. Even with this assumption, $\eta_{\textrm{destr}} \geq 1$ is still excluded by our MCMC analysis of the FDS at 97.8\% confidence.

\section{Conclusions}
\label{conclusions}

We studied the tidal susceptibility of dwarf galaxies in the Fornax Cluster to gravitational effects of the cluster environment in both $\Lambda$CDM and MOND. In both theories, we found cluster tides to be the main effect. Thus, cluster tides should be able to explain the observed morphological disturbance of some Fornax dwarfs and the lack of low surface brightness dwarfs towards the cluster centre (Fig.~\ref{fig:tid_edge}). By constructing a test mass simulation of the Fornax system and performing a statistical analysis using the MCMC method, we constrained the tidal susceptibility ($\eta \equiv r_h/r_{\textrm{tid}}$) value at which a Fornax dwarf should get destroyed in order to match the observations, which we call $\eta_{\textrm{destr}}$. We found that $\eta_{\textrm{destr}} = 0.25_{-0.03}^{+0.07}$ in $\Lambda$CDM and $1.88_{-0.53}^{+0.85}$ in MOND.

The $\eta_{\textrm{destr}}$ value in $\Lambda$CDM falls significantly below analytic expectations (Equation~\ref{rtid_LCDM}) and is in $>5\sigma$ tension with \textit{N}-body simulation results, which indicate that $\eta_{\textrm{destr}} \approx 1$ \citep{Penyarrubia_2009, Van_den_Bosch_2018}. In other words, the very low $\eta$ values of FDS dwarfs imply that they should be unaffected by cluster tides, contradicting the observed signs of tidal disturbance. We also found that the other major environmental influence of interactions with individual massive galaxies in the cluster should not be a significant process in $\Lambda$CDM \citep[see also section~7.3.3 of][]{Venhola_2019}. We discarded the possibility that the above-mentioned discrepancy is due to the minimum allowed density of the simulated sample of dwarfs being too low, the deprojection parameters being different from our nominal ones, the resolution of the test mass simulation not being high enough to get reliable results, and the dwarfs having less dark matter than we assumed (Section~\ref{discussion}). In particular, the velocity dispersions of nearby isolated dwarfs suggest a slightly lower dark matter fraction (dashed line in Fig.~\ref{fig:M_dyn/M_star}). Using this only slightly raises $\eta_{\textrm{destr}}$ to $0.33^{+0.04}_{-0.05}$. Even if we conservatively assume that the FDS dwarfs have only $10\times$ as much dark matter as stars within their optical radius, we still get a $2.29\sigma$ tension with expectations (Equation~\ref{rtid_LCDM}). Therefore, our results reliably show that the $\Lambda$CDM paradigm is in serious tension with observations of perturbed dwarf galaxies in the Fornax Cluster \citep[observations which are strongly suggestive of tidal effects, see also section~7.4 of][]{Venhola_2022}.

An alternative model that assumes different properties for the dark matter particles could perhaps reconcile the basics of the $\Lambda$CDM cosmology with the observed morphological disturbances of some Fornax dwarfs. One of the most popular alternatives is the `superfluid dark matter' model \citep{Berezhiani_2015, Hossenfelder_2020}. Like most hybrid models, it attempts to reconcile the successes of MOND on galaxy scales with the advantages of dark matter on larger scales, especially with regards to the CMB anisotropies and galaxy cluster dynamics. However, this model also presents its own problems, including orbital decay of stars in the Galactic disc from Cherenkov radiation \citep{Mistele_2022_Cherenkov} and that the LG satellite planes extend beyond the estimated superfluid core radii of the MW and M31, making it difficult to explain the high observed internal velocity dispersions of the satellites in these planes \citep[see section~5.6 of][]{Roshan_2021_disc_stability}. There are also difficulties explaining the observed regularities in rotation curves consistently with gravitational lensing results in a theory where baryons feel extra non-gravitational forces that do not affect photons \citep*{Mistele_2022_SFDM}. Another possibility is that the dark matter particles are fuzzy with a low mass and thus a long de Broglie wavelength, reducing their density in the central region of a dwarf galaxy. However, ultralight bosons \citep{Hu_2000, Hui_2017} are in significant tension with observations of the Lyman-$\alpha$ forest \citep{Rogers_2021}. More generally, reducing the ability of dark matter to cluster on small scales would make it difficult to form dwarf galaxies at high redshift and to explain their high Newtonian dynamical $M/L$ ratios.

This brings us to the MOND case, in which the inferred $\eta_{\textrm{destr}}$ is much more consistent with analytic expectations (Equation~\ref{rtid_MOND}). In order to compare $\eta_{\textrm{destr}}$ with the results of \textit{N}-body simulations as we did for $\Lambda$CDM, we had to perform numerical MOND simulations ourselves \citep[though one pioneering study exists, see][]{Brada_2000_tides}. From our simulations tailored to the properties of a typical dwarf galaxy in the Fornax Cluster, we obtained that $\eta_{\textrm{destr}} = 1.70 \pm 0.30$, in excellent agreement with the value required to fit the observational data according to the MCMC method. We therefore conclude that MOND performs significantly better than $\Lambda$CDM and is clearly the preferred model in all the tests that we conducted throughout this work, even though it was not designed with the FDS in mind. Nevertheless, MOND still needs an additional ingredient to explain some of the observations on larger scales, especially the temperature and pressure profiles in galaxy clusters and the CMB power spectrum \citep{Famaey_McGaugh_2012}. For this, several models have been proposed that complement MOND. Some of the most promising ones are the relativistic MOND theory which can fit the speed of gravitational waves and the CMB anisotropies but likely cannot explain the dynamics of virialized galaxy clusters \citep{Skordis_2021}; and the $\nu$HDM model that assumes MOND gravity and 11~eV sterile neutrinos \citep{Angus_2009}. These proposed particles would play the role of a collisionless component that only aggregates at the scale of galaxy clusters, helping to explain the Bullet Cluster \citep{Angus_2007} and other virialized galaxy clusters \citep{Angus_2010}, where the MOND corrections to Newtonian gravity are generally small. MOND has also proved capable of explaining several physical phenomena that $\Lambda$CDM has been failing to describe, including the planes of satellite galaxies in the LG and beyond \citep{Pawlowski_2021_Nature_Astronomy, Pawlowski_2021}, the weakly barred morphology of M33 \citep{Sellwood_2019, Banik_2020_M33}, and the pattern speeds of galaxy bars \citep{Roshan_2021_disc_stability, Roshan_2021_bar_speed}. Using the $\nu$HDM extension, MOND can also explain the CMB \citep{Angus_Diaferio_2011}, the KBC void and Hubble tension \citep{Haslbauer_2020}, and the early formation of the interacting galaxy cluster El Gordo \citep{Katz_2013, Asencio_2021}. Therefore, this later model is capable of explaining both the CMB and the dynamics of galaxy clusters while preserving the successes of MOND at galaxy scales \citep[][and references therein]{Banik_Zhao_2022}. In this study, we have shown that it should also be capable of resolving the problem faced by $\Lambda$CDM with regards to the observed signs of tidal disturbance in Fornax Cluster dwarf galaxies.

\section*{Acknowledgements}

EA is supported by a stipend from the Stellar Populations and Dynamics Research Group at the University of Bonn. IB is supported by Science and Technology Facilities Council grant ST/V000861/1, which also partially supports HZ. IB acknowledges support from a ``Pathways to Research'' fellowship from the University of Bonn. PK acknowledges support through the Deutscher Akademischer Austauschdienst-Eastern European Exchange Programme. EA would like to thank Prof. Xufen Wu for providing the initial conditions templates of the dwarf galaxy used in the MOND \textit{N}-body simulations. The authors are grateful to Sara Eftekhari for providing the table of literature data shown in Fig.~\ref{fig:M_dyn/M_star}. They are also grateful to the referee for comments which substantially improved this paper.

\section*{Data availability}

The results presented can be reproduced by using the data available in the Vizier catalogue\footnote{\url{https://vizier.u-strasbg.fr/viz-bin/VizieR-3?-source=J/A\%2bA/620/A165/dwarf\&-out.max=50\&-out.form=HTML\%20Table&-out.add=_r\&-out.add=_RAJ,_DEJ\&-sort=_r\&-oc.form=sexa}} and following the methods described in this paper. For a user guide describing how to install \textsc{por} and providing links from which it can be downloaded, we refer the reader to \citet{Nagesh_2021}.

\bibliographystyle{mnras}
\bibliography{Fornax_dwarfs}

\begin{appendix}

\section{Deprojecting distances in the sky plane to 3D distances}
\label{deproj}

In order to convert an observed 2D projected distance $R_{\textrm{sky}}$ into a 3D distance $R$, we use a simplified version of the deprojection method applied in \citet{Venhola_2019}. For convenience, we normalize distances to $d_{\textrm{Fornax}} = 20$~Mpc, the distance to the Fornax Cluster \citep{Blakeslee_2009}. Thus, we define
\begin{eqnarray}
    \theta_{\textrm{2D}} ~\equiv~ \frac{R_{\text{sky}}}{d_{\textrm{Fornax}}} \, , \quad \theta_{\textrm{3D}} ~\equiv~ \frac{R}{d_{\textrm{Fornax}}} \, .
\end{eqnarray}
Fig. 6 of \citet{Venhola_2019} shows the relation between these quantities for nucleated and non-nucleated dEs.\footnote{Results are also shown for dwarf irregulars, but we removed these from our sample.} The relation for nucleated dwarfs is almost parallel to the line of equality, but with an offset of $\approx 0.4^\circ$. Therefore, we deproject a dwarf labelled as `nucleated' using
\begin{eqnarray}
	\theta_{\textrm{3D}} ~=~ \theta_{\textrm{2D}} + \textrm{offset} \, ,
\end{eqnarray}
with offset~$=0.4^\circ$ in our nominal analysis.

In the case of non-nucleated dwarfs, $\theta_{\textrm{3D}}$ has a constant floor value of $\approx 1.2^\circ$ until it joins the relation between $\theta_{\textrm{3D}}$ and $\theta_{\textrm{2D}}$ followed by nucleated dwarfs at $\theta_{\textrm{2D}} > \textrm{nnuc}_{\textrm{floor}} - \textrm{offset}$. Therefore, for the non-nucleated dwarfs, we apply the following deprojection:
\begin{eqnarray}
	\theta_{\textrm{3D}} =
	\begin{cases}
    	\textrm{nnuc}_{\textrm{floor}}, & \textrm{if}\ \theta_{\textrm{2D}} \leq \textrm{nnuc}_{\textrm{floor}} - \textrm{offset} \, , \\
   		\theta_{\textrm{2D}} + \textrm{offset}, & \textrm{if}\ \theta_{\textrm{2D}} \geq \textrm{nnuc}_{\textrm{floor}} - \textrm{offset} \, ,
    \end{cases}
\end{eqnarray}
where $\textrm{nnuc}_{\textrm{floor}} = 1.2^\circ$ in our nominal analysis. As with the nucleated dwarfs, we use offset~$=0.4^\circ$.

\section{Obtaining $R_{\textrm{per}}$ from a 3D distance}
\label{Rper}

Assuming a thermal eccentricity distribution \citep{Jeans_1919, Ambartsumian_1937, Kroupa_2008}, we have that the probability distribution of eccentricities is $P_e = 2e$. If the orbits are approximately Keplerian, the pericentre distance $R_{\textrm{per}} = a \left( 1 - e \right)$, where $a$ is the semi-major axis and $e$ is the eccentricity. The time-average distance can be calculated as $\langle R \rangle = a \left(1 + e^2/2 \right)$ \citep[section~3 of][]{Mendez_2017}. To obtain the relation between $\langle R \rangle$ and $R_{\textrm{per}}$, we integrate over the whole eccentricity distribution:
\begin{eqnarray}
	\frac{R_{\textrm{per}}}{\langle R \rangle} = \int \left. \frac{R_{\textrm{per}}}{\langle R \rangle} \right|_e P_e \; de = \int_{0}^{1} \left( \frac{1 - e}{1 + \frac{e^2}{2}} \right) 2e \; de = 0.29.
\end{eqnarray}
We assume that the 3D distance of a dwarf inferred from its observed projected distance (Appendix~\ref{deproj}) is about the same as its time-average distance. We therefore obtain that for the FDS dwarfs, $R_{\textrm{per}} = 0.29 \, R$.

\section{Do two experiments have the same proportion of successes?}
\label{Binomial_significance}

In Section~\ref{comparison_disturbance}, we encountered the problem that one experiment gives $S_{\textrm{obs, 1}}$ `successes' out of $T_1$ trials while another experiment gives $S_{\textrm{obs, 2}}$ successes out of $T_2$ trials, with a success defined as a dwarf galaxy that appears disturbed. The problem is to test the null hypothesis that the proportion of successes ($x$) is the same in both experiments assuming that $T_1$ and $T_2$ are set in advance independently of the actual number of successes. We consider this problem in two stages as follows:
\begin{enumerate}
    \item Keeping $x$ fixed, we evaluate the likelihood $P_x$ of obtaining data as bad as or worse than the observed combination ($S_{\textrm{obs, 1}}$, $S_{\textrm{obs, 2}}$) for the null hypothesis; and
    \item We then obtain a weighted mean value for $P_x$ by considering all plausible $x$, each time weighting by the likelihood that the observed ($S_{\textrm{obs, 1}}$, $S_{\textrm{obs, 2}}$) arises with that $x$.
\end{enumerate}

If we know $x$, we can use binomial statistics (Equation~\ref{binomial}) to find the likelihood of obtaining any combination ($S_1$, $S_2$). We obtain $P_x$ by adding the probabilities of all ($S_1$, $S_2$) combinations which are as likely as or less likely than the observed combination ($S_{\textrm{obs, 1}}$, $S_{\textrm{obs, 2}}$). This follows the usual principle that if the data seems unlikely given the null hypothesis, we should consider all the ways in which it could look as bad or even worse.

If the null hypothesis were true, the probability distribution of its parameter $x$ can be found more accurately by combining the two experiments to obtain a single experiment with $\left( S_{\textrm{obs, 1}} + S_{\textrm{obs, 2}} \right)$ successes out of $\left( T_1 + T_2 \right)$ trials. We use Equation~\ref{Bernoulli_mean_stdev} to calculate the mean $x_0$ and uncertainty $\sigma_x$ of the resulting posterior inference on $x$ assuming a uniform prior. We then consider all values of $x$ within the range $x_0 \pm 5\sigma_x$ provided this does not go outside the mathematically allowed range ($0-1$). Within the considered range of $x$, we weight each $P_x$ determination by the binomial likelihood $P_{\textrm{obs}} \left( x \right)$ of obtaining the observed combination ($S_{\textrm{obs, 1}}$, $S_{\textrm{obs, 2}}$), so $P_{\textrm{obs}} \left( x \right)$ is a product of the binomial likelihood from each of the experiments. The idea is that each $P_x$ should be weighted by how plausible the corresponding $x$ is given the data in the context of the null hypothesis. This leads to our estimated $P$-value:
\begin{eqnarray}
    P ~=~ \frac{\int P_x P_{\textrm{obs}} \left( x \right) \, dx}{\int P_{\textrm{obs}} \left( x \right) \, dx} \, .
\end{eqnarray}
Since it is possible that no value of $x$ matches the observations very well because the null hypothesis is wrong, $P_{\textrm{obs}} \left( x \right)$ might not integrate to 1.

In the particular case of Section~\ref{comparison_disturbance}, calculating the significance $P$ in this way only tells us how plausible it is that $f_d$ is the same in the low $\eta$ and high $\eta$ subsamples, which is the null hypothesis. Our alternative hypothesis specifies that $f_d$ should be higher in the high $\eta$ subsample on physical grounds, not merely that $f_d$ should have some sort of correlation with $\eta$. Since the inferred $f_d$ indeed rises with $\eta$, we should bear in mind that the low likelihood of the null hypothesis is caused by a deviation in just the sense expected on physical grounds under the alternative hypothesis where tides are relevant. On the other hand, we tried all possible choices of $\eta_t$ to maximize the significance of the signal, leading to a look-elsewhere effect.

\section{Tidal susceptibility of Newtonian TDG\lowercase{s}}
\label{tidal_sus_newton}

As discussed in Section~\ref{discussion}, our results indicate a higher level of tidal susceptibility than is expected in $\Lambda$CDM. This could be a sign that the Fornax dwarfs lack dark matter altogether, which is possible in this framework if the FDS dwarfs are mostly TDGs. These are expected to be rather rare in $\Lambda$CDM, so the scenario is not very plausible \citep{Haslbauer_2019_TDG}. We nonetheless consider it for completeness.

If the dwarfs are of tidal origin, they would be free of dark matter \citep{Barnes_1992, Wetzstein_2007}. However, the massive cluster galaxies would still be surrounded by a dark matter halo. In this scenario, the mass ratio between the dwarfs and the massive galaxies would be rather extreme, suggesting a serious problem with the stability of the dwarfs.

\begin{figure} 
	\includegraphics[width = 8.5cm]{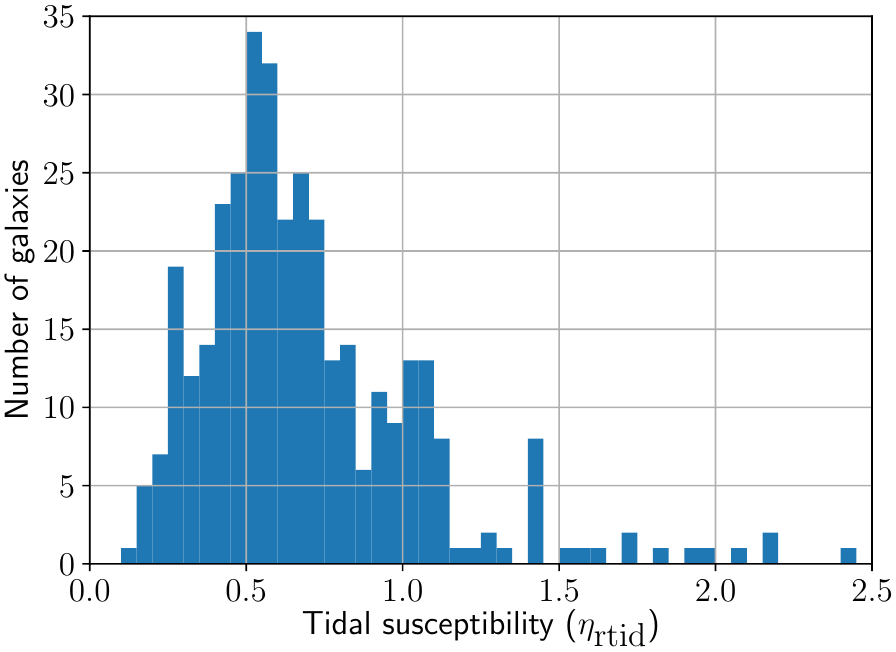}
	\hspace{0.07cm}
	\includegraphics[width = 8.5cm]{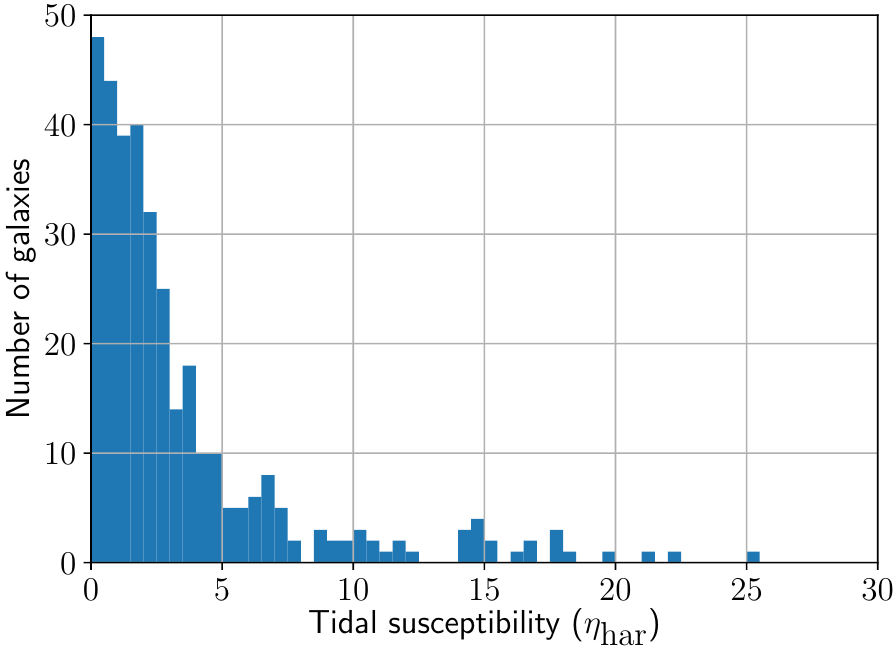}
	\caption{The distribution of tidal susceptibility values of Fornax Cluster dwarfs in a Newtonian TDG scenario to cluster tides (top) and harassment (bottom), with a bin width of 0.05 and 0.5, respectively.}
	\label{fig:hist_tidal_sus_newton}
\end{figure}

To quantify this, we obtain the tidal radius of a dwarf by applying Equation \ref{rtid_LCDM} considering only its baryonic mass. Similarly, we can obtain the disruption time-scale by applying Equation~\ref{td_LCDM} and accounting for the fact that the terms referring to the dwarf (those labelled with a subindex `dwarf') should be purely baryonic while the terms referring to the large galaxies (labelled with a subindex `p') should still account for the dark matter contribution to the mass and half-mass radius. We can then substitute in these results to obtain the susceptibility to cluster tides (Equation \ref{eta_rtid}) and galaxy-galaxy harassment (Equation \ref{eta_har}). The results are shown in Fig.~\ref{fig:hist_tidal_sus_newton}.

As expected, the dwarfs are now much more susceptible to cluster tides (higher $\eta_{\textrm{rtid}}$ than in Fig.~\ref{fig:hist_tidal_sus}). The distribution of $\eta_{\textrm{rtid}}$ becomes very similar to MOND, suggesting that maybe the Newtonian TDG scenario is plausible. However, the tidal susceptibility to harassment ($\eta_{\textrm{har}}$) is very large in this scenario and greatly exceeds 1 for the vast majority of the dwarfs. The high $\eta_{\textrm{har}}$ values arise because the dwarfs are completely unprotected: They do not have a boost to their self-gravity either from MOND or from a dark matter halo. Given their low surface brightness, this leads to very weak self-gravity. However, in a $\Lambda$CDM universe, the large galaxies must still have dark matter haloes. As a result, purely baryonic dwarfs governed by Newtonian gravity should have already been destroyed by encounters with the massive cluster galaxies. Therefore, we can consider that the TDG scenario in $\Lambda$CDM is extremely unlikely. Note that in MOND, our analysis is not sensitive to whether the dwarfs are TDGs or formed primordially $-$ they are purely baryonic in either case.



\section{Distribution of dwarf densities in $\Lambda$CDM}
\label{dwarf_dens_LCDM}

\begin{figure}
	\centering
	\includegraphics[width = 8.5cm]{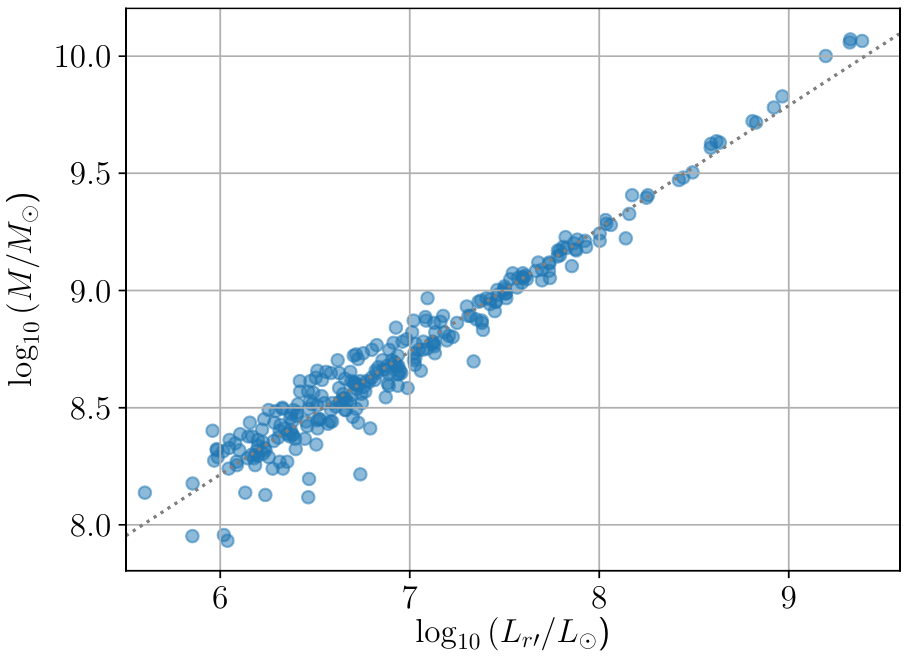}
	\caption{Similar to Fig.~\ref{fig:M_L}, but for $\Lambda$CDM instead of MOND and showing only the linear regression.}
	\label{fig:M_L_LCDM}
\end{figure}

\begin{figure}
	\centering
	\includegraphics[width = 8.5cm]{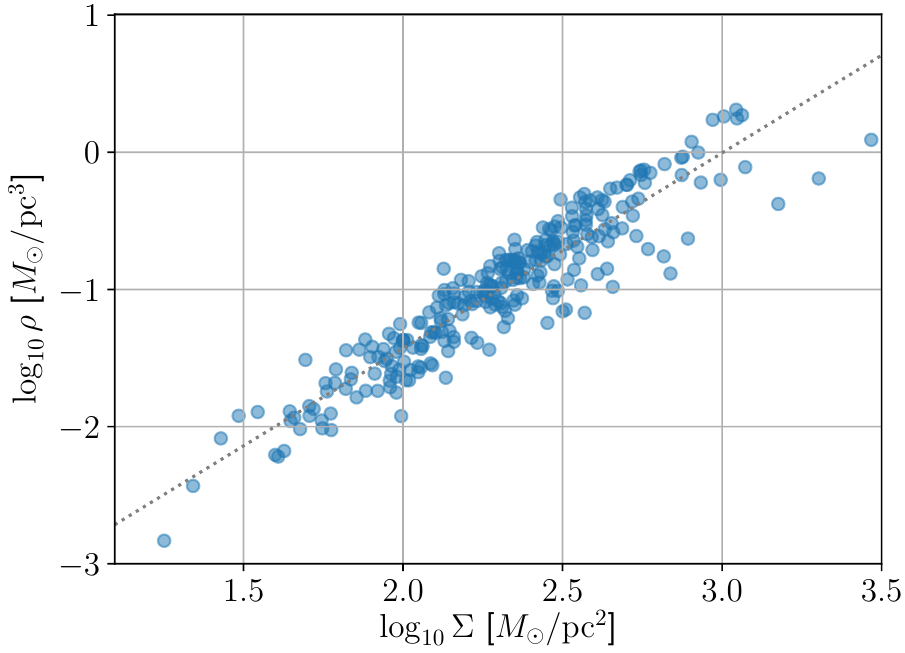}
	\caption{Similar to Fig.~\ref{fig:surfdens_voldens}, but for $\Lambda$CDM instead of MOND and showing only the linear regression.}
	\label{fig:surfdens_voldens_LCDM}
\end{figure}

\begin{figure}
	\centering
	\includegraphics[width = 8.5cm]{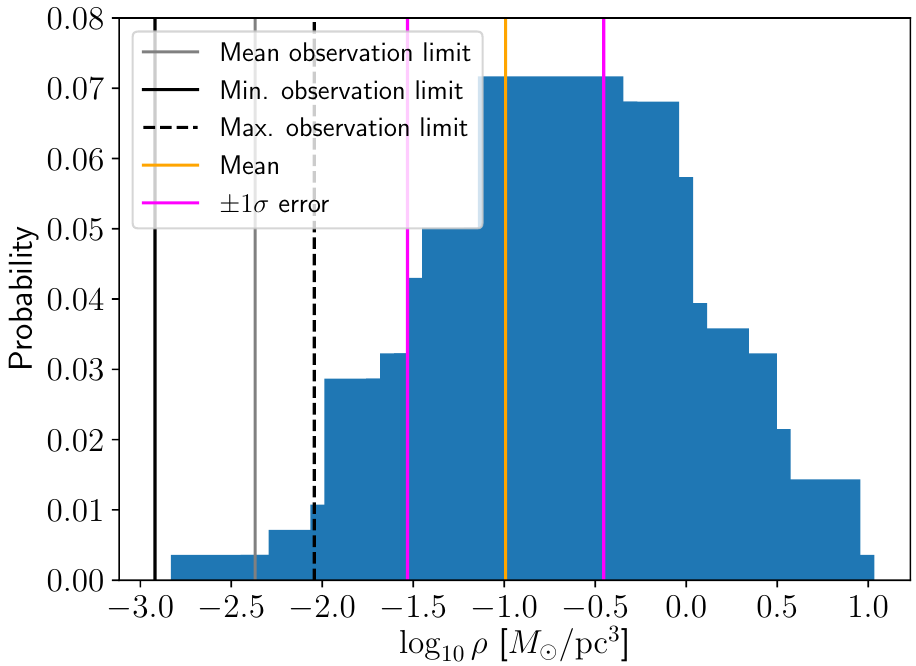}
	\caption{The distribution of each dwarf galaxy's mean density within its half-mass radius, accounting for both baryonic and dark matter. The orange vertical line at $-0.99$ shows the sample mean, while the magenta lines offset by $\pm 0.54$~dex show the standard deviation around it. Other lines have a similar meaning to Fig.~\ref{fig:hist_dens} and have been obtained similarly to the MOND case to allow a fair comparison (see the text).}
	\label{fig:hist_dens_LCDM}
\end{figure}

Our MCMC analysis relies on an assumed distribution for the dwarf densities, which are crucial to their tidal stability. We therefore need to repeat the steps discussed in Section~\ref{subsubsec:dwarf_density} for the case of $\Lambda$CDM. For this model, we show the mass-luminosity relation (Fig.~\ref{fig:M_L_LCDM}), the surface density-volume density relation (Fig.~\ref{fig:surfdens_voldens_LCDM}), and the histogram of volume densities of the dwarfs in the FDS catalogue (Fig.~\ref{fig:hist_dens_LCDM}). The main difference is that the mass of the dwarfs is higher since it includes the contribution of the dark matter component within the optical radius (Equation~\ref{M_dwarf_rule}). This raises their surface and volume density. We found that $M/L_{r'} = 74.92 \pm 52.38~M_{\odot}/L_{\odot, r'}$, indicating a rather high dispersion. Moreover, we can no longer approximate that the slope of the relation is 1 on logarithmic axes, indicating non-linearity.

Due to these difficulties, we found that it would be unsuitable to repeat the steps described in Section~\ref{subsubsec:dwarf_density}. To enable a fair comparison with MOND, we nonetheless used as similar a procedure as possible. For this, we fixed the logarithmic offset between the density of the least dense dwarf in our sample ($\rho_{\textrm{min,FDS}}$) and the adopted density threshold of the survey ($\rho_t$). As a result, the minimum observational limit (black line in Fig.~\ref{fig:hist_dens_LCDM}) is 0.09~dex below $\rho_{\textrm{min,FDS}}$, the mean observational limit (grey line in this figure) is 0.46~dex above $\rho_{\textrm{min,FDS}}$, and the maximum observational limit (dashed black line in this figure) is 0.79~dex above $\rho_{\textrm{min,FDS}}$. As in the MOND case, we choose the minimum observational limit (black line in Fig.~\ref{fig:hist_dens_LCDM}) as our nominal density limit for the distribution since it is the only one of these three choices that implies $\rho_t < \rho_{\textrm{min,FDS}}$, which is required of a realistic detection threshold. Assuming instead the mean observational limit would make us lose 2 observed dwarf galaxies from the low-density tail of the distribution. Note also that these dwarfs have a clear tidal morphology because we removed any dwarfs where this is unclear (Section~\ref{data_sel}).

To summarize, our nominal $\rho_t$ in $\Lambda$CDM is $3.58\sigma$ below the mean logarithmic density, while $\rho_{\textrm{mean}}$ is $2.56\sigma$ below.

\section{Triangle plots with alternative modelling choices}
\label{sec:triang}

In this appendix, we rerun our MCMC analysis with different modelling assumptions and show their impact using triangle plots similar to Fig.~\ref{fig:triang_MONDLCDM}. Instead of showing $\Lambda$CDM and MOND results on the same graph as done there, our approach will be that each graph shows results for different modelling assumptions but within the context of the same theory. We will use different panels for the different theories. As before, we show only the $1\sigma$ contour for each pair of parameters, though the full probability distribution is shown when considering the posterior on one parameter marginalized over all others. The results presented here are discussed in more detail in Section~\ref{discussion}.

\begin{figure}
	\includegraphics[width = 8.5cm]{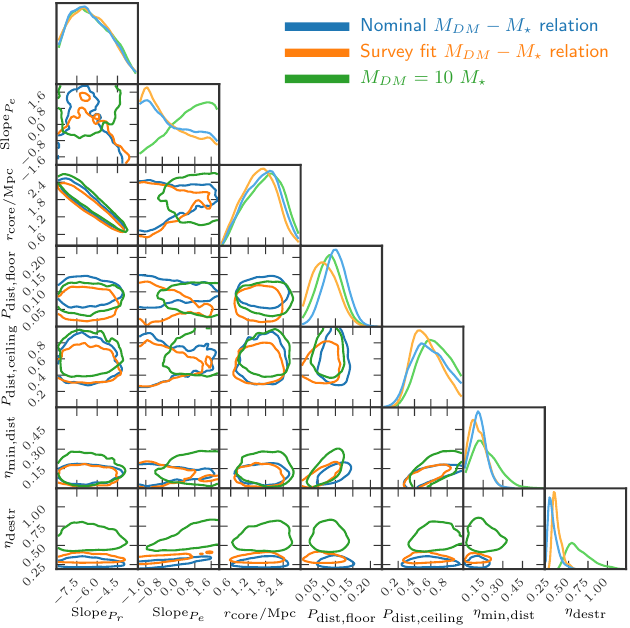}
	\caption{Triangle plot showing the inferred parameter values in the $\Lambda$CDM model constrained by our statistical analysis assuming the nominal dark matter fraction (Equation~\ref{M_dwarf_rule}; blue), our fit to the empirically determined dark matter fractions of nearby isolated dwarfs (Equation~\ref{revised_DM_frac}; orange), and a very conservative scenario in which the mass of dark matter within the optical radius of each dwarf is only $10\times$ that of the baryons (green).}
	\label{fig:triang_newDMfrac}
\end{figure}

In Fig.~\ref{fig:triang_newDMfrac}, we check how decreasing the dark matter fraction within the optical radius of the FDS dwarfs affects the results. In particular, we consider the revised dark matter fraction given in Equation~\ref{revised_DM_frac} based on the observed velocity dispersions of nearby dwarfs (Section~\ref{newDMfrac}). As discussed there, we also consider the very conservative case $M_{\textrm{DM}} = 10 \, M_{\star}$. The main impact is on the parameters $\eta_{\textrm{destr}}$ and $\eta_{\textrm{min, dist}}$. The inference on the slope of the eccentricity distribution is rather different for the case $M_{\textrm{DM}} = 10 \, M_{\star}$, but otherwise the posteriors are not much different to the nominal $\Lambda$CDM case in both revised analyses shown here.

\begin{figure*}
	\includegraphics[width = 8.5cm]{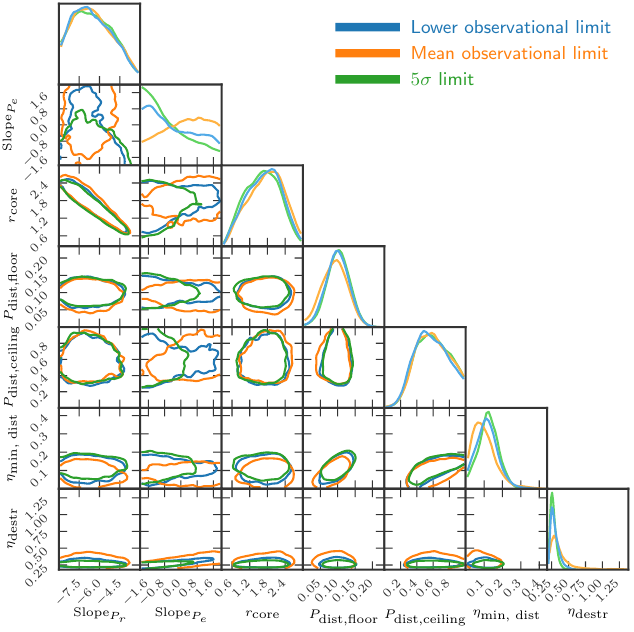}\hspace{0.07cm}
	\includegraphics[width = 8.5cm]{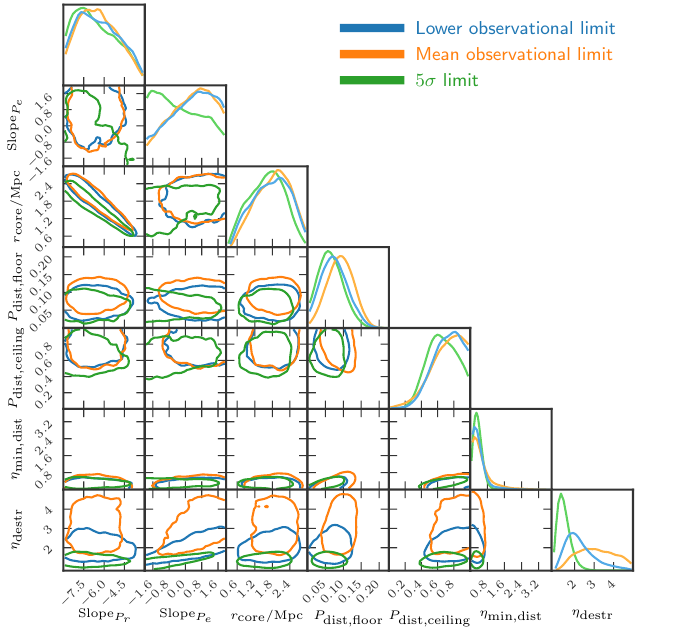}
	\caption{Triangle plot similar to Fig.~\ref{fig:triang_MONDLCDM}, but this time showing $\Lambda$CDM (left) and MOND (right) in separate panels. Each panel shows the difference in the final results when using three different lower limits to the density distribution of the simulated Fornax dwarfs. The results for the $5\sigma$ lower limit are shown in green, those with the nominal density limit are shown in blue, and results for the mean observational limit ($\rho_{\textrm{mean}}$) are shown in orange.}
	\label{fig:3dens}
\end{figure*}

In Fig.~\ref{fig:3dens}, we compare the parameter inferences resulting from the MCMC analysis assuming three different lower limits ($\rho_t$ values) to the density distribution of the dwarfs:
\begin{enumerate}
    \item The lowest considered $\rho_t$ is set at $5\sigma$ below the mean logarithmic density;
    \item The second-lowest considered $\rho_t$ is the nominal value used in the main analysis; and
    \item The highest considered $\rho_t$ is the mean observational limit ($\rho_{\textrm{mean}}$), which we obtained in Section~\ref{subsubsec:dwarf_density} and Appendix~\ref{dwarf_dens_LCDM} for MOND and $\Lambda$CDM, respectively.
\end{enumerate}

\begin{figure*}
	\includegraphics[width = 8.5cm]{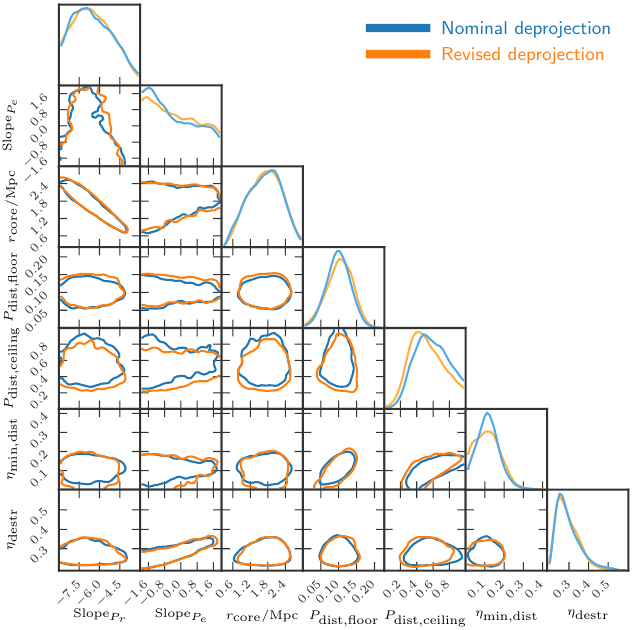}\hspace{0.07cm}
	\includegraphics[width = 8.5cm]{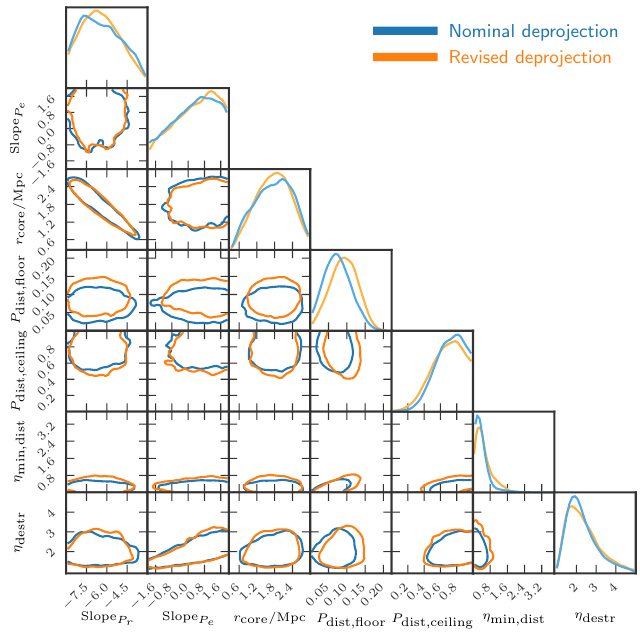}
	\caption{Similar to Fig.~\ref{fig:3dens}, but showing the difference in the final results for $\Lambda$CDM (left) and MOND (right) when two different sets of values for the parameters `offset' and `non-nucleated floor' are used to deproject distances (Appendix~\ref{deproj}). Results for the nominal deprojection (offset~$=0.4^\circ$, non-nucleated floor~$= 1.2^\circ$) are shown in blue, while results with the revised deprojection parameters (offset~$= 0.5^\circ$, non-nucleated floor~$= 1.5^\circ$) are shown in orange.}
	\label{fig:deproj}
\end{figure*}

In Fig.~\ref{fig:deproj}, we compare $\Lambda$CDM and MOND while assuming two different values for the deprojection parameters `offset' and `non-nucleated floor' (see Appendix~\ref{deproj}). In addition to the nominal values used in the main analysis, we also consider a higher set of values corresponding to the highest plausible 3D distance given the sky-projected distance \citep[see fig.~6 of][]{Venhola_2019}. This entails setting $\textrm{nnuc}_{\textrm{floor}} = 1.5^\circ$ and offset~$=0.5^\circ$ instead of the nominal $\textrm{nnuc}_{\textrm{floor}} = 1.2^\circ$ and offset~$=0.4^\circ$.

\begin{figure*}
	\includegraphics[width = 8.5cm]{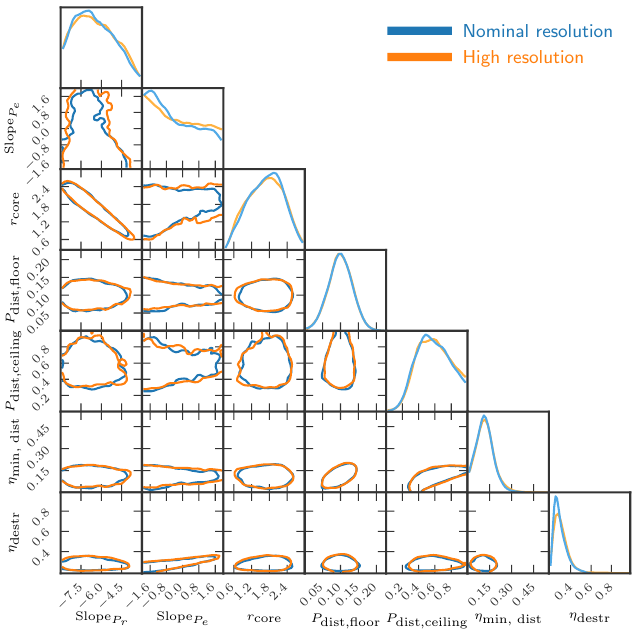}\hspace{0.07cm}
	\includegraphics[width = 8.5cm]{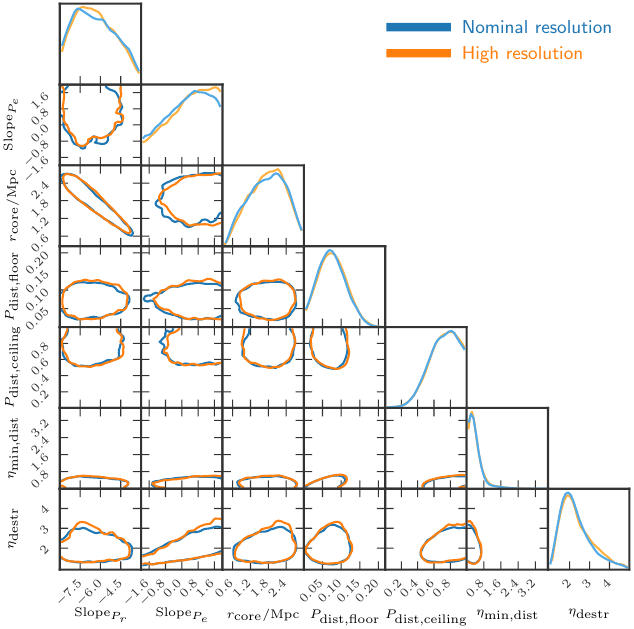}
	\caption{Similar to Fig.~\ref{fig:3dens}, but showing the difference in the final results for $\Lambda$CDM (left) and MOND (right) with two different resolutions in orbital elements. The nominal resolution case (blue) uses a grid of $100 \times 100$ bins to generate orbits with different eccentricities and initial positions/semi-major axes. The high-resolution case (orange) uses a grid of $200 \times 200$ bins.}
	\label{fig:highres}
\end{figure*}

In Fig.~\ref{fig:highres}, we check if increasing the resolution of the orbital elements in the test mass simulation affects the results for $\Lambda$CDM and MOND. The nominal resolution used is a grid of size $100 \times 100$ for the eccentricity $e$ and initial distance from the cluster centre ($R_i$), which also corresponds to the semi-major axis. In the higher resolution case, this is raised to $200 \times 200$.

\end{appendix}

\bsp
\label{lastpage}
\end{document}